\documentclass[11pt, a4paper,eqsecnum,nofootinbib,floats,article,aps,prd,floatfix,titlepage,superscriptaddress]{article} 
\usepackage{epsfig}
\usepackage{graphics}
\usepackage{bm}
\usepackage{amssymb}
\usepackage{amsfonts}
\usepackage{jheppub}
\usepackage{color}
\newcommand\fverb{\setbox\pippobox=\hbox\bgroup\verb}
\newcommand\fverbdo{\egroup\medskip\noindent%
			\fbox{\unhbox\pippobox}\ }
\newcommand\fverbit{\egroup\item[\fbox{\unhbox\pippobox}]}
\newbox\pippobox

\title{Random Field Theories in The Mirror Quintic Moduli Space}

\author[a]{Kate Eckerle}
\author[b]{and Brian Greene}

\affiliation[a]{Center for Theoretical Physics, Columbia University, \\
New York, New York 10027, USA}
\affiliation[b]{Center for Theoretical Physics and Department of Mathematics, Columbia University, \\
New York, New York 10027, USA}

\emailAdd{ke2176@columbia.edu}
\emailAdd{greene@physics.columbia.edu}

\abstract{We investigate the distribution of field theories that arise from the low energy limit of flux vacua built on type IIB string theory compactified on the mirror quintic. For a large collection of these models, we numerically determine the distribution of Taylor coefficients in a polynomial expansion of each model's scalar potential to fourth order, and show that they differ significantly from potentials generated by random choices of such coefficients over a flat measure.}

\begin{document}

\maketitle
\section{Introduction}

The enormous number of vacua scattered throughout the string landscape poses one of the most significant challenges for making contact between string theory and observable physics.
Generally speaking, three approaches for tackling this challenge have been advanced. The first focuses on the distribution of string vacua within the moduli space of geometrical compactificaitons, seeking mathematical structure that might entail patterns in physical observables emerging from the low energy sector of the theory. Seminal papers of this sort are \cite{Douglas,Denef:2004ze, denef-nonsusy, Denef:2004dm}. A second approach
has been to model the collection of low energy models arising from string compactifications as a space of random quantum field theories, with coefficients of all renormalizable terms drawn from a random distribution of perturbatively sensible values over a flat measure. The papers \cite{Arkani-Hamed, Distler, Chen:2011ac, tumbling, Marsh:2011aa, Marsh:2014} illustrate the types of conclusions that can be drawn using this approach. The third approach combines aspects of the the first two by
investigating the space of low energy field theories arising from string compactifications, and determining the degree to which this space is well modeled by random field theories drawn from a flat measure. That is, the third approach seeks nontrivial structure in the space of low energy string dynamics which can then be used to sharpen conclusions drawn from the first two approaches noted above. Examples of this approach include \cite{Danielsson:2006xw, Podolsky:2008du, Dine-Paban, Dine}. In this paper, we push forward on this third approach.

The third approach involves two competing influences. The starting point for low energy string dynamics, for definiteness type IIB supergravity in ten dimensions, is surely a theory which enjoys a great deal of structure. As we compactify this theory, preserving various amounts of low energy supersymmetry, we inject additional structure such as the rich  topology and geometry of Calabi-Yau compactifications. At the same time, the distribution of low energy string dynamics arising from such compactifications becomes substantially broader as masses and couplings depend sensitively on the detailed  topological and geometrical data of the compact manifold. This becomes all the more apparent when we adorn our compactifications with branes and fluxes, which introduce yet more degrees of freedom on which low energy properties depend. Collectively, then, one might imagine that notwithstanding the structure of supergravity and string geometry, the impact of varying the choices of fluxes in conjunction with the distinct locations of the vacua in moduli space associated with each such flux choice, would result in the low energy field theories arising from string theory being essentially random. As above, various authors, (those following ``approach two") have indeed relied on this perspective.

Nevertheless, papers in approach three have noted that even with the statistical tendency toward flattened distributions, residual structure in the the low energy dynamics can persist. For example, \cite{denef-nonsusy,  Marsh:2014} have shown that the distribution of mass eigenvalues fill out a non-random, and by now well-understood, mathematical pattern. The purpose of the current paper is to push this perspective further by considering higher order coefficients beyond the Hessian, and to determine the degree to which these terms are well-modeled -- or not -- by a random distribution over a flat measure. We will argue that, much as was found for mass eigenvalues, the third and fourth order terms in the low energy scalar potential retain non-random structure.

The analysis leading to this result is conceptually straightforward, albeit computationally technical. We begin in section \ref{CY-geom} by providing background on the various elements required for understanding low energy string dynamics arising from flux compactifications, focusing for definiteness on the mirror of the quintic hypersurface in $\mathbb{CP}^4$. We include this material for completeness and to set up notation; the reader familiar with the geometrical machinery of flux compactifications can skip the first three subsections of this discussion. The fourth and final background subsection, \ref{hessiansection}, reviews how flux compactifications give rise to a particular pattern of mass eignenvalues, illustrating in the simplest setting the kind of mathematical features relevant for the third approach. In section \ref{calc-approach} we lay out our calculational approach for generating a sample set of flux vacua, for computing the form of the low energy Lagrangian describing small fluctuations about such vacua up to fourth order, and for comparing these Lagrangians to those emerging from a random set of theories drawn from a flat measure. In section \ref{results} we provide our results and, in particular, reveal nontrivial structure in the third and fourth order coefficients. Finally in section \ref{Discussion} we summarize our results and suggest further directions for study.

\section{Background}\label{CY-geom}
\subsection{Review of Flux Compactifications}\label{compactification}

The low energy dynamics of the type IIB string is governed by the type IIB supergravity action, which provides our starting point \cite{Giddings:2001yu},
\begin{equation}
S_{IIB}=\frac{2\pi}{\ell_s^8}\left[\int_{}^{} d^{10} x \sqrt{-g^{10}}R^{10}-\frac{1}{2}\int_{}^{}\frac{d\tau\wedge\ast d\bar{ \tau}}{(\text{Im}(\tau))^2}+\frac{G_{(3)}\wedge\ast\bar{G}_{(3)}}{\text{Im}(\tau)}+\frac{\tilde{F}_{(5)}^2}{2}+C_{(4)}\wedge H_{(3)}\wedge F_{(3)}\right]+S_{loc}\label{10d-action},
\end{equation}
where $R^{10}$ is the $10$d Ricci Scalar in the Einstein frame, $G_{(3)}$ is the combined $3$-form flux, 
\begin{align}
G_{(3)}=F_{(3)}-\tau H_{(3)},
\end{align}
$\tau$ is the axio-dilaton related to the dilaton, $\phi$, by 
\begin{equation}
\tau=C_{(0)}+i e^{-\phi}
\end{equation}
and $F_{(p)}$ and $H_{(3)}$ are obtained from potentials $C_{(p-1)}$ and $B_{(2)}$,
\begin{align}
F_{(p)}&=dC_{(p-1)}\\
H_{(3)}&=dB_{(2)}.
\end{align}

This theory can be compactified on a Calabi-Yau $3$-fold to yield an effective action for the moduli fields, which describe how the compact manifold $\mathcal{M}$ varies from one spacetime location to another in the four large dimensions. Such parameters are complex valued and change continuously across the given family of Calabi-Yau, so they enter the 4d theory as complex scalar fields. It is instructive to sketch the derivation of the effective action, and give a very brief review of the geometry of Calabi-Yau moduli spaces.  In the process we introduce notation and summarize our strategy for generating an ensemble of random effective field theories. Experienced readers may wish to skip this section. 

Calabi-Yau moduli come in two different types: those associated with deformations of the manifold's complex structure, and those associated with deformations of its K{\"a}hler form, $J$. The former are in one-to-one correspondence with elements of the $(2,1)$-de Rham cohomology group, $H^{(2,1)}(\mathcal{M})$, and the latter with $H^{(1,1)}(\mathcal{M})$. We denote the dimension of these vector spaces by their Hodge numbers, $h^{2,1}$ and $h^{1,1}$, respectively.

Complexifying the K{\"a}hler form, we deal with a  moduli space of complex dimension $h^{(2,1)}+h^{(1,1)}$,  itself a K{\"a}hler manifold that factors locally into the direct product of two separate K{\"a}hler manifolds: one spanned by the complex structure moduli and the other spanned by the complexified K{\"a}hler moduli, with K{\"a}hler potential of the form,
\begin{equation}
\mathcal{K}^{cs}(z^1,z^2,...,z^{h^{2,1}})+\mathcal{K}^{ka}(w^1,w^2,...,w^{h^{1,1}}).
\end{equation}
Lower case indices ($a,b,c,...$) will refer to Calabi-Yau moduli. They are ordered from $1$ to $h^{2,1}+h^{1,1}$ running through all the complex structures first, followed by those of K{\"a}hler type. However, their range in certain expressions may be restricted to moduli of one of the two types. Most often this will be obvious from the context, but where there is the possibility for ambiguity we will state which if any moduli are excluded. 

The K{\"a}hler potential for the complex structure moduli is,
\begin{equation}
\mathcal{K}^{cs}(z^1,...,z^{h^{2,1}})=-\log\left(-i\int_{\mathcal{M}}\Omega\wedge\bar{\Omega}\right)
\end{equation}
where $\Omega$ is the holomorphic three-form of the Calabi-Yau manifold. It can be shown that differentiating $\Omega$ with respect to any of the complex structure moduli yields a component proportional to $\Omega$, and a remaining closed $(2,1)$-form. That is,
\begin{equation}
\frac{\partial \Omega}{\partial z^a}=k_a \Omega+\chi_a
\end{equation}
 with $\chi_a\in H^{(2,1)}(\mathcal{M})$. In particular the proportionality constant $k_a$ turns out to be,
\begin{equation}
k_a=-\mathcal{K}_a=-\partial_a \mathcal{K} .
\end{equation}

This allows us to construct a basis for $H^{(2,1)}$ by acting on the holomorphic $3$-form with a \emph{K{\"a}hler covariant} derivative, $D_a$, whose action on $\Omega$ is defined by,
\begin{equation}
D_a \Omega\equiv \chi_a=\partial_a \Omega+\mathcal{K}^{cs}_a\Omega.
\end{equation}
Furthermore, since 
\begin{equation}
\int_{\mathcal{M}}\Omega\wedge\frac{\partial \Omega}{\partial z^a}=0,
\end{equation}
such $(2,1)$-forms are orthogonal to $\Omega$.

We can now compute the term proportional to $ G_{(3)}\wedge \ast\bar{G}_{(3)}$ in the supergravity action upon compactification. This process amounts to taking the 10d spacetime to be the direct product of a four dimensional (noncompact) Lorentzian manifold, $M_4$, and a compact Riemannian one, $\mathcal{M}$, which for us is a Calabi-Yau 3-fold. We write,
\begin{equation}
M_{10}=M_4\times\mathcal{M}(z^1,...,z^{h^{2,1}},w^1,...,w^{h^{1,1}}),
\end{equation}
and perform the integration over $\mathcal{M}$ in the action. Since $\mathcal{M}$ is parameterized by the aforementioned moduli, and since the Calabi-Yau are allowed to vary across locations in $M_4$, performing the integral over $\mathcal{M}$ will yield an effective field theory involving moduli fields, $\phi^{a}(x_\mu)$.

Requiring Poincar\'e invariance in $M_4$ implies only $G_{(3)}$'s components with all indices in the compact dimensions may be nontrivial, and so
\begin{equation}
\int_{\mathcal{M}} G_{(3)}\wedge\ast\bar{G}_{(3)}=\int_{\mathcal{M}} G_{(3)}\cdot\thickspace\bar{G}_{(3)}.
\end{equation}
This is essentially a norm of a (for now general) closed $3$-form on $\mathcal{M}$. We may expand $G_{(3)}$ and $\bar{G}_{(3)}$ in an orthogonal basis for 
\begin{equation}
H^{(3)}(\mathcal{M})=H^{(3,0)}(\mathcal{M})\oplus H^{(2,1)}(\mathcal{M})\oplus H^{(1,2)}(\mathcal{M})\oplus H^{(0,3)}(\mathcal{M}),
\end{equation}
namely,
\begin{equation}
\Omega\thickspace \thickspace\text{,}\thickspace\thickspace \{\chi_a \}_{_{a=1}}^{h^{2,1}}\thickspace\thickspace \text{,}\thickspace \{\bar{\chi}_a \}_{_{a=1}}^{h^{2,1}}\thickspace\thickspace\text{ and }\thickspace\thickspace\bar{\Omega}
\end{equation}
allowing us to write,
\begin{equation}
\int_{\mathcal{M}} G_{(3)}\cdot\thickspace\bar{G}_{(3)}=\frac{i}{\int_{\mathcal{M}} \Omega\wedge\bar{\Omega}} \left(\int_{\mathcal{M}} G_{(3)}\wedge\bar{\Omega}\int_{\mathcal{M}} \bar{G}_{(3)}\wedge \Omega +\mathcal{K}^{a\bar{b}}\int_{\mathcal{M}} G_{(3)}\wedge\bar{\chi}_a \int_{\mathcal{M}} \bar{G}_{(3)}\wedge \chi_b\right).
\label{G3-expand}
\end{equation}

Each of these can be expressed in terms of covariant derivatives of the Gukov-Vafa-Witten superpotential, $W$, defined in terms of the $(3)$-form flux as,
\begin{equation}
W(z,\tau)=\int_{\mathcal{M}}\Omega\wedge G_{(3)}.\label{superpotential}
\end{equation}
The second term on the right hand side of eq. \ref{G3-expand} involves K{\"a}hler covariant derivatives of the superpotential with respect to the complex structure moduli (because it is built out of $(2,1)$-forms). It is proportional to,
\begin{equation}
{\mathcal{K}^{}}^{a\bar{b}}D_aW \bar{D}_{\bar{b}}\bar{W}.\label{DWsq-moduli}
\end{equation} 

As is standard, we can define a ``K{\"a}hler potential" for the axio-dilaton such that the first term in on the right hand side of eq. \ref{G3-expand} has the same form as eq. \ref{DWsq-moduli}, i.e. so it is $\sim|D_\tau W|^2$. Specifically, we choose
\begin{equation}
\mathcal{K}^{ax}=-\log(-i(\tau-\bar{\tau}))\label{axioKpotential},
\end{equation}
and 
\begin{equation}
{\mathcal{K}^{ax}}^{\tau \bar{\tau}}=(\mathcal{K}^{ax}_{\tau \bar{\tau}})^{-1}=(\partial_\tau\partial_{\bar{\tau}}\mathcal{K}^{ax})^{-1}.
\end{equation}
One makes this choice because 
\begin{equation}
\frac{1}{(\bar{\tau}-\tau)}\int_{}^{} G_{(3)}\wedge\bar{\Omega}=\left(\partial_\tau-\frac{i}{\tau-\bar{\tau}}\right)W=\left(\partial_\tau+\partial_\tau \mathcal{K}^{ax}\right)W\equiv D_\tau W,
\end{equation}
and so first term in \ref{G3-expand} is proportional to
\begin{equation}
|D_\tau W|^2={\mathcal{K}^{ax}}^{\tau \bar{\tau}}D_\tau W\bar{D}_{\bar{\tau}}\bar{W},
\end{equation} 
which parallels the form arising for the other complex structure moduli, reflecting the relationship of type IIB string theory to F-theory in which the axio-dilaton explicitly becomes another complex structure modulus.  

Notationally, to include the  axio-dilaton as as additional modulus we use new indices -- capital letters -- that begin from zero, the index value reserved for the axio-dilaton. We denote the full K{\"a}hler potential by $\mathcal{K}$. It is the sum of all three pieces, $\mathcal{K}^{cs}$, $\mathcal{K}^{ax}$ and $\mathcal{K}^{ka}$.
The result, then, of dimensionally reducing the $3$-form flux term is
\begin{equation}
\frac{2\pi}{\ell_s^8}\frac{i}{2\text{Im}(\tau) \thickspace \int_{\mathcal{M}} \Omega\wedge\bar{\Omega}}\mathcal{K}^{I \bar{J}}D_I W \bar{D}_{\bar{J}}\bar{W}=\frac{2\pi}{\ell_s^8}e^{\mathcal{K}^{cs}+\mathcal{K}^{ax}}\mathcal{K}^{I \bar{J}}D_I W \bar{D}_{\bar{J}}\bar{W}
\end{equation}
where the K{\"a}hler moduli are excluded.
 
The kinetic terms for all the Calabi-Yau moduli come from the Einstein Hilbert term in the 10d action. They are noncanonical. We identify where they come from as well compute the total relative factor between the kinetic and potential terms which will involve one remaining expression given in terms of the volume of the compactification manifold. This is meant as a qualitative description. 
First we decompose the 10d curvature scalar into the trace of the noncompact component of the Ricci tensor, that of the compact component (which is zero because Calabi-Yaus are Ricci flat), and the remaining terms which will involve products of the metric and its derivatives with indices in both the compact manifold and large dimensions, which we label $R^{\text{mix}}$,
\begin{equation}
R^{10}=R^4+R^6+R^{\text{mix}}.
\end{equation}

Since $R^4$ is a constant over the Calabi-Yau, integration of it over $\mathcal{M}$ yields a factor of the volume of the Calabi-Yau. The factor then in front of the resulting 4d Einstein-Hilbert term is 
\begin{equation}
\frac{2\pi}{\ell_s^8}\text{Vol}(\mathcal{M})=\frac{2\pi}{\ell_s^2}\mathcal{V}_0=\frac{M_p^2}{2}4\pi \mathcal{V}_0
\end{equation}
where we've defined the dimensionless constant $\mathcal{V}_0$, the volume of the Calabi-Yau manifold in string units. Since the string length is the fundamental length scale at which one will see string modes, the volume of the Calabi-Yau must be large compared to $\ell_s^6$ for the direct compactification procedure we are employing to be valid. 

In order to have a canonical Einstein-Hilbert term in the effective action one must rescale the 4d metric so that the curvature rescales precisely with a factor of $\frac{1}{4\pi\mathcal{V}_0}$. The new curvature term also comes with kinetic terms for the volume modulus because the volume, and thus the factor by which the 4d spacetime metric is rescaled, may be expressed in terms of the volume modulus, $\rho$. 
\footnote{{The imaginary part of $\rho$ cubed is proportional to the volume squared.}}

Incidentally, the term in eq. \ref{10d-action} which clearly yields kinetic terms for the axio-dilaton, namely,
\begin{equation}
\sim\int_{}\frac{ d\tau\wedge\ast d\bar{\tau}}{(\text{Im}(\tau))^2},
\end{equation}
arises in precisely the same manner; specifically from transforming from the string metric to the Einstein metric by rescaling the string metric by $e^{\phi/2}$. The resulting kinetic terms for $\rho$ and $\tau$ are noncanonical, specifically given by,
\begin{equation}
 \frac{M_p^2}{2}\int d^4x\thickspace \frac{3}{(\rho-\bar{\rho})^2}\partial_\mu \rho \partial^\mu \bar{\rho}+ \mathcal{K}^{ax}_{\tau \bar{\tau}}\partial_\mu \tau \partial^\mu \bar{\tau}
 \label{rho-tau-kin}
\end{equation}
$\rho$ is not itself one of the Calabi-Yau moduli denoted by our indices $a,b,..$. Rather it is a specific function of \emph{all} the K{\"a}hler moduli. We shall shortly see that their noncanonical kinetic terms (involving the contraction with their K{\"a}hler metric) reside in that for $\rho$ in eq. \ref{rho-tau-kin}.

The kinetic terms for the complex structure moduli come from integration of $R^{\text{mix}}$ over $\mathcal{M}$, and so are also generally noncanonical involving contraction with their respective K{\"a}hler metric as follows,
\begin{equation}
\frac{M_p^2}{2}\int d^4 x {\mathcal{K}}_{a\bar{b}}\partial_\mu \phi^a \partial^\mu \bar{\phi}^b\label{kinetic-term-moduli}
\end{equation}

We can thus write the effective action describing the moduli,
\begin{equation}
S_{eff}=\frac{M_p^2}{2}\int_{}^{}d^4 x \mathcal{K}_{I \bar{J}}\partial_\mu \phi^I\partial^\mu \bar{\phi}^J  -\frac{M_p^2}{4\pi }\frac{e^{\mathcal{K}^{cs}+\mathcal{K}^{ax}}}{\mathcal{V}_0^2}\left( \mathcal{K}^{I \bar{J}}D_{I} W \bar{D}_{\bar{J}}\bar{W}\right).\label{eff-action}
\end{equation}
Any consistent flux compactification of type II string theories on Calabi-Yau manifolds requires the addition of negative tension localized objects. This is necessary in order to satisfy $\tilde{F}_{(5)}$'s equation of motion which, when integrated over the Calabi-Yau, yields the following tadpole condition,
\begin{equation}
\frac{1}{\ell_s^4}\int_{\mathcal{M}}F_{(3)}\wedge H_{(3)}+Q^{loc}_{3}=0\label{tadpole}.
\end{equation}
This is effectively a statement of the consistency of the configuration of field lines wrapping the Calabi-Yau $3$-cycles (i.e. field lines in the small dimensions curl and close onto themselves, while those in large dimensions end on mathematically valid sources). 

It can be shown that the term in eq. \ref{tadpole} involving the R-R and NS-NS fluxes is positive definite. Of 
the allowed localized objects that preserve Poincar\'e invariance in the four large dimensions and can act as sources for the fluxes, only the O3-planes contribute a negative charge to the total $Q^{loc}_{3}$, thus they must be included in the compactifation in order to cancel all the remaining positive definite terms in eq. \ref{tadpole}. Though the O3-planes are not dynamical, they do in general impact the moduli space geometry. We assume a model in which such effects are negligible. O3-planes also reduce the $\mathcal{N}=2$ supersymmetry we began with to $\mathcal{N}=1$. 

Generic theories with $\mathcal{N}=1$ supersymmetry involve complex scalars described by a potential of the form,
\begin{equation}
V=e^\mathcal{K}\left( \mathcal{K}^{a \bar{b}}D_a W \bar{D}_{\bar{b}}\bar{W}-3|W|^2\right),\label{ScalarPotential}
\end{equation}
where the superpotential, $W$, is a holomorphic function of the complex scalars $\{\phi_a\}$. 
Notice that the $-3|W|^2$ term is absent in our effective action. This `no-scale' form arises
from a simple but general cancellation inherent to Calabi-Yau compactifications at the classical level. Namely, because the classical superpotential is independent of the K{\"a}hler moduli, the K{\"a}hler dependence of the scalar potential arises solely from the contribution to $\sim|DW|^2$ from
\begin{equation}
\mathcal{K}^{a \bar{b}}\mathcal{K}_{a}W\mathcal{K}_{\bar{b}}\bar{W}=\mathcal{K}_a\mathcal{K}^a |W|^2
\end{equation}
with indices running over the K{\"a}hler moduli only. The classical expression for their K{\"a}hler potential (i.e. that which comes from the special geometry) is,
\begin{align}
\mathcal{K}^{ka}&=-2\log\left(\int_{\mathcal{M}}J\wedge J\wedge J\right)\\
&=-2\log\left(\frac{1}{\ell_s^6}\int_{\mathcal{M}} dV\right)=-\log(\mathcal{V}_0^2)\\
&=-3\log\left(-i(\rho-\bar{\rho})\right).\label{kahler-potential}
\end{align}

So,
\begin{align}
\mathcal{K}_a \mathcal{K}_{\bar{b}}\mathcal{K}^{a \bar{b}}&= \partial_a \rho \partial_{\bar{b}} \bar{\rho} \mathcal{K}_{\rho} \mathcal{K}_{\bar{\rho}} \times\frac{1}{\partial_{\rho}\partial_{\bar{\rho}} \mathcal{K} }(\partial_{\rho} \phi^a \partial_{\bar{\rho}} \bar{\phi}^{b} )\\
&=\mathcal{K}_{\rho} \mathcal{K}_{\bar{\rho}} \mathcal{K}^{\rho\bar{\rho}}
\end{align}
which gives,
\begin{equation}
\mathcal{K}_a\mathcal{K}^a=\frac{-3}{\rho-\bar{\rho}}\frac{+3}{\rho-\bar{\rho}}\frac{-(\rho-\bar{\rho})^2}{3}=+3.
\end{equation}
This then yields the cancellation
\begin{equation}
\mathcal{K}^{a \bar{b}}\mathcal{K}_{a}W\mathcal{K}_{\bar{b}}\bar{W} - 3|W|^2 = 0.
\end{equation}

\begin{figure}
\centering
\includegraphics[width=.65\textwidth]{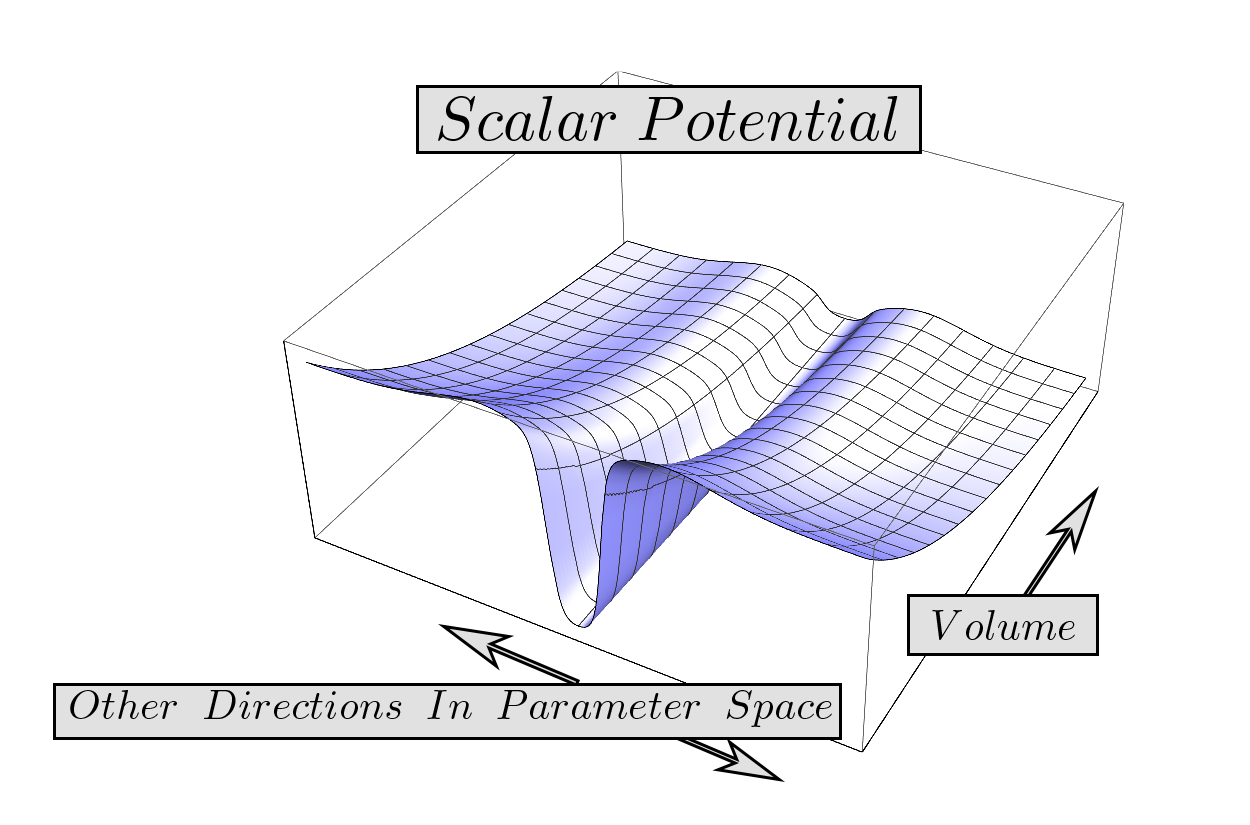}
\caption{The qualitative dependence of the scalar potential on the volume of a Calabi-Yau manifold.}  
\label{ski-slope}
\end{figure}

The resulting scalar potential is positive semi-definite, and so its zeros are its global minima. Solutions of the SUSY condition, $D_IW=0$ for all $I=0,1,...,h^{2,1}$, are the only zeros because the metric and $e^{\mathcal{K}}$ are positive definite. In general, when $V\neq0$ the scalar potential depends on the K{\"a}hler moduli through its overall dimensionful factor, $M_p^2/4\pi \mathcal{V}_0^2$, but when $V = 0$, all such dependence drops out. The flattening of the potential at a zero in the volume direction of parameter space  is shown schematically in Figure \ref{ski-slope}.

We see too that the $\rho$-dependent factor in the kinetic term for the volume modulus in eq. \ref{rho-tau-kin} is indeed its K{\"a}hler metric, $\mathcal{K}_{\rho\bar{\rho}}$. We may identify this term as the net kinetic term for the K{\"a}hler moduli in eq. \ref{eff-action}, similarly by the chain rule. Finally, we recognize $1/\mathcal{V}_0^2$ in eq. \ref{eff-action} as $e^{\mathcal{K}^{ka}}$, and thus, the effective action we obtain upon compactification as that of a theory with $1+h^{2,1}+h^{1,1}$ complex scalars and $\mathcal{N}=1$ supersymmetry with an additional/non-generic feature. Namely, the cancellation of the $-3|W|^2$ in the potential by the contribution from a subset of the scalars, specifically $h^{1,1}$ of them. The feature is entirely due to the fact we've compactified on a Calabi-Yau manifold and used only classical expressions.

These are, generally speaking, subject to both  $\alpha'$ and $g_s$ corrections.  In type IIB, the complex structure K{\"a}hler potential is protected from both types, but the K{\"a}hler moduli  are not shielded from either. However, as is well known, such corrections are suppressed in the large volume limit, \footnote{This limit is one in which not only the $6$-volume but all subvolumes are large compared to the natural sizes (involving the dimensionful constants). } as are the instanton corrections the superpotential receives. This setting also ensures backreaction of the fluxes on the geometry of the manifold is subdominant. We will work in this regime and so now use the formulae we've reviewed to set up explicit calculations on the mirror quintic.

\subsection{Period Integrals}\label{periods-sect}
To search for local minima of the effective potential and compute its Taylor coefficients in the expansion about these minima one must express the quantities in eq. \ref{eff-action} as explicit functions of the complex structure(s) and axio-dilaton. To accomplish this we need only express $W$ and $\mathcal{K}^{cs}$ in this fashion, as all terms in eq. \ref{eff-action} are obtained from them. Generally, the integrals over the compactification manifold need not be computed directly. Rather they can be expressed in terms of a basis of systematically calculable functions, the \emph{period integrals} of the Calabi-Yau manifold, which are solutions to differential equations (specific to the compactification manifold) known as the Picard-Fuchs equations. 

By the Poincar\'e dulaity $H^{(3)}(\mathcal{M})$ is isomorphic to $H_{(3)}(\mathcal{M})$, the space of nontrivial $3$-cycles. Thus, for any two closed $3$-forms $\alpha_{(3)}$ and $\beta_{(3)}$ there exist two $3$-cycles $A$ and $B$ such that,
\begin{equation}
\int_\mathcal{M}\alpha_{(3)}\wedge\beta_{(3)}=\int_A \beta_{(3)}=\int_B \alpha_{(3)}.\label{poincare}
\end{equation}
If $\{ C_i\}$ are a basis of $3$-cycles, the right hand sides of eq. \ref{poincare} are a linear combination of the integrals of the relevant $3$-form over the basis cycles. 
\begin{align}
\int_A \beta_{(3)}&=\sum_{i=1}^{h^3} A^i \int_{C_i}\beta_{(3)}\\
\int_B \alpha_{(3)}&=\sum_{i=1}^{h^3} B^i \int_{C_i}\alpha_{(3)}
\end{align}
where the $A^i$ and $B^i$ are real numbers (the components of $A$ and $B$ in the $C_i$ basis). Thus, $W$ can be expressed as a linear combination of the integrals of the holomorphic $3$-form over the basis cycles for $H_{(3)}(\mathcal{M})$. These are known as the \emph{period integrals}, or \emph{period functions}. They are functions of the complex structure moduli only. 

We note the existence of an integral and symplectic basis. The first of these properties means $C_i$ is a geometrical cycle (that is, a submanifold, not merely a formal object defined as the dual to a $3$-form). The second means each basis cycle intersects only one other basis cycle, and does so exactly one time. We denote the period functions in this basis as follows,  
\begin{equation}
\Pi_i(z^1,...z^{h^{2,1}})=\int_{C_i}\Omega
\end{equation}
where the index $i$ ranges from zero to $2h^{2,1}+1$ for a total of $h^3$ different periods. The symplectic basis is the one most natural for us because the period functions have well defined expansions about special points in the moduli space where vacua accumulate, as we shall discuss at greater length shortly.

The intersection form allows us to express the effective action in terms of these natural period functions. Two $3$-cycles intersect at points in a $6$-dimensional manifold. Since cycles are oriented such points will have multiplicity $\pm 1$. The intersection form, in the context where the $(3)$-homology groups are the domain, takes in two $3$-cycles and sums the intersection multiplicities. In light of the Poincar\'e duality this is equally viewed as a map from two copies of the $(3)$-cohomology groups. That is, we write
\begin{align}
Q_{ij}&=Q(C_i,C_j)=\langle C_i \smile C_j[\mathcal{M}] \rangle\\
&=\tilde{Q}_{ij}=\tilde{Q}(\alpha_i,\alpha_j)=\int_\mathcal{M}\alpha_i\wedge\alpha_j.
\end{align}
In an integral and symplectic basis $Q_{ij}$ are the entries of a symplectic $h^3\times h^3$-matrix. 

The superpotential, $W,$ can now be expressed as follows,
\begin{align}
W=\sum_{i=0}^{4}G^i \Pi_i(z)\\
= (F-\tau H)\cdot\Pi(z)\label{FluxVectors}
\end{align}
where $F$ and $H$ are row vectors whose four entries indicate the quantity of R-R and NS-NS flux wrapping the basis cycles, and $\Pi(z)$ is a column vector containing the $h^3$ period functions. It can be shown that the $3$-form fluxes wrapping the integral and symplectic basis cycles are integrally quantized in units of $4\pi ^2\alpha'$,
\begin{equation}
\frac{1}{2 \pi\alpha'}\int_{C_i} F_{(3)}\in 2\pi \mathbb{Z}
\end{equation}
and similarly for $H_{(3)}$. Since the overall dimensionful factor has been pulled outside the potential, this amounts to requiring the entries of the $F$ and $H$ vectors in eq. \ref{FluxVectors} be integers.  

The K{\"a}hler potential for the complex structure modulus is expressed in terms of the period functions as follows,
\begin{align}
\mathcal{K}^{cs}(z,\bar{z})=-\log(i\Pi^\dag(\bar{z}) Q^{-1} \Pi(z))\label{KahlerPotentialCS}
\end{align}
In evaluating these functions, one can avoid performing an integration over the compactification manifold because the periods are solutions to particular differential equations, the Picard-Fuchs equations (associated with the given Calabi-Yau). Given the above expressions for the superpotential and K{\"a}hler potential, one need only find the solutions to these differential equations to write down an explicit effective action for the moduli. 

The complexity of the Picard-Fuchs equations quickly mounts as the number of moduli increase. We consider the simplest case, where $h^{2,1}=1$, and so there are a total of four period functions. There is a complete list 14 such compactifications, the most well known being the mirror quintic. For these 14 models the Picard-Fuchs equation takes the following form,
\begin{equation}
\left[\delta^4-z(\delta+\alpha_1)(\delta+\alpha_2)(\delta+\alpha_3)(\delta+\alpha_4)\right]u(z)=0\label{PF}
\end{equation}
where $\delta\equiv z\frac{d}{dz}$, and the $\alpha_j$ are rational numbers specific to the compactification (the mirror quintic has $\alpha_j=j/5$). 

A convenient basis for expressing solutions to this ODE, which we shall label $\{ U_i(z)\}_{i=0}^{3}$, is as follows \cite{Ahlqvist:2010ki}
\begin{align}
&U_0(z)=c\thickspace G_{4,0}^{1,3}(-z;\{1-\alpha_1,1-\alpha_2,1-\alpha_3,1-\alpha_4\},\{0,0,0,0\})\label{MeijerGs0}\\
&U_1(z)=\frac{c}{2\pi i}G_{4,0}^{2,2}(z;\{1-\alpha_1,1-\alpha_2,1-\alpha_3,1-\alpha_4\},\{0,0,0,0\})\label{MeijerGs1}\\
&U_2^-(z)= \frac{c}{(2 \pi i)^2}G_{4,0}^{3,1}(-z;\{1-\alpha_1,1-\alpha_2,1-\alpha_3,1-\alpha_4\},\{0,0,0,0\})\nonumber\\
 &U_3(z)=\frac{c}{(2 \pi i)^3}G_{4,0}^{4,0}(z;\{1-\alpha_1,1-\alpha_2,1-\alpha_3,1-\alpha_4\},\{0,0,0,0\})\label{MeijerGs3}\\
&U_2(z)=\left\{
\begin{array}{ll}
      U_2^-(z) &  \text{Im}(z)\leq 0 \\
      U_2^-(z) - U_1(z)& \text{Im}(z) > 0\\
\end{array} \label{MeijerGs2}\right.
\end{align}
The $G^{m,n}_{p,q}$ are Meijer-G functions defined in terms of contour integrals in the complex, say, $s$-plane,
\begin{equation}
G^{m,n}_{p,q}(z;\{a_1,...a_p\},\{b_1,...,b_q\})=\frac{1}{2\pi i}\int_{L}ds \frac{\Pi_{j=1}^{m}\Gamma(b_j-s)\Pi_{j=1}^{n}\Gamma(1-a_j+s)}{\Pi_{j=m+1}^{q}\Gamma(1-b_j+s)\Pi_{j=n+1}^{p}\Gamma(a_j-s)} z^s \label{meijerGdefinition}
\end{equation}
where $c$ is a constant specific to the given Calabi-Yau (one of the 14 models),
\begin{equation}
c=\frac{1}{\Gamma(\alpha_1)\Gamma(\alpha_2)\Gamma(\alpha_3)\Gamma(\alpha_4)}.
\end{equation}

The particular linear combinations of the $U_i(z)$ that yield the periods in the integral symplectic basis, the $\Pi_i(z)$'s, are fixed by the calculable monodromy transformations of the homology cycles when transported about certain special points in the moduli space. For the case of $h^{2,1}=1$ there are three such special points: the large complex structure point (which corresponds to $z=0$ in our coordinates), the conifold point ($z=1$) and the Landau-Ginsburg point ($z=\infty$). These nontrivial monodromy transformations of the $3$-cycles in turn yield nontrivial transformations for the corresponding period functions. 

For instance, if we donate the shrinking sphere as the conifold is approached by $C_3$, and the cycle it intersects by $C_0$, then
\begin{align}
Q_{03}=\langle C_0 \smile C_3,[\mathcal{M}] \rangle\rightarrow&\langle C_0+n C_3 \smile C_3,[\mathcal{M}]\rangle\\
&= \langle C_0 \smile C_3,[\mathcal{M}] \rangle+n \langle C_3 \smile C_3,[\mathcal{M}] \rangle\\
&=Q_{03}+n*0=Q_{03}.
\end{align}
The integer $n$ is specified by requiring mutual consistency between all the monodromy transformations in the mirror quintic's moduli space, and as is well-known, this requires $n= 1$.
The monodromy transformations imply that the linear combinations of the aforementioned solutions to the Picard-Fuchs equation, $\{ U_i\}$, that correspond to the period integrals in a symplectic basis for the mirror quintic are given by, 
\begin{align}
 \Pi_0(z)&= U_0(z)\label{periods-0}\\
 \Pi_1(z)&=- U_1(z)\label{periods-1}\\
 \Pi_2(z)&=3 U_1(z)-5 U_2(z)\label{periods-2}\\
 \Pi_3(z)&= 5 U_1(z)+5 U_3(z)\label{periods-3}
\end{align}
where $\Pi_3$ is the (analytic) period that vanishes at the conifold point, it's partner, $\Pi_0$, picks up a copy of $\Pi_3$ for each revolution about the conifold point, and the remaining periods are analytic and nonvanishing. For a detailed derivation including the general form for any of the 14 one parameter models see, for instance, Appendix A of \cite{Ahlqvist:2010ki}.

The transformations of the periods upon circling a given special point in the moduli space fix their expansions in the neighborhood of the special point. In the case of the conifold point the transformation,
\begin{equation}
\Pi_0(z)\rightarrow \Pi_0(z)+\Pi_3(z)
\end{equation}
for each revolution $z\rightarrow (z-1)e^{2 \pi i} +1$, implies 
\begin{equation}
\Pi_0(z)=\Pi_3(z)\frac{\log (z-1)}{2 \pi i}+f(z)
\end{equation}
where $f(z)$ is analytic and nonvanishing at the conifold point. The expansions of the period functions are discussed in detail in the following section. For now we remark that the branch cut for $\Pi_0$ introduces a branch point singularity in first derivative of $\Pi_0$ which in turn results in a singularity in the K{\"a}hler metric at the conifold point.

We also adopt the standard convention (see, e.g.,  \cite{Ahlqvist:2010ki} for details) where the entries of the period vector, $\Pi(z)$, are given in descending order,
\begin{equation}
\Pi(z)= \left( \begin{array}{c}
\Pi_3(z) \\
\Pi_2(z) \\
\Pi_1(z)\\
\Pi_0(z) \end{array} \right)
\end{equation}
while those in the flux vectors are labeled in ascending order,
\begin{align}
F&= \left( \begin{array}{cccc}
F_0,& F_1,&F_2,&F_3  \end{array} \right)\\
H&= \left( \begin{array}{cccc}
H_0,& H_1,&H_2,&H_3  \end{array} \right).
\end{align}

With this review of notation and conventions, all functions in the effective action have now been specified. The only free parameters are the fluxes, which for us consist of eight integers (four R-R and four NS-NS fluxes). So, by randomly selecting a set of eight integers, constructing the corresponding scalar potential, searching for local minima (in the $z-\tau$ field space), and evaluating the potential's Taylor coefficients about the local minima so identified, one obtains the masses and couplings of a sample of effective field theories in the landscape of the mirror quintic. Since this model's only dynamical fields are the axio-dilaton and the mirror quintic's sole complex structure modulus there are a total of four real degrees of freedom. 

For the masses and couplings to have physical significance one must trade the $\{z,\tau,\bar{z},\bar{\tau}\}$ basis for one that simultaneously diagonalizes the Hessian of the scalar potential, and yields kinetic terms that are canonical (i.e. the K{\"a}hler metric evaluated at the vacuum is the identity). This transformation and several other technicalities are discussed in detail in section \ref{calc-approach}, but here we finish outlining the strategy in broad strokes. 

To minimize a function numerically we must begin by providing a guess for the vacuum location. Vacua are known to accumulate near the aforementioned special points in the moduli space, especially near the conifold point. We focus our search there. Moreover, we look specifically for zeros of the scalar potential, which are solutions to the SUSY condition $D_z W=D_\tau W=0$. This restriction both dramatically reduces the computational expense of searching for vacua by decreasing the (real) dimension of the space over which the function needs to be minimized from four to two, as well as enables us to compute a guess location given a choice of fluxes (which is essential for numerical minimization). These are not SUSY vacua in the traditional sense because we do not require that the superpotential itself vanish at the vacua. 

The first of these simplifications is due the fact that the two SUSY conditions imply that the vacuum value of the axio-dilaton $\tau_{SUSY}$ for a given choice of fluxes is an explicit function of the complex structure vacuum location. In particular,
\begin{equation}
\tau_{SUSY}=\frac{F\cdot \bar{\Pi}(\bar{z}_{SUSY})}{H\cdot \bar{\Pi}(\bar{z}_{SUSY})}\label{taususy}
\end{equation}
So, we evaluate the axio-dilaton in the function we seek to minimize, $|D_z W|^2$, at $\tau=\tau_{SUSY}(z)$. We then need only minimize over the variation of two real fields (the real and imaginary parts of $z$). The guess location, $z_{guess}$, for the given set of fluxes can be computed straightforwardly by using the near conifold period expansions in the period vectors and the K{\"a}hler potential below,
\begin{equation}
D_z W(z, \tau_{SUSY}(z))\negthickspace=\left(F-\frac{F\cdot \bar{\Pi}(\bar{z})}{H\cdot \bar{\Pi}(\bar{z})}H\right)\cdot(\Pi'(z)+\mathcal{K}_{z}\Pi(z))=0\label{minimization},  
\end{equation}
We compute the leading order solution to the above, which amounts to retaining the $\log(z-1)$ and constant terms, and dropping everything $\mathcal{O}(z-1)$. The resulting $z_{guess}$ is given in terms of the flux integers and period expansion coefficients in section \ref{calc-approach}.

To proceed further, it is essential to have high accuracy approximations to the period functions near the conifold point. The Meijer-G functions, with respect to which the periods and their derivatives can be expressed, are generally slow to evaluate numerically. As the singularities of the Meijer-G's are approached (for example the branch point singularity for $\Pi_0'(z)$, and terms $\sim \frac{1}{(z-1)^{k-1}}$ for its $k^{\rm th}$ order derivative) this becomes a significant obstacle. We not only evaluate such expressions multiple times while searching for a single vacuum, but we then must compute the Taylor coefficients at the near conifold vacuum found. This will involve many additional evaluations of increasingly divergent (due to the derivatives taken) Meijer-G's near their singularities. The tremendous number of times the search algorithm needs to be run to find a sufficiently large random sample of vacua, and the subsequent computation of the Taylor coefficients makes it essential to have high accuracy fast approximations to the period functions near the conifold point.

Additionally, we note that such approximations are also needed near the large complex structure point ($z=0$). The minima we are searching for typically have basins of attraction that narrow sharply near the minimum. Though a particular set of fluxes may yield a guess in the neighborhood of the conifold point, and so be worthy of pursuing, the guess may lie far up the minimum's basin outside the basin's thin throat. Iterative minimization procedures work by taking a steps in the direction of the gradient of the function being minimized. A narrow basin that then flattens out can result in significant overshooting of the minimum during the first steps. The search region needs to be large enough to contain these initial sweeps as it ping-pongs around the minimum, and eventually spiral into it. 
 
\begin{figure}
\centering
\includegraphics[width=.65\textwidth]{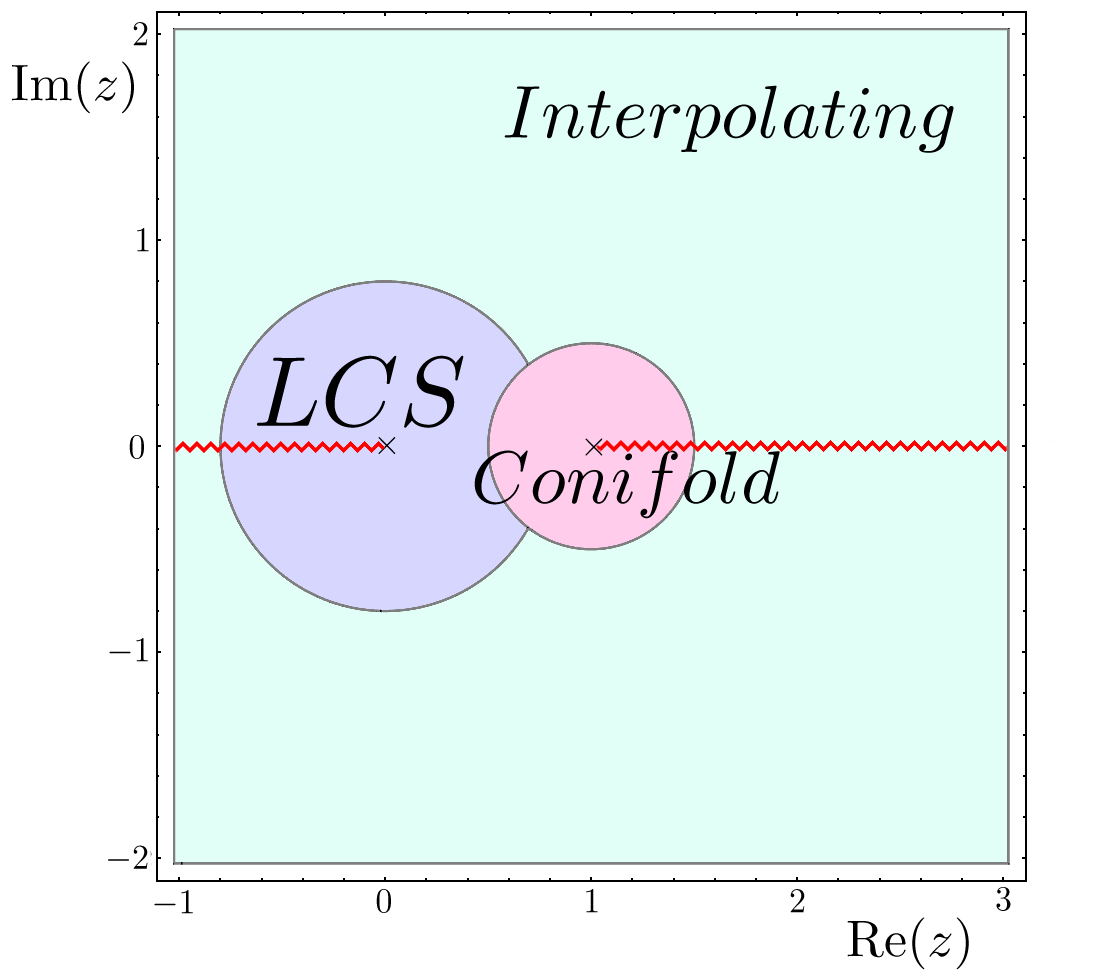}
\caption{We search for no scale vacua in the square portion of the complex plane for $z$ depicted above. The three regions where we use different fast approximations to the period functions are shown using different colors. The near-conifold patch consists of the disk of radius $0.5$ centered at the conifold, $z=1$. The portion of the disk of radius $0.8$ centered at the LCS point, $z=0$, that is not contained within the near-conifold region is shown in purple. Here we use the $12$th order expansions about $z=0$ obtained directly from Mathematica. Lastly, an interpolating function built from discrete Meijer-G data is used in the remaining portion of the square search region, shown in light green. Branch cuts are indicated by the red zigzag lines, with the one emanating from the conifold point along the positive real axis applying to $\Pi_0$, and those emanating along the negative real axis from the LCS point of relevance to all periods excluding $\Pi_0$.} 
\label{patched}
\end{figure}

The surrounding buffer area we need includes the large complex structure point. The behavior of the periods here is well known, $\Pi_i\rightarrow (z\log(z))^i$. Due to the branch cuts in the periods, many of the Meijer-G's in the expressions we seek to minimize are singular. So, when searching for vacua we use ``patched period functions"-- piecewise defined fast approximations to the exact expressions in terms of the Meijer-G's. Outside the neighborhoods of both the large complex structure point and conifold point (where the expansions are used), we build interpolating functions by evaluating the exact periods on a grid. The entire search region showing the neighborhoods where each of the three type of approximations to the period functions are used is found in Figure \ref{patched}. We postpone further discussion of the search algorithm until the Calculational Approach section, and now turn to the computation of the fast approximations to the period functions. 

\subsection{Period Expansions for the Mirror Quintic}\label{periods-subsection}
Three of the mirror quintic's four period integrals (in the integral and symplectic basis) are analytic in the neighborhood of the conifold point. These are the intersecting pair $\Pi_1(z)$ and $\Pi_2(z)$ which are nontrivial at the conifold point, and $\Pi_3(z)$ which vanishes because it is an integral over the collapsing three cycle. These can be approximated straightforwardly by truncating their Taylor series. We write,
\begin{align}
\Pi_1(z)&=\sum_{n=0}^{q}b_n (z-1)^n\label{Pi1expansion}\\
\Pi_2(z)&=\sum_{n=0}^{q}c_n (z-1)^n\\
\Pi_3(z)&=\sum_{n=1}^{q}d_n (z-1)^n.
\end{align}

The periods and their first derivatives enter the scalar potential. Since we seek to collect up to fourth order Taylor coefficients of the scalar potential at the vacua located, we will be evaluating fifth order derivatives of the periods near $z=1$. For the sake of accuracy we take $q=8$. The expansion coefficients can be found in Table \ref{period-coeffs-123} of Appendix \ref{appendix}.

The remaining period, $\Pi_0$, picks up one factor of its intersecting partner, $\Pi_3$, for each loop about the conifold point. This transformation property of $\Pi_0$ restricts its form to
\begin{equation}
\Pi_0(z)=\Pi_3(z)\frac{\log(z-1)}{2\pi i}+f(z)\label{Pi0expansion}
\end{equation}
for some function $f(z)$ that is analytic at the conifold point. Before proceeding we note that we shall henceforth include an overall minus sign in front of the argument in the logarithm in the expansion of $\Pi_0$ so that all explicit values of the expansion coefficients correspond to a consistent choice of branch cuts in Mathematica. Specifically, the expressions given for the periods in terms of the Meijer-G's use the convention of branch cuts emanating from the conifold point along the positive real axis, and from the large complex structure point along the negative real axis. Since Mathematica's logarithm function places the branch cut along the argument's negative real axis, it is necessary to include a minus sign in front of the log's argument in the expansion, eq. \ref{Pi0expansion}, to flip it from $(-\infty,1]$ to $[1,+\infty)$. 

Note that the argument of the logarithm in $\Pi_0$'s expansion may be rescaled freely because this amounts to a relabeling of analytic terms. The righthand side of eq. \ref{Pi0expansion} is equivalently written as 
\begin{equation}
\Pi_3(z)\frac{\log(-(z-1))}{2\pi i}+f(z)-\frac{\Pi_3(z)}{2}=\Pi_3(z)\frac{\log(-(z-1))}{2\pi i}+\tilde{f}(z).
\end{equation}
The shifted function, $\tilde{f}(z)$, is still analytic because $\Pi_3$ is. Relabeling $\tilde{f}(z)$ by $f(z)$ we have the same expression as eq. \ref{Pi0expansion}, only with a negative sign in front of the $(z-1)$. We choose however to keep the ``extra" analytic term, $-\Pi_3/2,$ separate and take the form,
\begin{equation}
\Pi_0(z)=\Pi_3(z)\left(\frac{\log(-(z-1))}{2\pi i}-\frac{1}{2}\right)+f(z).
\end{equation}
It is a convenient choice for performing checks of the accuracy of the $\Pi_0$ approximation because the factor multiplying $\Pi_3$ does not change sign (the range of the imaginary part of the logarithm function in Mathematica is $[-\pi,\pi]$).

Since we have a polynomial expansion for $\Pi_3$, the task of obtaining a fast approximation for $\Pi_0$ amounts to finding one for the unknown $f(z)$. Since we have no special restrictions to this function's properties aside from analyticity, the simplest approximation is a Taylor series about $z=1$. We write
\begin{equation}
f(z)=\sum_{n=0}^{q}a_n (z-1)^n.
\end{equation}
The zeroth coefficient is the value of $\Pi_0$ at the conifold point, which is trivial to compute. The higher order coefficients are more difficult. 

Although each
\begin{equation}
a_n=\frac{1}{n!}\frac{d^n f}{dz^n}\bigg|_{z=1}=\frac{1}{n!}\left(\frac{d^n \Pi_0}{dz^n}\bigg|_{z=1}-\frac{1}{2\pi i}\frac{d^n }{dz^n}\left[\Pi_3(z)\left(\log(-(z-1))-\frac{1}{2}\right)\right]\bigg|_{z=1}\right)
\end{equation}
is finite, the fact that the divergences between the two terms on the righthand side cancel exactly at each order is lost if one attempts to evaluate (numerically) the righthand side exactly the conifold point. Mathematica's ``Limit" function cannot be used to remedy this. However, the next coefficient, $a_1$, is nonetheless easily obtained numerically by exploiting the weakness of the divergences that cancel in the first derivative, which are logarithmic. 

In particular, to leading order in $s\equiv (z-1)$, $a_1$ is given by, 
\begin{equation}
a_1=\left( \frac{d \Pi_0}{d z}\bigg|_{z=1}-d_1\frac{\log(-s)}{2 \pi i}\right)+\frac{i d_1}{2\pi }+\frac{d_1}{2}+\mathcal{O}(s \log(s)).
\end{equation}
We obtain an approximate value for $a_1$ by dropping the $\mathcal{O}(s)$ terms which involve higher order (unknown as of now) $a_i$'s and evaluating the remaining known expressions on the righthand side sufficiently close to the conifold point that errors due to the truncation are negligible. Taking the form $s=e^{-t}$, the negligibility of such errors at a finite $t=t^*$ is ensured if the value of $\tilde{a}_1(t)$ defined by,
\begin{equation}
\tilde{a}_1(t)=\Pi'_0(1+e^{-t})+\frac{i d_1}{2 \pi }+\frac{d_1}{2}-d_1\frac{\log(-e^{-t})}{2 \pi i}
\end{equation}
converges within the relevant precision one is using for $t\rightarrow t^*$. Such convergence is exhibited in Table \ref{a1-convergence}.
\begin{table}[h!]
  \centering
  \begin{tabular}{|l|l|}
    \hline
   $t$ & $\tilde{a}_1(t)$\\
       \hline
      $2$ & $0.0082657465 - 0.1488734062 i$\\
      $3$& $0.0143193561 - 0.1662234075 i$  \\
       $4$&$0.0193211265 - 0.1734738528 i$\\
      $5$&$0.0221685901 - 0.1762752502 i$\\
      $6$&$0.0235728366 - 0.1773249249 i$\\
      $7$&$0.0242171435 - 0.1777137131 i$\\
      $8$&$0.0245004660 - 0.1778570993 i$\\
      $9$&$0.0246216047 - 0.1779098968 i$\\
      $10$&$ 0.0246723700 - 0.1779293266 i$\\
      $12$ & $0.0247018694 - 0.1779391052 i$\\
      $14$& $0.0247067045 - 0.1779404287  i$  \\
       $16$&$0.0247074729 - 0.1779406078 i$\\
       $20$&$0.0247076106 - 0.1779406353 i$\\
       $25$&$0.0247076138 - 0.1779406359 i$\\      
    \hline
  \end{tabular}
   \caption{Depiction of the the convergence of the expansion coefficient $a_1$ computed numerically.}
   \label{a1-convergence}
\end{table}

The third coefficient in $f$'s expansion, $a_2$, can be obtained in a similar fashion. We write
\begin{equation}
\tilde{a}_2(t)\equiv\frac{1}{2}\left[\Pi_0''(1+e^{-t})+d_2+\frac{3 i d_2}{2\pi}+\frac{i d_1 e^{t}}{2\pi}-\frac{d_2\log(-e^{-t})}{\pi}\right]
\label{a2tilde}
\end{equation}
However, here it is essential to use high-digit accuracy computations when evaluating the righthand side for a given value of $t$. This is because we're extracting a small number by taking the difference of two large numbers, $\Pi_0''(1+e^{-t})$ and the term proportional to $e^t$ in eq. \ref{a2tilde}. Table \ref{a2-convergence} displays the convergence of $a_2$. 
\begin{table}[h!]
  \centering
  \begin{tabular}{|l|l|}
    \hline
   $t$ & $\tilde{a}_2(t)$\\
       \hline
      $2$ &$ \thickspace \thickspace\thickspace0.00213219053 + 0.09247047120 i$\\
      $3$& $-0.00142768524 + 0.11114853496 i$  \\
      $4$&$-0.00594277215 + 0.11937168288 i$\\
      $5$&$-0.00880198201 + 0.12261391639 i$\\
      $6$&$-0.01027763172 + 0.12383805148 i$\\
      $7$&$-0.01097135473 + 0.12429273911 i$\\
      $8$&$-0.01128102108 + 0.12446060393 i$\\
      $10$&$-0.01147126928 + 0.12454519737 i$\\
      $12$&$-0.01150432390 + 0.12455665210 i$\\
      $16$&$-0.01151065730 + 0.12455841226 i$\\
       $20$&$-0.01151081435 + 0.12455844450  i$\\
       $24$&$-0.01151081798 + 0.12455844509 i$\\
       $28$&$-0.01151081806 + 0.12455844510 i$\\
    \hline
  \end{tabular}
   \caption{Depiction of the convergence of the expansion coefficient $a_2$ computed numerically.}
   \label{a2-convergence}
\end{table}

Clearly this strategy is limited to the lowest expansion coefficients. At each higher order the righthand side will involve evaluating increasingly divergent terms near the conifold point and extracting an ever (comparatively) smaller difference. To obtain the higher order coefficients we instead derive a recursion relation for the $a_n$'s by using the fact that both $\Pi_0$ and $\Pi_3$ are solutions to the Picard-Fuchs equation. Specifically, since the Picard-Fuchs equation is linear, $f(z)$ must satisfy
\begin{equation}
\hat{O}_{PF}[f(z)]=-\hat{O}_{PF}\left[\Pi_3(z)\left(\frac{\log(-(z-1))}{2\pi i}-\frac{1}{2}\right)\right].\label{PFonf}
\end{equation}

The righthand side is a known, albeit messy, analytic function due to the fact that $\Pi_3$'s near conifold expansion coefficients are known. Note that because $\mathcal{O}_{PF}$ is a fourth order differential operator the righthand side will contain terms that are individually divergent (from derivatives acting on the $\log$ times the lower order terms in $\Pi_3$ so as to yield contributions $\sim s^{-1}$ and $s^{-2}$). The divergences, though, exactly cancel due to the strict relationship among $\Pi_3$'s coefficients, owing to the fact that it satisfies
\begin{equation}
\hat{O}_{PF}[\Pi_3(z)]=0.
\end{equation}

Thus, we need only express the lefthand side of eq. \ref{PFonf} as a power series in $s$ (whose coefficient at a given order is a linear combination of a subset of the $\{a_i\}$) and ensure we have enough of the lowest order coefficients to generate the rest. It will turn out that the zeroth order term on the lefthand side involves $f$'s four lowest order coefficients, and all those of order $n>0$ involve the $n-1^{\text{th}}$ and subsequent four coefficients: $\{a_{n-1},..,a_{n+3}\}$. Hence, the $a_0$, $a_1$ and $a_2$ obtained numerically as described above will be sufficient to start off the recursive procedure, and provide us as many coefficients as we need.

To that end, we rewrite the Picard-Fuchs differential operator as follows,
\begin{align}
\hat{O}_{PF}&=-s\delta^4-s (k_1\delta^3-k_2\delta^2-k_3\delta-k_4)-(k_1\delta^3-k_2\delta^2-k_3\delta-k_4)\\
&=-s\sum_{i=0}^{4}k_i\delta^{4-i}-\sum_{i=1}^{4}k_i \delta^{4-i}\nonumber
\end{align}
where the $k_i$ are the constants,
\begin{align}
k_0&=1\\
k_1&=\sum_{i=1}^{4}\alpha_i\\
k_2&=\sum_{i=1}^{4}\sum_{j=i+1}^{4}\alpha_i\alpha_j\\
k_3&=\alpha_1\alpha_2\alpha_3+\alpha_1\alpha_2\alpha_4+\alpha_2\alpha_3\alpha_4\\
k_4&=\alpha_1\alpha_2\alpha_3\alpha_4.
\end{align}
Next note that $\delta$ acts on $s^n$ as,
\begin{align*}
\delta s^n&=(s+1)\frac{d}{dz}(z-1)^n\\
&=(s+1)ns^{n-1}\\
&=n(s^n+s^{n-1}).
\end{align*}
By repeatedly applying this rule each of the $\delta^js^{n}$ terms can be computed. For instance,
\begin{align*}
\delta^2 s^n&=\delta[n(s^n+s^{n-1})]\\
&=n(n(s^n+s^{n-1})+(n-1)(s^{n-1}+s^{n-2}))\\
&=n^2s^n+n(2n-1)s^{n-1}+n(n-1)s^{n-2}.
\end{align*}
After similarly obtaining $\delta^3$ and $\delta^4$ on $s^n$, collecting terms and shifting indices of summation, the Picard-Fuchs operator's action on $f(z)$ can be expressed in the form, 
\begin{equation}
\hat{O}_{PF}\left[f(z)\right]=\sum_{n=0}^{\infty}\bigg(C_{-1}(n) a_{n-1}+C_0(n) a_{n}+ C_{1}(n) a_{n+1}+C_{2}(n) a_{n+2}+C_{3}(n) a_{n+3}\bigg)s^n,
\end{equation}
where $a_{-1}\equiv0$ and the constants $C_j(n)$ are also functions of the $k_i$. Though a tedious exercise, the  $C_j(n)$ can be obtained straightforwardly with the aid of Mathematica. 
A similar procedure yields the expansion of the righthand side of eq. \ref{PFonf} thus completing the recursion relation. The resulting values for the $a_n$ are given in Table \ref{an-coeffs} located in Appendix \ref{appendix}.

The analogous expansions about the large complex structure point are far easier to obtain numerically, despite the fact that three as opposed to one of the cycles transform nontrivially upon circling it. Being finite but multiple-valued, the form of the corresponding three periods involve linear combinations of powers of $z\log{z}$. As mentioned at the end of subsection \ref{periods-subsection}, the leading order behavior of the $j$th period in our notation (that is the term that contributes the most divergent term to the period's derivative) is  $(z\log{z})^j$. This form includes the behavior of the analytic period, $\Pi_0$, which is $\approxeq 1$.

Since each of the other three periods has its own residual analytic term (analogous to $f(z)$ in eq. \ref{Pi0expansion}) as well as an additional subleading logarithmic term at each period index $j$ higher, the near large complex structure expressions are more complicated. Nonetheless, the approximations can be obtained using Mathematica's ``Series" function, due to the special properties of Meijer-Gs. Essentially, one can expand the $G^{m,n}_{p,q}$ in eq. \ref{meijerGdefinition} about $z=0$ to yield a series of integrals whose individual terms are easy to evaluate. 

\subsection{Structure of the Hessian}\label{hessiansection}
The masses of the moduli in the effective field theory associated with a given vacuum are contained in the Hessian of the scalar potential specified by the particular flux configuration. When evaluated at the vacuum location in the moduli space, the eigenvalues of the Hessian in canonically normalized field coordinates are the squares of the masses in the effective theory. No scale vacua have additional structure built in from the outset as compared to ordinary general $\mathcal{N}=1$ supersymmetric theories. 

We begin by expressing the general $\mathcal{N}=1$ scalar potential -- that which includes the K{\"a}hler moduli and does not assume cancellation of the 3$|W|^2$ term -- and its partial derivatives in terms of the appropriately invariant quantities. It is convenient to adopt the standard notation for the K{\"a}hler and geometrically covariant derivatives of the superpotential, up to third order,
\begin{equation}
F_I\equiv D_I W;\thickspace \thickspace Z_{IJ}\equiv D_ID_J W;\thickspace \thickspace U_{IJK}\equiv D_ID_J D_KW
\end{equation}
Note that $F_I$ is not to be confused with the amount of R-R flux wrapping a particular $3$-cycle of the compact manifold. We express general $\mathcal{N}=1$ scalar potential then as,
\begin{equation}
V=e^{\mathcal{K}}(F_I \bar{F}^{I}-3|W|^2).\label{scalarVgeneral}
\end{equation}
where indices run over all moduli.

Due to the K{\"a}hler invariance of eq. \ref{scalarVgeneral} we may trade partial derivatives for covariant ones and obtain the following covariant expressions \cite{denef-nonsusy}.
\begin{align}
\partial_I V&=e^{\mathcal{K}} \left((D_I D_J W)\bar{F}^J-2F_I \bar{W}\right)=e^{\mathcal{K}} \left(Z_{IJ}\bar{F}^J-2F_I \bar{W}\right)\\
\partial_{I}\partial_{J} V&=e^{\mathcal{K}} \left((D_I D_J D_K W)\bar{F}^K-D_ID_J\bar{W}\right)=e^{\mathcal{K}}\left(U_{IJK}\bar{F}^K-Z_{IJ}\bar{W}\right)\\
\partial_{I}\partial_{\bar{J}} V&=e^{\mathcal{K}} \left(-R_{I\bar{J}K\bar{L}}\bar{F}^K F^{\bar{L}}+ K_{I\bar{J}} F_K \bar{F}^K-F_I \bar{F}_{\bar{J}}+(D_I D_K W)(\bar{D}_{\bar{J}}\bar{D}^K \bar{W})-2 K_{I\bar{J}} |W|^2\right)\\
&=e^{\mathcal{K}} \left(-R_{I\bar{J}K\bar{L}}\bar{F}^K F^{\bar{L}}+ K_{I\bar{J}} F_K \bar{F}^K-F_I \bar{F}_{\bar{J}}+Z\bar{Z}_{I\bar{J}}-2 K_{I\bar{J}} |W|^2\right).
\end{align}

When the no scale cancellation takes place, and the SUSY condition for the remaining dynamical moduli in the theory is imposed the nontrivial components of the Hessian reduce to,
 \begin{align}
 \partial_{I}\partial_{J}  V&=2e^{\mathcal{K}}\bar{W}Z_{IJ}\\
 \partial_{I }\partial_{\bar{J}}V&=e^{\mathcal{K}}\left(Z\bar{Z}_{I\bar{J}}+\mathcal{K}_{I\bar{J}}|W|^2\right).
 \end{align}
where $Z\bar{Z}$ is defined with the contraction of one holomorphic and one anti-holomorphic index using the (inverse) K{\"a}hler metric, and indices now run over only the axio-dilaton and complex structure moduli. 

Next choose a basis for the complex moduli fields that is orthonormal with respect to the vacuum K{\"a}hler metric,
\begin{equation}
\mathcal{K}_{I\bar{J}}|_{vac}=\delta_{I\bar{J}}.
\end{equation}
Such a basis is only unique up to unitary transformations. An arbitrary choice will not in general simultaneously diagonalize $Z^{can}\bar{Z}^{can}$. Note that while $Z$ and $\bar{Z}$ are complex and symmetric, $Z\bar{Z}$ is Hermitian and positive definite, and is thus related to the diagonal matrix containing $N=1+h^{2,1}$ eigenvalues by a unitary transformation. In canonical coordinates we write, 
\begin{align}
Z^{can}\bar{Z}^{can}&=U \Sigma^2 U^{\dagger}
\end{align}
and express the eigenvalues as the squares of real positive numbers $\lambda_i$. The columns of the unitary matrix, $U$, are of course the corresponding eigenvectors of the canonical $Z\bar{Z}$.
(An excellent resource for understanding no scale structure and its implications is \cite{Marsh:2014}. We've adopted their notation in our abridged calculation here in order to facilitate its use to readers seeking greater detail).

Thus, the Hessian in canonical coordinates,
\begin{equation}
\mathcal{H}^{\text{can}}=e^{\mathcal{K}}\mathcal{U}^\dagger\left( \begin{array}{cc}
(Z\bar{Z}) ^{can}_{I\bar{J}} +1_{2\times2}|W|^2& 2\bar{Z}^{can}_{\bar{I}\bar{J}}W\\
 2 Z^{can}_{IJ}\bar{W}&(Z\bar{Z}) ^{can}_{\bar{I}J}+1_{2\times2}|W|^2 \end{array} \right)\mathcal{U}\label{Hess-2}
\end{equation}
can be diagonalized by a unitary transformation defined in terms of the 2N$\times$2N-matrix $\mathcal{U}$,
\begin{equation}
\mathcal{U}=\left( \begin{array}{cc}
U& 0\ \\
0&U^\dagger\end{array} \right).
\end{equation}
In particular, one can rewrite eq. \ref{Hess-2} as,
\begin{equation}
\mathcal{H}^{\text{can}}=e^{\mathcal{K}}\mathcal{U}^\dagger\left( \begin{array}{cc}
\Sigma^2 +1_{2\times2} |W|^2& 2\Sigma W\ \\
2\Sigma\bar{W} &\Sigma^2+1_{2\times2}|W|^2 \end{array} \right)\mathcal{U}\label{Hess-3}
\end{equation}

A permutation of the rows and columns of matrix between $\mathcal{U}$ and $\mathcal{U}^\dagger$ in eq. \ref{Hess-3} casts it as block diagonal, with each of the $N$ 2$\times$2 blocks having the form,
\begin{equation}
\left( \begin{array}{cc}
\lambda_i^2+|W|^2& 2\lambda_iW\\
2\lambda_i\bar{W}&\lambda_i^2+|W|^2\end{array} \right).\label{Hess-blocks}
\end{equation}
The eigenvalues of the Hessian then come in pairs, namely those of each block times the overall factor of $e^\mathcal{K}$,
\begin{equation}
m_{i\pm}^2=e^{\mathcal{K}}\left(\lambda_i\pm|W|\right)^2\label{mass-pairs}. 
\end{equation}

The fact that the scalar masses in no scale supergravity take the form of eq. \ref{mass-pairs} does not ensure
a discernible pattern among the masses of an ensemble of vacua will emerge. Which pattern is present, if any, depends on the relative scales of the $\lambda_i$ as well as how they compare to the magnitude of the superpotential at vacua. We shall see that a pronounced hierarchy and splitting of the field space \emph{consistent across the ensemble} arises due to the special features of the conifold point, where our vacua accumulate. This is discussed at length in section \ref{results}. It is worth remarking that no association of one particular kind of moduli field (or a particular linear combination) with a heavy or light mass pair, nor the existence of separated mass pairs, is imposed by eq. \ref{mass-pairs}. 

\section{Calculational Approach}\label{calc-approach}
There are two components to our procedure for generating a random sample of effective field theories in the mirror quintic's moduli space. In this section we discuss each of these in turn.

\subsection{Generating a Random Sample of Vacua}\label{vacuumsearch}
Recall that we make the assumptions that the effect of O3-planes on the compact geometry is negligible at the level of the 4d action for the moduli, that all K{\"a}hler moduli are stabilized, and that the backreaction from fluxes (warping) can be ignored thus preserving the no scale structure given by compactifying type IIB supergravity on a Calabi-Yau. 

The effective action for the two remaining complex scalars takes the form, 
\begin{equation} 
S_{eff}=\frac{M_p^2}{2}\int{}{}d^4x\thickspace \mathcal{K}^{cs}_{z \bar{z}}\partial_\mu z\partial^\mu \bar{z} +\mathcal{K}^{ax}_{\tau \bar{\tau}}\partial_\mu \tau\partial^\mu\bar{\tau}-V(z,\tau,\bar{z},\bar{\tau})\label{model},
\end{equation}
where both components of the field space metric are obtained by taking one holomorphic and one antiholomorphic derivative of the relevant K{\"a}hler potential. The K{\"a}hler potential  for the complex structure is given by eq. \ref{KahlerPotentialCS}, which is known explicitly in terms of Meijer-G functions via eqs. \ref{periods-0}--\ref{periods-3} and \ref{MeijerGs0}--\ref{MeijerGs2}. That for the axio-dilaton is obtained from eq. \ref{axioKpotential}. 

The scalar potential, $V$, is not a holomorphic function of $z$ and $\tau$. It is however defined in terms of the holomorphic superpotential, $W(z,\tau)$,
\begin{equation}
V=\frac{M_p^2}{4 \pi \mathcal{V}_0^2}e^{\mathcal{K}^{cs}(z,\bar{z})+\mathcal{K}^{ax}(\tau,\bar{\tau})}\left(\mathcal{K}^{z\bar{z}}D_zW\bar{D}_{\bar{z}}\bar{W}+{\mathcal{K}}^{\tau\bar{\tau}}D_\tau W\bar{D}_{\bar{\tau}}\bar{W}\right).
\end{equation}
The superpotential is parameterized by eight integers indicating the amount of R-R and NS-NS fluxes wrapping/piercing each of the mirror quintic's four 3-cycles. In particular, we write the superpotential as in eq. \ref{FluxVectors}; a linear combination of the mirror quintic's period integrals in a symplectic basis (the $\Pi_i$'s).

Solutions to the SUSY condition, $D_zW =D_\tau W=0$, are zeros of the scalar potential and thus are global minima of the theory. We search specifically for such solutions by randomly scanning through models defined by eq. \ref{model}, that is by randomly drawing eight flux integers. For simplicity, we assume a flat measure for the fluxes, and draw from the interval, $[-20,20]$. Once the set of fluxes is drawn, the corresponding superpotential can be built, and the zeros of $D_IW$ can be searched for numerically.  

The SUSY condition for the axio-dilaton implies it is an explicit function of the complex structure vacuum location, specifically that in eq. \ref{taususy}. By evaluating the SUSY condition for the complex structure, $D_z W=0$, at $\tau=\tau_{SUSY}(z)$ we accomplish an important reduction in the computational expense of finding vacua numerically -- we need only minimize a single (semipositive definite) function of two real variables, namely,
\begin{equation}
u(x,y;\{F_i,H_i\})=\left|\left(F-\tau_{SUSY}(x+i y) H\right)\cdot \left(\Pi'(x+i y)+\mathcal{K}_z \Pi(x+i y) \right)\right|^2\label{minfunction}.
\end{equation}
where $\mathcal{K}_z$ is of course also a function of the real and imaginary parts of the complex structure, $x$ and $y$. Specifically,
\begin{equation}
\mathcal{K}_z=-\frac{\Pi^\dagger(x-i y) Q^{-1}\Pi'(x+i y)}{\Pi^\dagger(x-i y) Q^{-1}\Pi(x+i y)}.
\end{equation}

For a given set of fluxes the function, $u$, defined in eq. \ref{minfunction} can be assembled and minimized directly in Mathematica using its FindMinimum function, provided that an initial starting point (for $x$ and $y$) is specified. Vacua are known to accumulate near the conifold point, $z=1$, so it is reasonable to target our search here. Since we have simple expansions for the period functions here, namely eqs. \ref{Pi1expansion}--\ref{Pi0expansion}, we may expand the K{\"a}hler covariant derivative of the superpotential with respect to the complex structure in $z-1$. The term involving $\Pi_0'$ in $D_zW$ will yield a logarithm of $z-1$. This is the most divergent term. By retaining only the logarithmic and $\mathcal{O}(1)$ terms in $D_zW=0$ we can solve for $z$ in terms of the fluxes. The result is, 

\begin{align}
&z_{guess}=1-e^\varphi\\
&\varphi=-1+2 \pi i \left(\frac{\beta a_0 -a_1-\frac{d_1}{2}}{d1}+\frac{F_2 - t H_2}{F_3 - t H_3} \frac{\beta b_0 - b_1}{
   d_1} + \frac{F_1 - t H_1}{F_3 - t H_3} \frac{\beta c_0 - c_1}{
   d_1} - \frac{F_0 - t H_0}{F_3 - t H_3}\right)\label{guess}\\
&\beta=-\frac{\bar{a_0}d_1-\bar{c_0}{b_1}+b_0\bar{c_1}}{\bar{b_0}c_0-\bar{c_0}b_0}\\
&t=\frac{F_3 \bar{a_0}+F_2\bar{b_0}+F_1\bar{c_0}}{H_3 \bar{a_0}+H_2\bar{b_0}+H_1\bar{c_0}}
\end{align}
Note that the axio-dilaton has been evaluated at $\tau_{SUSY}(z;F,H)$ and expanded as well. It is the the $\mathcal{O}(1)$ constant, $t$, above.

There is no guarantee that a random choice of fluxes will have a near conifold minimum that satisfies the SUSY condition. In fact, the vast majority do not. Whether this is the case can be determined from the initial guess. If $z_{guess}-1$ is so large that $\mathcal{O}(z-1)$ terms dominate $\log(z-1)$, then the expansion that lead to $z_{guess}$ was not valid to begin with, and so the eight fluxes are redrawn. 

To summarize,  then, the steps of our search algorithm are:\

1. Randomly draw eight integers independently from the interval $[-20,20]$.\

2. Compute the guess via eq. \ref{guess}. If it is more than $0.5$ away from conifold redraw the fluxes. Otherwise construct $u(x,y;F,H)$ using the patched period functions. \

3. Minimize $u$ using FindMinimum, with the real and imaginary parts of the guess as the starting point. (We further invoke the option that limits the search region to avoid Mathematica searching far away from the region of interest when there is no near conifold minimum, and/or extrapolating the periods so as to give artificial minima).\

4. If a minimum is found, and that minimum is sufficiently close to the conifold, we  collect a variety of useful information including the list of fluxes, the minimum's location, the magnitude of $u$, and the axio-dilaton's location $\tau_{SUSY}(z_{min})$.\

5. Repeat.\

\

Lastly, it is necessary to filter these local minima of $u$. Since the imaginary part of the axio-dilaton is proportional to the inverse of the string coupling, any minima found during the search  that have negative $\text{Im}(\tau_{SUSY}(z_{min}))$ are unphysical. Removing these, the list of potential vacua is approximately cut in half. Next, only the zeros of $u$ should be kept, as the evaluation of the axio-dilaton at \ref{taususy} assumes vacua solve $D_I W=0$. We eliminate minima whose $u$ is above a threshold of $10^{-6}$. 

To compare vacua properly, particularly with regard to their locations in the the $\tau$ field space, we need to perform an $SL(2,\mathbb{Z})$ transformation that maps each vacua's $\tau$ value into the fundamental domain. A set of four integers $\{a,b,c,d\}$ is found such that the transformation,
\begin{equation}
\tau\rightarrow\frac{a\tau+b}{c\tau+d}
\end{equation}
maps the axio-dilaton into the strip in the upper half-plane with both $\text{Re}(\tau)<\frac{1}{2}$ and $|\tau|>\frac{1}{2}$. The vacuum's fluxes are then mapped as follows,
\begin{align*}
F_i\rightarrow aF_i+bH_i\\
H_i\rightarrow cF_i+dH_i.
\end{align*}

This is the four dimensional incarnation of the original $SL(2,\mathbb{Z})$ symmetry enjoyed by the 10d type IIB supergravity action. In fact, one reason for formulating the theory in terms of the axio-dilaton is so this symmetry is made manifest. The transformation is stated for the total $3$-form flux as,
\begin{equation}
G_{(3)}\rightarrow c G_{(3)}+d.
\end{equation}
Performing the transformation also enables us to check for duplicate vacua. The transformation is not $1$-to-$1$, so vacua with different fluxes prior to mapping may actually correspond to the same vacuum in the fundamental domain. Though possible, no instances of double counting were found among the vacua identified with our algorithm.

Finally, we impose the type IIB tadpole condition on the fluxes, eq. \ref{tadpole}. This integral expression can be restated conveniently in terms the number of orientifold planes and D3-branes, and the dimensionless fluxes wrapping the mirror quintic's 3-cycles as
\begin{equation}
F\cdot Q\cdot H=\frac{1}{4}N_{O3}-N_{D3},
\end{equation}
where $Q$ is the intersection matrix, and $F$ and $H$ are the vectors containing the four R-R and four NS-NS flux integers. The condition is often stated as an inequality by defining the maximum of the righthand side as the positive number $L_{max}$,
\begin{equation}
F\cdot Q \cdot H\leq L_{max}\label{tadpole-ineq}.
\end{equation}

We make an admittedly arbitrary choice of $L_{max}=300$, and dispose of vacua whose fluxes combine via $Q$ to violate this threshold. 
We are interested in the statistical features of flux vacua in the mirror quintic's moduli space as a probe of the landscape more broadly. So long as our results are not sensitive to the particular choice of $L_{max}$, we believe it is reasonable to relax the condition. We find this is the case and so proceed without concern for the actual maximum number of orientifold planes the mirror quintic can support.

\begin{figure}
\centering
   \includegraphics[width=\textwidth]{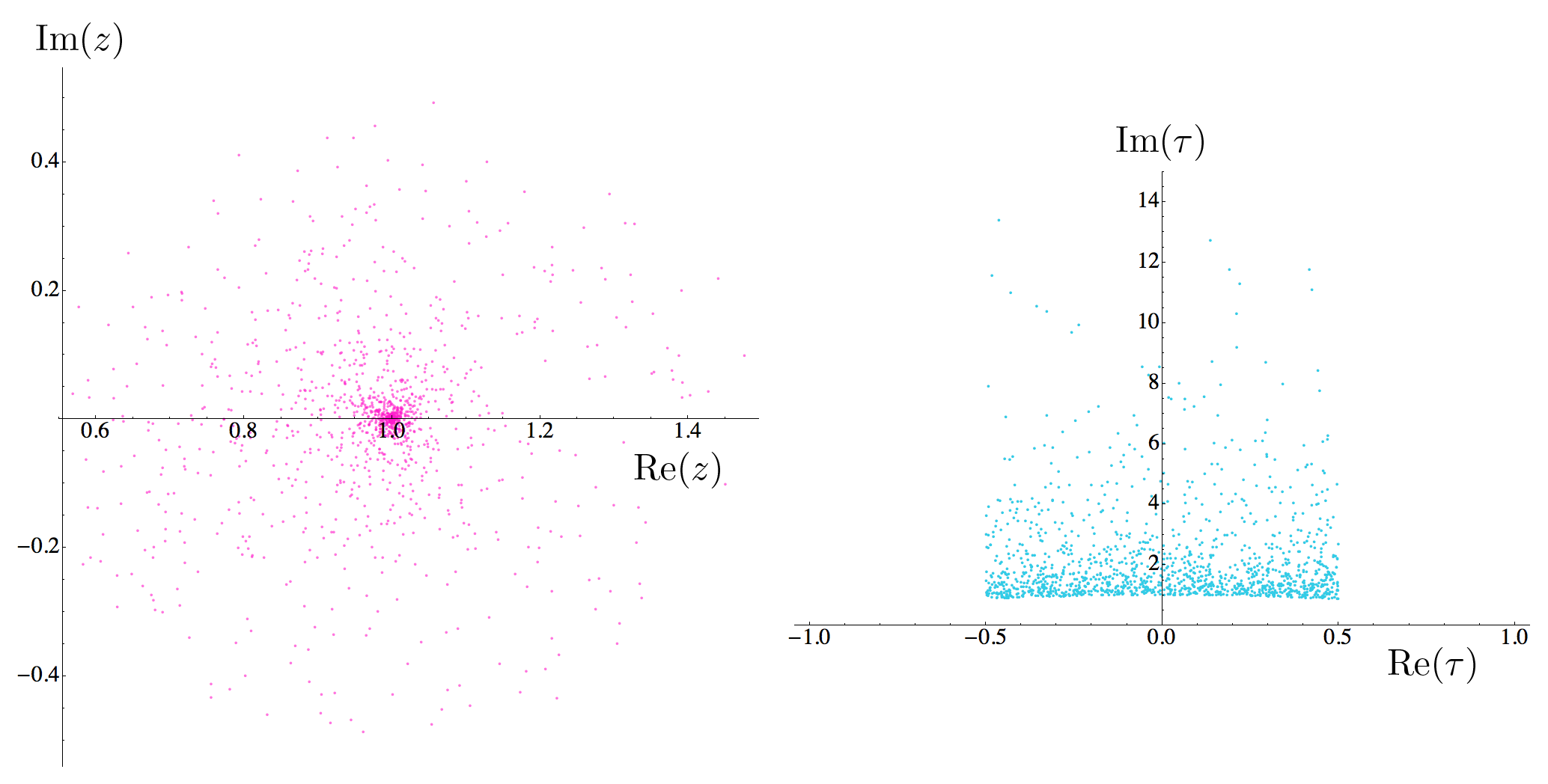}
    \caption{Complex structure and axio-dilaton vacuum locations for the random sample of 1358 vacua.}
    \label{vacuum-locations}
   \end{figure}

The results for the masses and couplings given in the following section are for the largest random sample of vacua we found using this search algorithm and filtering. It consists of 1358 near conifold vacua whose complex structure and axio-dilaton's locations are shown in Figure \ref{vacuum-locations}.
\subsection{Computing Coefficients}\label{taylor-coeffs-alg}
Now that we have a random sample of flux vacua, we turn to our second computational task -- obtaining the masses and couplings to quartic order in the corresponding ensemble of effective field theories. These are the data whose statistics we want to analyze. For a given model, the $n^{\text{th}}$ order couplings are the $n^{\text{th}}$ order Taylor coefficients of the corresponding scalar potential expanded about the model's vacuum and transformed appropriately so that the kinetic terms in all the effective field theories are canonical. 

Though the vacuum axio-dilaton coordinate location in each model is fixed in terms of the vacuum's complex structure location, the axio-dilaton is a full degree of freedom. We fixed it as an explicit function of $z$ in our search algorithm as a short-cut to finding the location of minima of the scalar potential. Here we leave $\tau$ as a variable in the scalar potential, and so have two complex degrees of freedom. The Hessian then is a $4\times4$-matrix, and the four masses come in two pairs due to the special structure of the mass matrix in no scale models, as we reviewed in section \ref{hessiansection}.

This structure is relevant to the higher order couplings because we need to report them in a basis which not only yields canonical kinetic terms but also diagonalizes the mass matrix. In this subsection we first discuss the field redefinitions,
\begin{align}
&\{ z,\tau,\bar{z},\bar{\tau}\}\rightarrow\{y_1,y_2,y_3,y_4\}\\
&(z,\tau)\in\mathbb{C}^2\text{,}\quad y\in\mathbb{R}^4
\end{align}
that accomplish this, and  we then the numerical algorithm for evaluating the Taylor coefficients. Though the original complex coordinates are the simplest in which to evaluate the Taylor coefficients because we have expansions for the periods in $z-1$, it will still be necessary to design an efficient algorithm. This somewhat tedious exercise is discussed in the second half of this section after defining the specific transformation that is applied to each tensor of Taylor coefficients calculated with the algorithm.

To that end, we define the column vector of fields in our original complex basis,
\begin{equation}
\Phi\equiv \left( \begin{array}{c}
z \\
\tau \\
\bar{z}\\
\bar{\tau} \end{array} \right).
\end{equation}
The effective field theory is obtained by expanding about a homogeneous background, $\Phi_{vac}$. We begin by writing, $\Phi=\Phi_{vac}+B\Psi$, where the matrix $B$ will serve to canonically normalize the kinetic terms. We have,
\begin{align}
\mathcal{L}(\Phi)&=\mathcal{L}(\Phi_{vac})+\partial_\mu \Psi^{\dagger} B^{\dagger}(G_{vac}+\mathcal{O}(\Psi))B\partial^{\mu}\Psi-V(B\Psi)\\
&=\mathcal{L}_{vac}+\partial_{\mu}\Psi^{\dagger}B^{\dagger}G_{vac}B\partial^{\mu}\Psi-\frac{1}{2}\Psi^{\dagger}B^{\dagger}MB\Psi+\mathcal{O}(\Psi^3)
\end{align}
where $G(\Phi)$ is the Hermitian matrix containing the components of the K{\"a}hler metrics. Specifically,
\begin{equation}
G(\Phi) =\left( \begin{array}{cccc}
\mathcal{K}_{z\bar{z}}&0&0&0 \\
0&\mathcal{K}_{\tau\bar{\tau}}&0&0 \\
0&0&\mathcal{K}_{\bar{z}z}&0\\
0&0&0&\mathcal{K}_{\bar{\tau}\tau}\end{array} \right).
\end{equation}
and $G_{vac}$ is that evaluated at the vacuum.

The effective field theory will have canonical kinetic terms provided
\begin{equation}
B^{\dagger}G_{vac}B=1.
\end{equation}
This is easily accomplished by rescaling the fields. A normalized complex basis, which we denote $\{\xi,\sigma\}$, will be useful in discussing our results so we define one during this otherwise intermediate step. It will be convenient, in addition, to shift the normalized complex structure field so that it is zero at the conifold. Thus we write,
\begin{equation}
z=1+C_1\xi\thickspace,\quad\tau=C_2 \sigma
\end{equation}
where $C_1$ and $C_2$ are the constants $1/\sqrt{\mathcal{K}_{z\bar{z}}|_{vac}}$ and $1/\sqrt{\mathcal{K}_{\tau\bar{\tau}}|_{vac}}$, respectively (we drop the superscripts on the K{\"a}hler potentials indicating the complex structure and axio-dilaton as its diagonal form in $z$ and $\tau$ renders them superfluous in the metric). Then,
\begin{equation}
\Psi\equiv\left( \begin{array}{c}
\xi-\xi_{vac}\\
\sigma-\sigma_{vac}\\
\bar{\xi}-\bar{\xi}_{vac}\\
\bar{\sigma}-\bar{\sigma}_{vac} \end{array} \right).
\end{equation}

The matrix $M$ contains the partial derivatives of the scalar potential (in the original coordinates) evaluated at the vacuum, but with the appropriate ordering so that it is Hermitian. We take the ordering $(z, \tau, \bar{z}, \bar{\tau})$ for the columns, so our rows have the barred ordering $(\bar{z}, \bar{\tau}, z, \tau)$. Using the like orderings for columns and rows will not yield a Hermitian matrix, only a symmetric one. We included the subscripts in the definition of $G$ in part to emphasize this. 

The ``rescaled" mass matrix, $B^{\dagger}M B,$ is not in general diagonal. This is because the scalar potential nontrivially mixes the complex structure with the axio-dilaton so that mixed partials, like $\partial_z\partial_\tau V$, do not vanish at the vacuum. We choose first to transform to a set of four real fields, and then to diagonalize the resulting real and symmetric matrix by an orthogonal transformation (for reasons that will become clear momentarily.)

We express $\Psi$ as
\begin{equation}
\Psi=TX\text{, where}\thickspace X=\frac{1}{\sqrt{2}}\left(\begin{array}{c}
\text{Re}(\xi-\xi_{vac}) \\
\text{Im}(\xi-\xi_{vac}) \\
\text{Re}(\sigma-\sigma_{vac}) \\
\text{Im}(\sigma-\sigma_{vac})  \end{array} \right).
\end{equation}
Note that $T$ it is unitary. Lastly, we take
\begin{equation}
X=OY
\end{equation}
where $O$ is the orthogonal matrix containing the (normalized) eigenvectors of $T^{\dagger}B^{\dagger}MBT$ as columns. That is,
\begin{equation}
Y^TO^T T^{\dagger}B^{\dagger}MBTOY=Y^TDY=\sum_{i=1}^{4}m_i^2 y_i^2
\end{equation}
where $D$ is diagonal.

The full transformation to the real basis that simultaneously diagonalizes the Hessian and canonically normalizes the kinetic terms (locally) is
\begin{align*}
\tilde{J}:(\Phi-\Phi_{vac})\rightarrow Y\\
\tilde{J}=(BTO)^{-1}
\end{align*}
So, if we compute the rank three and four symmetric tensors of partial derivatives of the scalar potential at the vacuum in the original complex coordinates,
\begin{equation}
A_{ijk}=\frac{\partial}{\partial\Phi^i}\frac{\partial}{\partial\Phi^j}\frac{\partial}{\partial\Phi^k}V\bigg|_{\Phi_{vac}},
\end{equation}
we need to transform according to the standard tensor transformation law,
\begin{align}
A_{i'j'k'}&=\frac{\partial\Phi^i}{\partial Y^{i'}}\frac{\partial\Phi^j}{\partial Y^{j'}}\frac{\partial\Phi^k}{\partial Y^{k'}} A_{ijk}\\
&=J^{i}_{i'}J^{j}_{j'}J^{k}_{k'}A_{ijk},
\end{align}
with $J$ defined as the inverse of $\tilde{J}$, and similarly for the fourth order coefficients, $A_{i'j'k'l'}$ as well. By choosing the particular unitary transformation that diagonalizes the rescaled mass matrix and yields a basis of real fields, $y_i$, we avoid concerning ourselves about reordering of the entries of the complex symmetric rank three and four tensors we compute numerically.

This final task, evaluating the Taylor coefficients in the original complex field coordinates, may seem trivial. After all, the vacua reside near the conifold point where the period functions are polynomial in $(z-1)$ and/or $(z-1)^n\log(z-1)$, so we never need evaluate the divergent Meijer-G's in the scalar potential resulting from derivatives of $\Pi_0$ (and $\bar{\Pi}_0$). However, the seemingly mundane exercise of symbolically simplifying the near conifold scalar potential resulting from plugging in the period expansions and  its partial derivatives proves prohibitive. 

This is mainly due to the cumbersome way the scalar potential mixes up the periods and  the relative factors of
\begin{equation}
\sim \frac{\partial^m\Pi^{\dagger}Q^{-1}\partial^n\Pi}{(\Pi^{\dagger}Q^{-1}\Pi)^{m+n}}
\end{equation}
between its summands. Since we are plugging in eighth order expressions for the periods, the number of terms that need to be expanded, collected and organized is large. Rather than try to undo the natural packaging of the period functions, we make use of it. 

Our strategy is to express each entry in the tensors we wish to evaluate, the $A_{ijk}$ and $A_{ijkl}$, in terms of simple combinations of a small number of blocks. Each block is built out of smaller elements, which include the periods, their derivatives and combinations thereof (the K{\"a}hler potential for the complex structure and its partial derivatives). For each model in the ensemble, we evaluate the periods and their derivatives up to fifth order once. The values in this array are then combined appropriately to obtain the rest of the elements needed to construct the blocks. The blocks are then assembled into each entry required by the tensors. Essentially, we are exploiting the fact that many expressions appear repeatedly within a given entry and across entries, and so we need not evaluate them repeatedly.

The blocks consist of all the K{\"a}hler covariant derivatives of the superpotential, $F_I$, and their partial derivatives up to third order taken with respect to any of the complex fields. Some of these are zero, for instance $\partial_{\bar{\tau}} F_z$, but note $\partial_{\bar{z}}F_z$ is in general non-vanishing. Since the scalar potential is 
\begin{equation}
V=e^{\mathcal{K}}\left(\mathcal{K}^{z \bar{z}}F_z \bar{F}_{\bar{z}} + \mathcal{K}^{\tau \bar{\tau}}F_{\tau} \bar{F}_{\bar{\tau}}\right)
\end{equation}
its third and fourth order partial derivatives involve several terms linear in $F_I$. All such contributions vanish however because the SUSY condition, $F_I=0$, is satisfied at all the vacua. By only retaining those terms in $A_{ijk}$ and $A_{ijkl}$ that have at least one partial derivative on $F_I$ and at least one on its conjugate we have more manageable expressions for each Taylor coefficient that need to be combined.

The individual blocks are compact when expressed in terms of the elements. For example,
\begin{align}
\partial_z\partial_{\bar{z}}\partial_{\tau}F_z&=\partial_z\partial_{\bar{z}}\partial_{\tau}(F-\tau H)\cdot (\Pi' + \mathcal{K}_z \Pi)\\
&=-H\cdot(\Pi''+\mathcal{K}_{z\bar{z}}\Pi'+\mathcal{K}_{zz\bar{z}}\Pi).
\end{align}
This approach enables us to compute both rank $3$ and $4$ tensors for the entire sample of $1358$ vacua in time of order tens of minutes.

Lastly, we note that the form of the Hessian outlined in subsection \ref{hessiansection} is confirmed by comparing that obtained by direct differentiation with that built from the metric and (separately constructed) $Z$ and $\bar{Z}$ matrices in the original noncanonical basis. The percent deviation between the eigenvalues of the two are on the order of $10^{-13}$.

\section{Results}\label{results}
In this section we analyze the distributions of masses and couplings for a collection of $1358$ vacua, found using the vacuum hunting algorithm described in subsection \ref{vacuumsearch}. There is a great deal of structure built in from the get-go. The task is to untangle the randomness that is present from that structure. As indicated in subsection \ref{hessiansection}, the no-scale structure for a theory with $N$ complex moduli is responsible for pairing the $2N$ scalar masses of the effective field theory. 

The association of each mass pair with a single one of the complex scalars (for us, either $z$ or $\tau$) is \emph{not} expected, a priori, because of the mixing between the axio-dilaton and the complex structure in the scalar potential. 

However, for near conifold flux vacua in the mirror quintic's moduli space that satisfy the SUSY condition, $D_IW=0$, the two scalar fields \emph{do} approximately separate; ever more so as the vacuum-to-conifold distance is diminished. Tied to this cleaving of the field space is also a scale separation between the associated axio-dilaton and complex structure mass pairs. We observe that such a hierarchy percolates through third and fourth order couplings. All this structure, nearly universal across our ensemble, can be traced back to a single quantity: the mirror quintic's Yukawa coupling. 

We will show that the larger the Yukawa coupling, the more exaggerated this structure becomes. Its singularity, located precisely where vacua accumulate -- at the conifold point -- is responsible for the expected pattern of near conifold mass and couplings dominating the ensemble. We use the qualifier ``expected" because there is one random ingredient: the vev of the superpotential. Depending on your point of view, it muddies otherwise sharply defined features, or provides the possibility of freedom from rigidity (albeit a vanishingly small possibility as the conifold is approached).   

In this section we first establish the distance of a vacuum from the conifold as the key quantity controlling the degree to which structure is amplified or diluted. Next, we build intuition for the mass pairs and their distributions. Finally we present the hierarchies and correlations observed in the data for cubic and quartic couplings, which similarly is attributable to the singular dependence on the vacuum-to-conifold distance. For ease of discussion, we will loosely refer to the magnitude of a vacuum's canonical complex structure coordinate, $|\xi_{vac}|$, as its distance to the conifold in moduli space. More precisely, this distance is a monotonically increasing fucntion of $|\xi|$, but is not identically equal to it. During our investigation we developed a Random Matrix Model that accurately captures this particular combination of both regularity and randomness. We comment on the possible generalizations of these results to models with more complex structure moduli in the Discussion section. 

\subsection{Masses}
Since the SUSY condition is satisfied at our vacua, the complex 2$\times$2-matrix $Z_{IJ}\equiv D_I D_J W$ is the matrix of vacuum values of the partial derivatives of $F_I$, specifically,
\begin{equation}
Z=\left(\begin{array}{cc}
\partial_{\tau}F_{\tau}& \partial_{z}F_{\tau}\\
\partial_{\tau}F_z&  \partial_z F_z\end{array} \right).
\end{equation}
Not all entries in this matrix are independent. It's form is restricted because there is no mixing between the complex structure and the axio-dilaton at the level of the K{\"a}hler potential ($\mathcal{K}_{z\tau}=0$) and also because $\mathcal{K}_{\tau\bar{\tau}}=-{\mathcal{K}_\tau}^2$. These two, together with the SUSY condition in $\tau$, imply $Z$ has the form,
\begin{equation}
Z=\left(\begin{array}{cc}
0&Z_{01} \\
Z_{10}&Z_{11}\end{array}\right),
\end{equation}
where the entries are complex valued (and $Z_{01}=Z_{10}$).

The two nontrivial entries, it turns out, are related by a known analytic function when the canonical basis is used (the fields we labeled $\xi$ and $\sigma$, whose corresponding $Z$ matrix is $Z^{can}$). For compactifications of type IIB on general Calabi-Yau the following equation is valid at solutions to $D_I W=0$,
\begin{equation}
Z_{I J}=\mathcal{F}_{I J K}\bar{Z}^{0 K}\label{yukawaZ}
\end{equation}
in a basis where the fields in the effective action are canonically normalized. The $\mathcal{F}_{I J K}$ in eq. \ref{yukawaZ} are the Yukawa couplings between the Calabi-Yau's complex structure moduli and their fermionic counterpart in the effective field theory. Since the mirror quintic has a single complex structure modulus we have the direct proportionality,
\begin{equation}
Z_{11}=\mathcal{F}_{111}\bar{Z}^{0 1}.
\end{equation}
where we've suppressed the ``can" superscripts.

The Yukawa coupling is singular at the conifold point. Its exact analytic form was found by Candelas and de la Ossa in their seminal papers on the manifold (see for e.g. \cite{Candelas}). Stated in terms of their complex structure field coordinate before canonical normalization, $\psi$, which is related to ours by $z=\psi^{-5}$ the Yukawa coupling takes the form
\begin{equation}
\kappa_{\psi\psi\psi}=\left(\frac{2\pi i}{5}\right)^3\frac{5\psi^2}{1-\psi^5}.\label{candelas-yukawa}
\end{equation}
Note that while zero and infinity switch under the coordinate transformation between $\psi$ and $z$, the conifold point is fixed. The conifold singularity in eq. \ref{candelas-yukawa} persists through the coordinate transformation to our $\mathcal{F}_{111}$, and manifests as one naively expects (as a $1/\xi$ divergence) with minor modification. This is because the prepotential from which $\kappa_{\psi\psi\psi}$ derives is the same as ours. 

It is useful to briefly sketch the calculation of Candelas and de la Ossa in order to understand the origin of the divergence, as well as its leading order form in our coordinates. They define a set of functions, ``Wronskians", in terms of derivatives of the prepotential. The $k$th Wronskian is given by
\begin{equation}
W_k=\mathcal{Z}^i\frac{d^k}{d\psi^k}\mathcal{G}_i-\mathcal{G}^i\frac{d^k}{d\psi^k}\mathcal{Z}_i
\end{equation}
where $\mathcal{Z}^i$ and $\mathcal{G}_i$ are an intersecting pair of periods (in an integral and symplectic basis). Since there are four nontrivial cycles, $i$ ranges from $1$ to $2$. The prepotential is
\begin{equation}
\mathcal{G}=\frac{1}{2}\mathcal{Z}^i\mathcal{G}_i.
\end{equation}

A crucial next step is to identify the Yukawa coupling,
\begin{equation}
\kappa_{\psi\psi\psi}=\int \Omega\wedge\frac{d^3 \Omega}{d\psi^3},
\end{equation}
with the third Wronskian. Together with the properties of Calabi-Yau, particularly the fact that the periods solve the Picard-Fuchs equation, they obtain an ordinary differential equation for $W_3$, whose solution is given in eq. \ref{candelas-yukawa}. 

When the fields in the effective action are canonically normalized, the Yukawa coupling receives a total factor of the inverse vacuum K{\"a}hler metric for the complex structure raised to the \emph{three halves}; one half power from the rescaling of the scalar field and one full power from the transformation of the fermion (one half power for each of the two factors of the fermion in the original interaction term $\sim\kappa\phi\chi\chi$). 

\begin{figure}
\centering
\includegraphics[width=.8\textwidth]{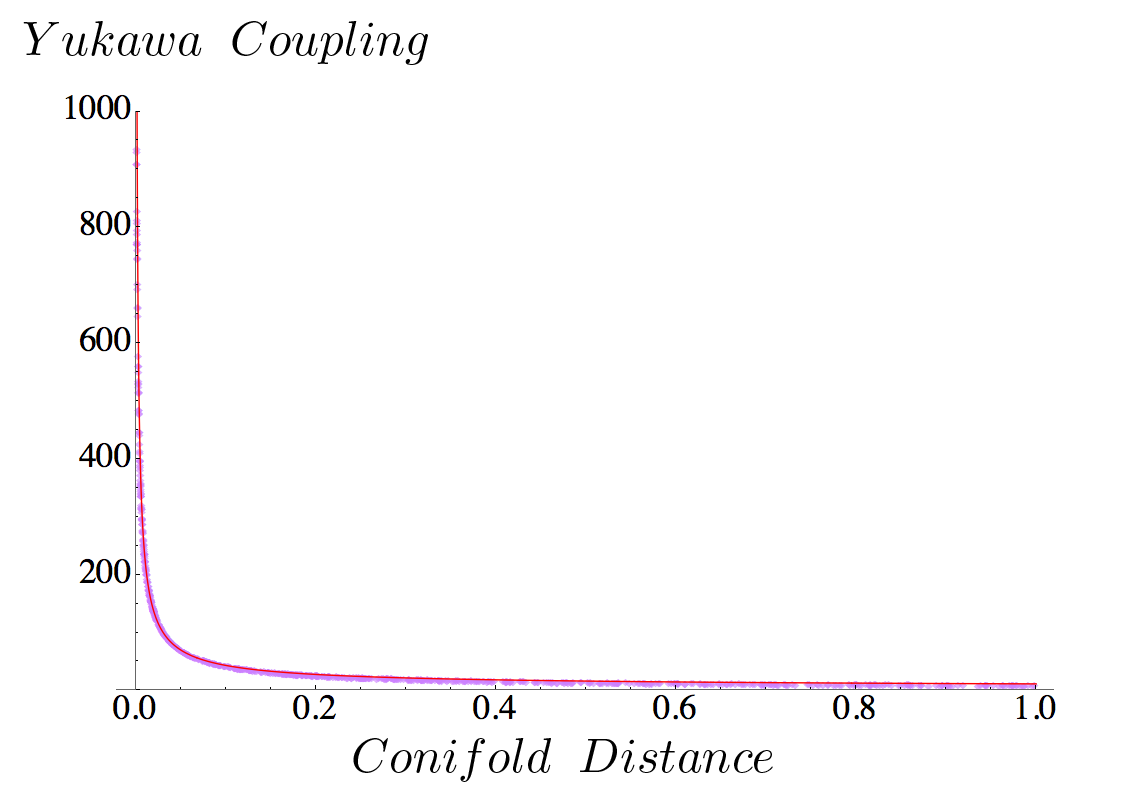}
\caption{Plotted in light purple is the ratio of the magnitudes of the entries of the $Z$ matrix for our vacua (the $11$ entry over the $01$ entry) on the vertical, against $|\xi_{vac}|$ which is a measure of the vacuum to conifold distance. Special geometry implies that this ratio ought to be the magnitude of the Calabi-Yau Yukawa coupling. The function plotted in red is a numerical fit of the data to the leading order form of the mirror quintic's $|\mathcal{F}|^2$. Note the extremely good agreement between the two, and the divergence at the conifold point, precisely where vacua accumulate.}  
\label{ratio}
\end{figure}

We've already accounted for one half of the total three halves by canonically normalizing our $z$ coordinate. This implies the ratio of our $Z^{can}_{11}$ to $Z^{can}_{01}$ will have a leading order behavior of $1/(\xi \log{\xi})$, since the K{\"a}hler metric goes like $\log(\xi)$ near the conifold point, $\xi=0$. In Figure \ref{ratio} we display the actual vacuum data for the magnitude of this ratio against the conifold distance. A numerical fit to the leading order form is overlaid in red. Note the exceedingly tight agreement between the two.
Essentially, the $\sim\frac{1}{\xi}$ dependence comes from a contribution $\sim \Pi_3\frac{d^3}{d\xi^3}\Pi_0$, since
\begin{align}
 \Pi_3\frac{d^3}{d\xi^3}\Pi_0&=\mathcal{O}(\xi)\frac{d^3}{d\xi^3}\left(\mathcal{O}(\xi)\log{\xi}+\text{analytic}\right)\\
 &=\mathcal{O}(\xi)\left(\mathcal{O}(\xi)\frac{d^3}{d\xi^3}\log{\xi}+\mathcal{O}(1)\frac{d^2}{d\xi^2}\log{\xi}+\mathcal{O}(1)\right)\\
 &=\mathcal{O}(\xi)\mathcal{O}(1/\xi^2)\\
 &=\mathcal{O}(1/\xi)
\end{align}

Our vacua live in a region where $|\mathcal{F}|>>1$, so $Z_{11}$ always dominates $Z_{01}$. This is consequential for the mass spectra and coordinate transformation that enters into the computation of the subsequent higher order couplings. Expressed in terms of the magnitude of the Yukawa coupling (where we've suppressed the ``$111$" indices), the $Z\bar{Z}$ matrix takes the form,
\begin{equation}
Z\bar{Z}=|Z_{01}|^2\left(\begin{array}{cc}
0&|\mathcal{F}|e^{-i\delta} \\
|\mathcal{F}|e^{+i\delta}&|\mathcal{F}|^2+1\end{array}\right),
\end{equation}
whose eigenvalues come in the pair,
\begin{equation}
\Lambda^2_{\pm}=\frac{|Z_{01}|^2}{2} \left(|\mathcal{F}|^2+2\pm|\mathcal{F}|\sqrt{|\mathcal{F}|^2+4}  \right)\label{ZZbareigs}.
\end{equation}
The larger of these is always $\Lambda_+^2$, so, in our labeling convention for the $\lambda_i$'s we identify $\lambda_1^2=\Lambda^2_+$, and $\lambda_2^2=\Lambda_{-}^2$.

Note that in either limit, $|\mathcal{F}|>>1$, or the reverse, we have $\Lambda_+^2>>\Lambda_-^2$. If $|\mathcal{F}|>>1$ the eigenvector associated with the larger eigenvalue is almost entirely contained within the span of the complex structure field, and in the opposite limit within that of axio-dilaton field. Since we always have the former case, the largest eigenvalue of $Z\bar{Z}$, $\lambda_1^2$, is associated always with the complex structure, and the smaller, $\lambda_2^2$, with the axio-dilaton. An immediate consequence of this is the cleaving of the eigenspace of the Hessian in two. 

One subspace is spanned almost entirely by the complex structure and is associated with the mass pair $m_{1\pm}^2$, while the other is spanned by the axio-dilaton and is associated with $m_{2\pm}^2$. This is because the 2$\times$2 blocks entering into the diagonalization of the Hessian, which would otherwise mix these two fields, are approximately equal to the identity. These 2$\times$2 blocks in the complex field coordinates of section \ref{hessiansection} are $U$ and its Hermitian conjugate. In the real field coordinates of section \ref{taylor-coeffs-alg} they form two by two blocks in $O$, upon a reordering of rows and columns. 

Whether or not the hierarchy in the $\lambda_i$ leads to a hierarchy between the two mass pairs -- a heavy complex structure pair and a light axio-dilaton pair -- depends on the relative sizes of $|W|$, $\lambda_1$ and $\lambda_2$. More precisely, we begin by noting there is no ambiguity about the heaviest mass. It is always $m_{1+}^2$, which for us is always associated with $z$. It's partner (still associated with $z$) need not be second heaviest, however. To see why note that
\begin{align*}
\left(\lambda_1-|W|\right)^2&<\left(\lambda_2\pm|W|\right)^2\\
\lambda_1^2-\lambda_2^2&<2|W|(\lambda_1\pm\lambda_2)\\
\lambda_1\mp\lambda_2&<2|W|.
\end{align*}
So, if half of the gap between the $\lambda_i$ is less than the magnitude of the vacuum superpotential the second heaviest of the four masses is the larger of the axio-dilaton masses, $m_{2+}$. If the average of the $\lambda_i$ is \emph{also} less than the magnitude of the superpotential then the third heaviest mass is the lighter of the axio-dilaton pair, and the lightest of the four masses is the lighter of the complex structure pair.
To summarize, the naive/expected ordering among the masses,
\begin{equation}
m_{1+}^2>m_{1-}^2>m_{2+}^2>m_{2-}^2\label{expected-hier}
\end{equation}
is realized if the difference condition is not met (large discrepancy between the $\lambda_i$'s). 

The middle two masses swap places,
\begin{equation}
m_{1+}^2>m_{2+}^2>m_{1-}^2>m_{2-}^2\label{middle-hier}
\end{equation}
if the gap condition is met but the average condition is \emph{not}. Lastly, if both conditions are met the lighter complex structure mass shuffles all the way to the bottom of the mass scale,
\begin{equation}
m_{1+}^2>m_{2+}^2>m_{2-}^2>m_{1-}^2.\label{last-hier}
\end{equation}

\begin{figure}
\centering
\includegraphics[width=\textwidth]{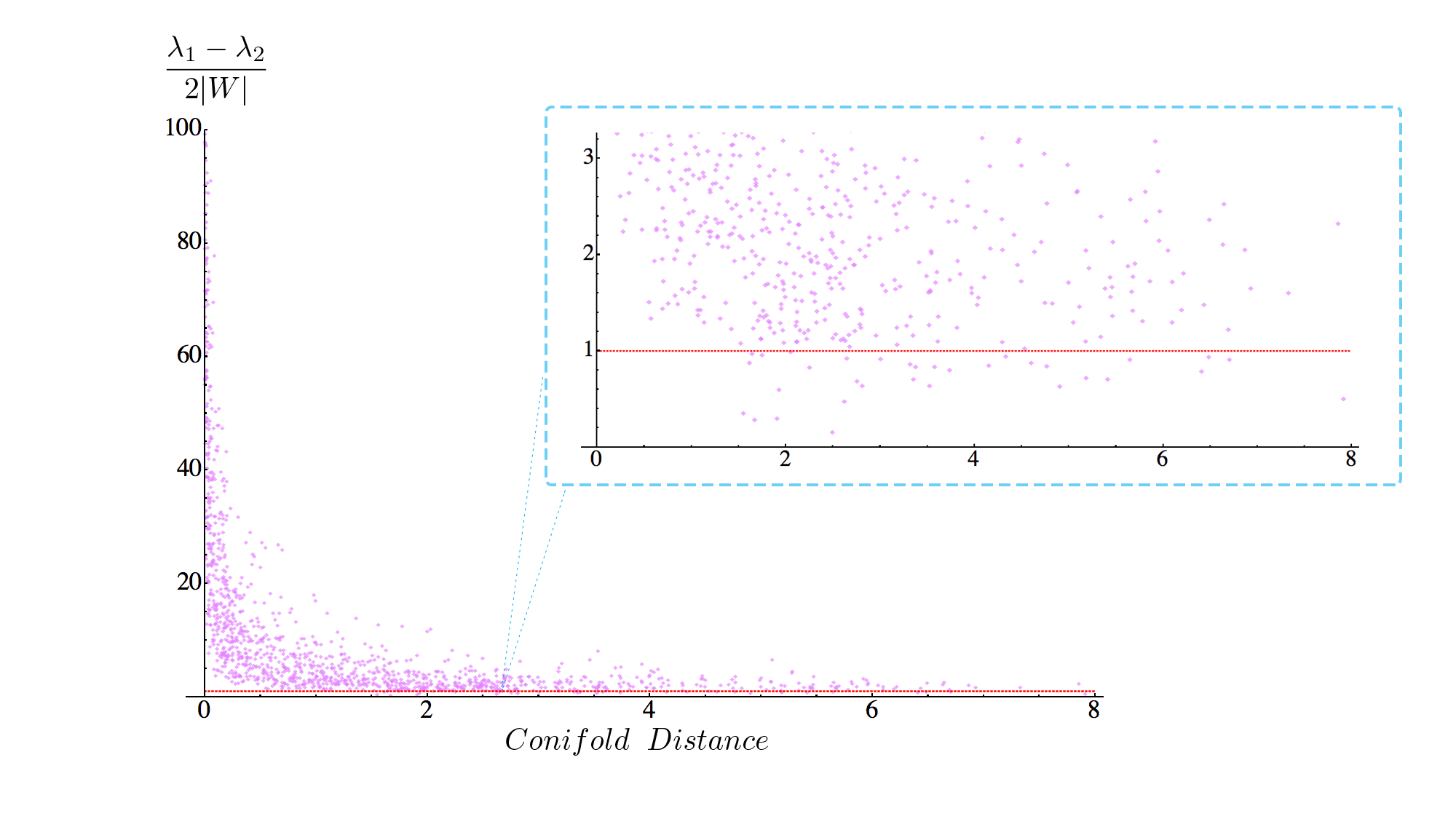}
\caption{The difference between the $\lambda_i$ divided by twice the vev of the superpotential vs. $|\xi_{vac}|$, illustrating the key quantity in the hierarchy condition. Data points that fall below the dashed pink line do not satisfy the condition and have a mass hierarchy that differs from the expected one by at least one swap. Note that as the conifold distance decreases the data points float upwards, confirming the expectation (based on the divergence of the Yukawa coupling) that the condition becomes ever more difficult to satisfy as the location of the vacuua approaches the conifold.} 
\label{masshierarchy}
\end{figure}

As we've seen, the difference in scale between the two distinct nontrivial entries of $Z$ diverges as the conifold point is approached. Since the larger the scale difference the larger the gap between the $\lambda_i$ will be, we expect the likelihood of the gap condition being met to diminish as the conifold point is approached. This is precisely what we find. In Figure \ref{masshierarchy} we plot 
$\lambda_1-\lambda_2$ divided by twice the magnitude of the superpotential against the conifold  distance for each vacuum. 
A horizontal line at $1$ is indicated by the dashed line, so points above this line fail the gap condition and the naive order exists, while those below have at least one rightward shift of $m_{1-}^2$ down the hierarchy in \ref{expected-hier}. 

\begin{figure}
\centering
\includegraphics[width=.6\textwidth]{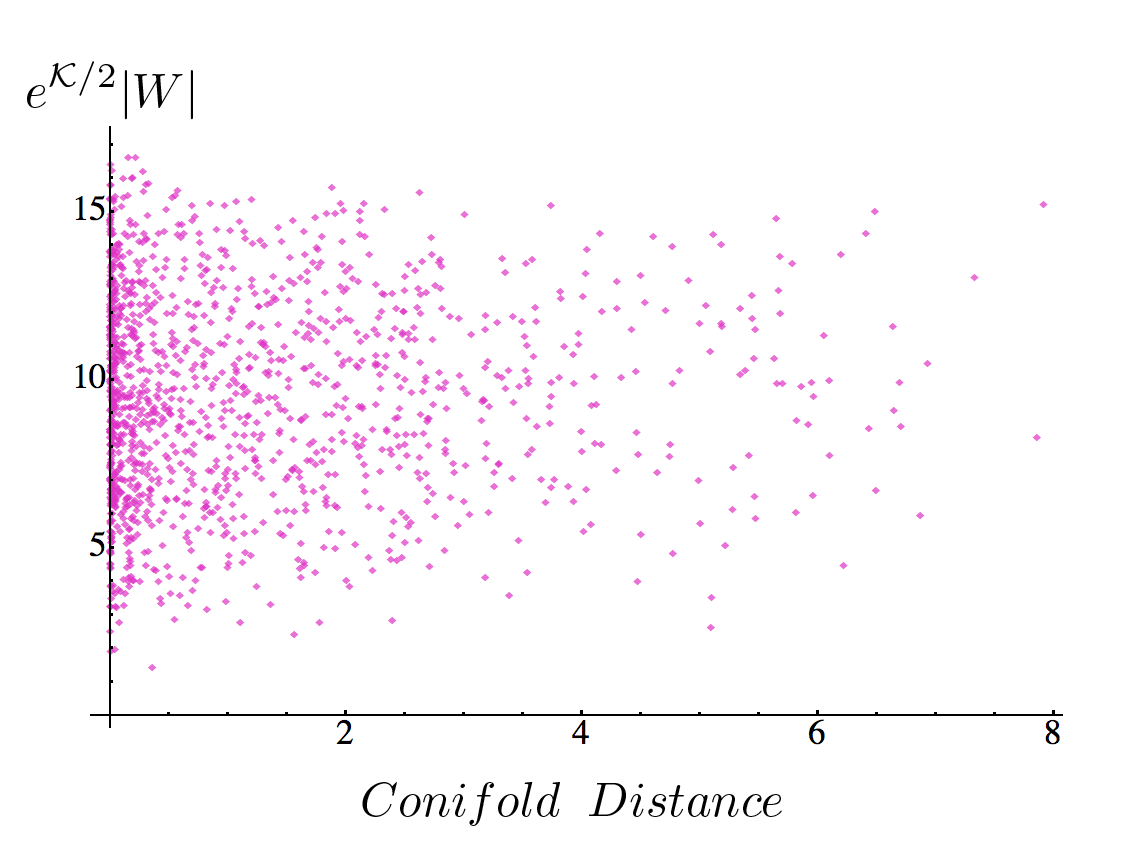}
\caption{A scatter plot of the magnitude of the vev of the superpotential vs. $\xi_{vac}$. Note that the two bear no significant dependence on one another. Data points become more clustered as one moves toward the peak or either quantity's distribution \emph{independently}.}  
\label{w-dist}
\end{figure}

There are two important observations. First, the vacua migrate upward as the conifold is approached making the condition ever more unlikely to be satisfied, verifying our expectation. Second, there are nonetheless a few vacua for whom the  condition is met. Specifically, we find $33$ out of $1358$ such instances, or $2.4\%$. The image toward the upper-right of Figure \ref{masshierarchy} shows the portion of the plot focused near the bottom (with exactly $33$ points below the dashed line). The random ingredient that allows for vacua to dip below the threshold for mass swaps is the magnitude of the superpotential. The vev shows no dependence on the distance of the vacuum from the conifold. This is shown in Figure \ref{w-dist}. 

\begin{figure}
\centering
\includegraphics[width=\textwidth]{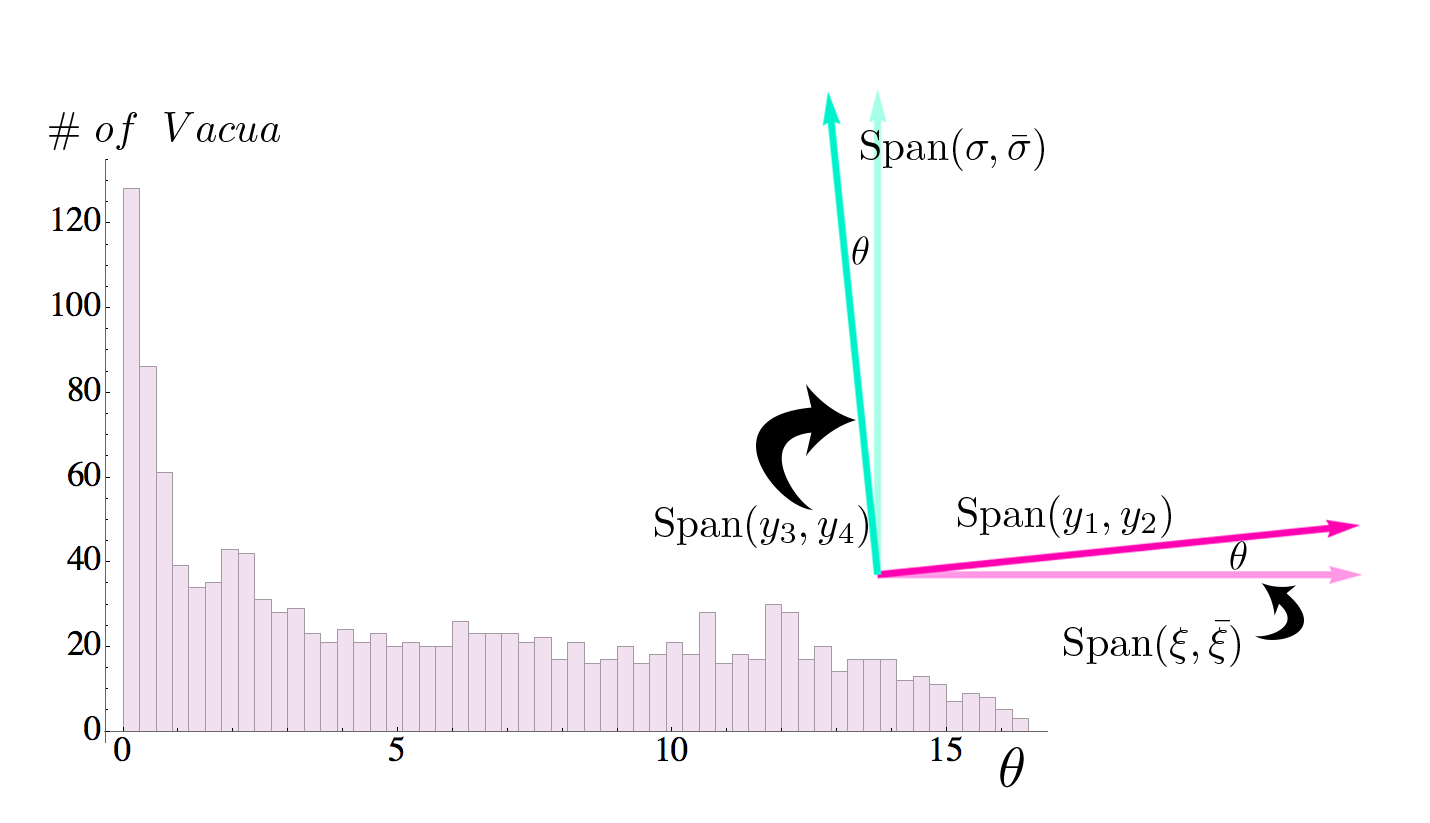}
\caption{A distribution of relative angle, $\theta$, between the complex structure subspace of the moduli space and the $m_{1\pm}^2$ eigenspace, which is identical to that between  the axio-dilaton subspace and the $m_{2\pm}^2$ eigenspace. The fact that the angles for all vacua are small indicates that the former pair are approximately parallel to each other, and likewise for the latter. A visual aid depicting this split of the eigenspace is shown to the right using the mean value of this angle, which is $5.85$ degrees for our vacua.}  
\label{eigen-split}
\end{figure}

We reiterate that in all cases, including these nonconformist $33$, the Hessian's eigenspace enjoys an approximate separation between the complex structure field space, and axio-dilaton field space. The angle between the subspace spanned by one of the moduli -- $\xi$ or $\sigma$ -- and that spanned by the two eigenvectors associated with one of the mass pairs -- $m_{1\pm}^2$ or $m_{2\pm}^2$ -- can be computed. In Figure \ref{eigen-split} we display the histogram of angles between the complex structure subspace and the $i=1$ mass pair for all vacua. The mean angle is 5.85 degrees, indicating that the subspaces are approximately parallel. The identical statement holds for the axio-dilaton subspace and the second mass pair. A visual depiction of the subspaces is included to the right of the histogram, and uses the mean angle.

\begin{figure}[h!]
\centering
\includegraphics[width=\textwidth]{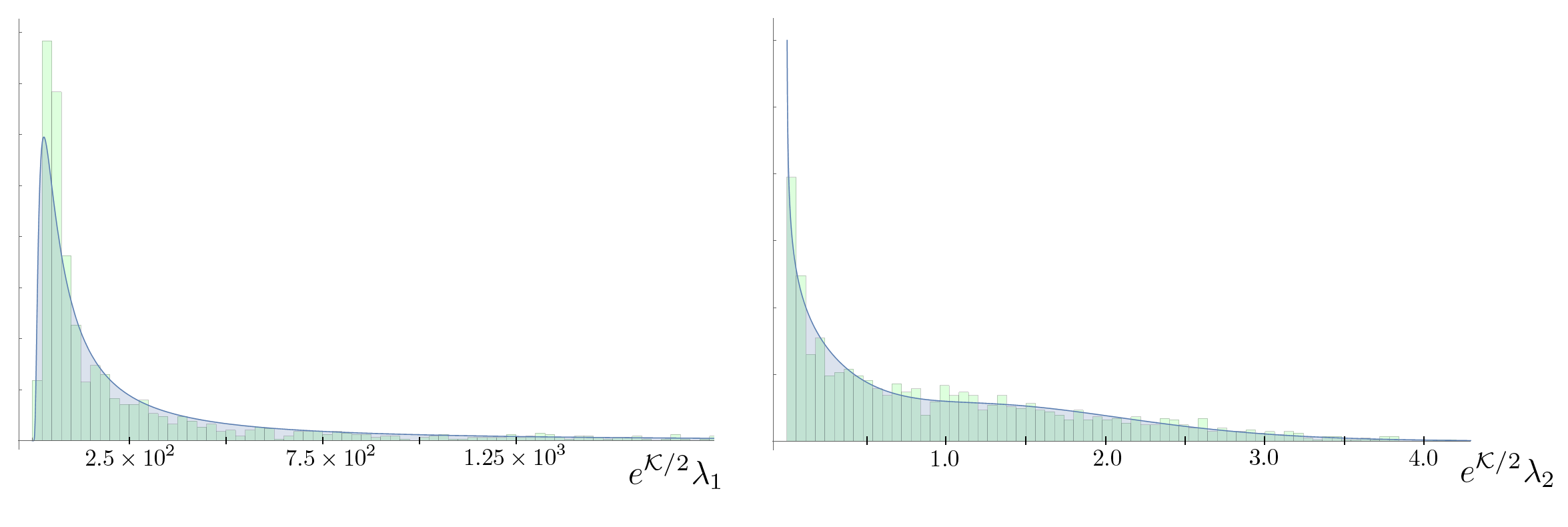}
\caption{Histograms of the K{\"a}ahler independent $\lambda_i$ for our ensemble of vacua.  Estimated distributions obtained numerically are plotted over each histogram in blue.}  
\label{lambda-dist}
\end{figure}

\begin{figure}[h!]
\centering
\includegraphics[width=.45\textwidth]{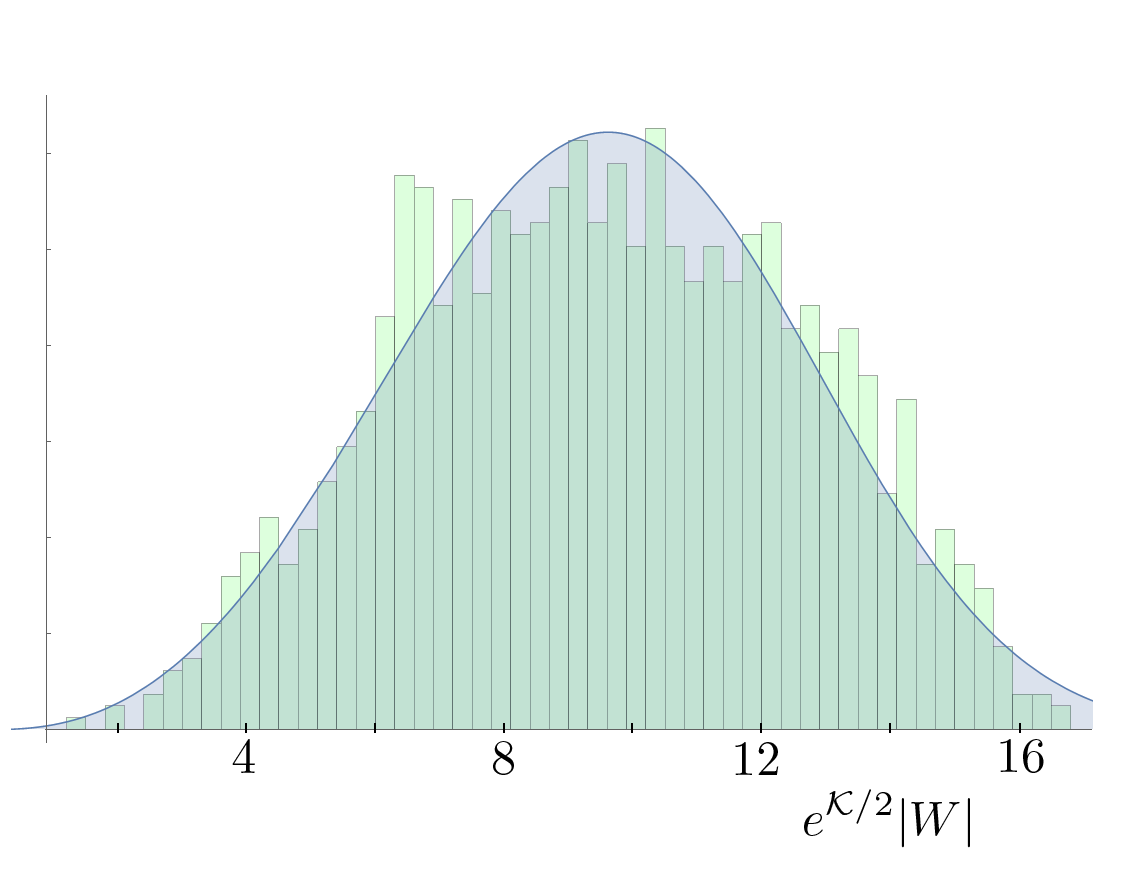}
\caption{The histogram of the K{\"a}ahler independent vevs of the superpotential for our ensemble. The estimated distribution obtained numerically is overlayed.}  
\label{W-dist}
\end{figure}

Now that we have established that this separation between $z$ and $\tau$ lines up with the half-way marker between the masses in virtually all cases, we turn to developing intuition for each pair. We display the distributions  of $\lambda_1$ and $\lambda_2$ in Figure \ref{lambda-dist}, and of $|W|$ in Figure \ref{W-dist}. We have absorbed a factor of the vev of $e^{\mathcal{K}/2}$ into the definitions of each of these three, as they are the correct K{\"a}hler invariant quantities, i.e. the physically relevant values to consider. 

As expected, the $\lambda_1$ distribution's scale is significantly larger than $\lambda_2$'s due to the accumulation of vacua where the Yukawa coupling diverges. 
Specifically, we find a difference of two to three orders of magnitude. The characteristics of the corresponding mass pairs will depend on the relative sizes of the $\lambda_i$ to $|W|$ individually. We find a superpotential that is approximately one order of magnitude larger than $\lambda_2$, but one order smaller than $\lambda_1$ (several orders smaller for vacua in the tail). 

\begin{figure}
\centering
\includegraphics[width=\textwidth]{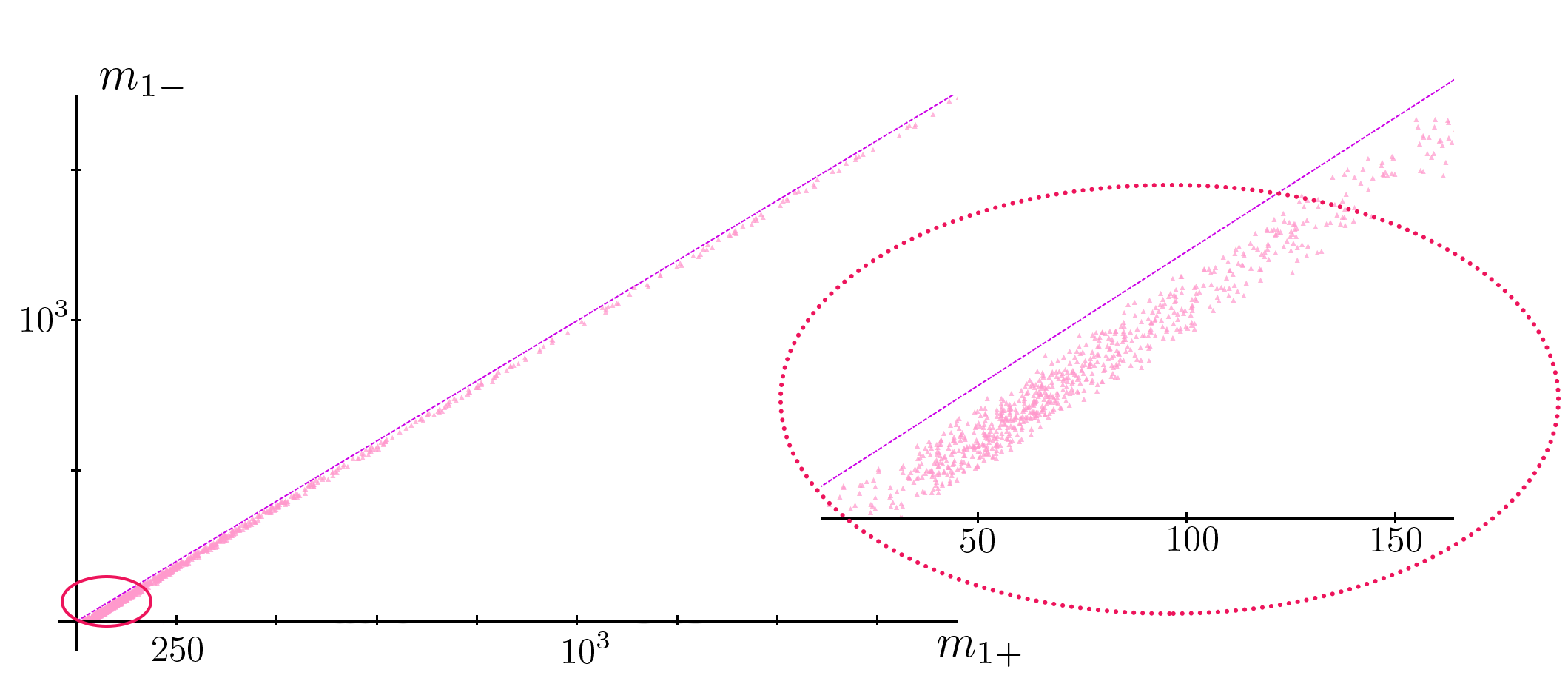}
\caption{A scatter plot for the mass pair associated with the complex structure modulus, with the heavier of the two, $m_{1+}$, on the horizontal and the lighter, $m^{1-}$on the vertical. A dashed line with slope one is plotted in purple. The portion of the plot focused where the masses distributions peak (i.e. where the data points cluster) is shown to the right.}  
\label{zmasses-scatter}
\end{figure}

\begin{figure}
\centering
\includegraphics[width=.6\textwidth]{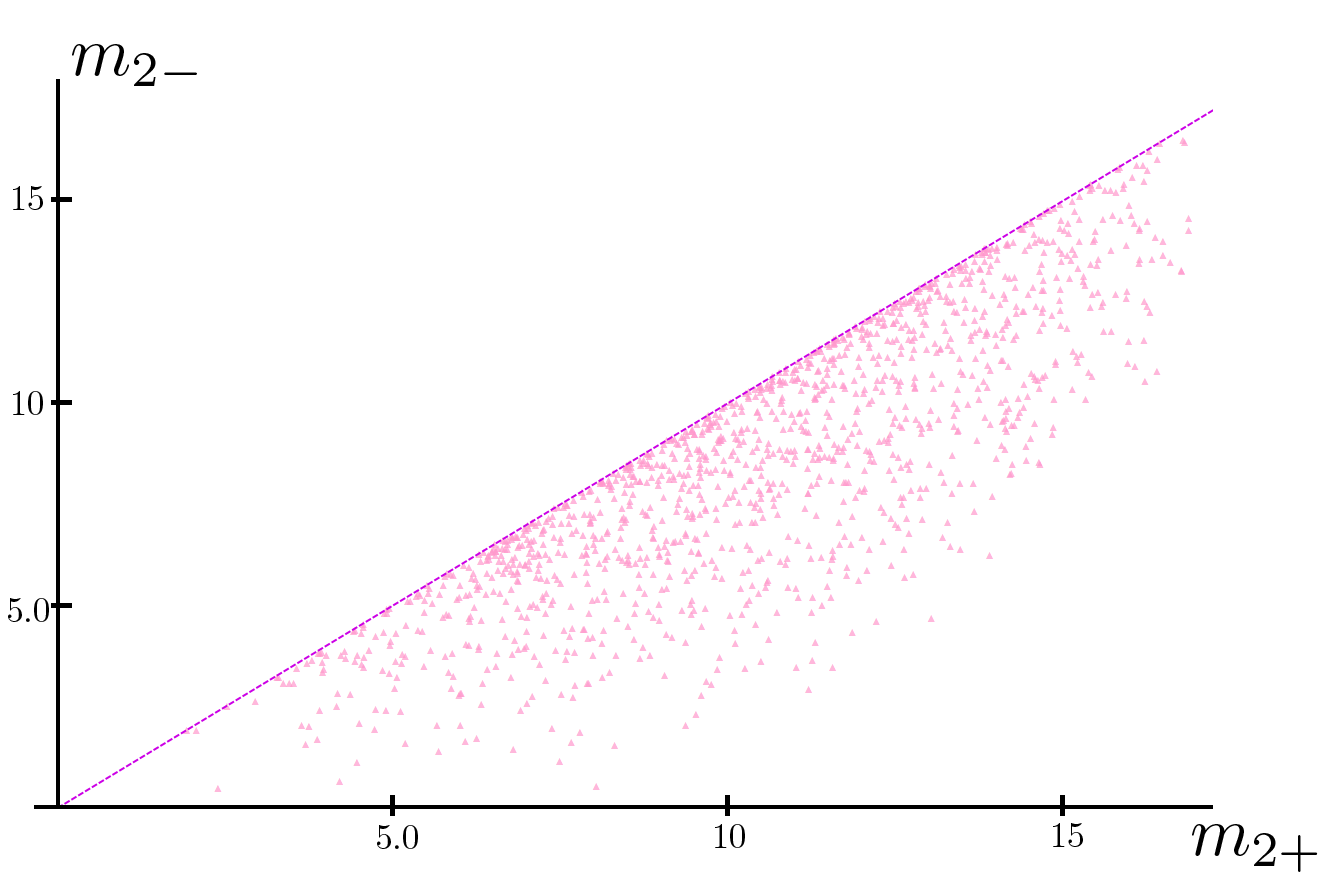}
\caption{The analogous scatter plot for the mass pair associated with the axio-dilaton as that in Figure \ref{zmasses-scatter}.}  
\label{taumasses-scatter}
\end{figure}

The resulting two mass pairs are displayed in Figures \ref{zmasses-scatter} and \ref{taumasses-scatter}, with the larger mass of each couple plotted on the horizontal. We immediately notice that the complex structure mass pair looks more tightly correlated than the axio-dilaton pair. This, as we'll analyze more precisely later, is entirely an artifact of the difference in scale between the two field's pairs; an effect that is exaggerated by the particularly wide range needed to include all of the $z$ mass data points in Figure \ref{zmasses-scatter}. The distribution (for both members of the $z$ pair) peaks at much lower values, around $100$.  
A fairer comparison with the $\tau$ masses, which are more widely/evenly distributed, would come from zooming in and excluding the $z$ masses' long tails. A partial zoom is shown in the ellipsoidal window on the right in Figure \ref{zmasses-scatter}. A more refined analysis will nonetheless reveal that the two fields have virtually identical levels of \emph{relative} degeneracy between the members of their respective pairs.

Turning to the $\tau$ data points shown in Figure \ref{taumasses-scatter}, note firstly that they fill in more of the triangular half below the diagonal \emph{including} the region immediately beneath the diagonal. This indicates that there is a larger variety among the dimensionful mass gaps for the axio-dilaton, than for the complex structure. There are more instances of near equality between the masses -- in an absolute/dimensionful sense -- as compared to those in the lower range of $z$'s distribution (there are far more data points along the diagonal boundary in the axio-dilaton's scatter plot than in the zoomed in complex structure's). There are also more instances of large differences for the $\tau$ pairs than the $z$'s. Clearly, the latter statement remains true when $z$'s tail is considered, but the former may not. These distinctions make sense given the distributions for the $\lambda_i$ and $|W|$. Essentially, the $\tau$ masses are dominated by the superpotential, which has a rather large spread and is not skewed (roughly Gaussian). This leads to a more uniform distribution horizontally throughout the triangle. 

The lack of space between data points and the diagonal is due to the fact that $\lambda_2$ peaks very near zero, and decays quickly before its decline steadies around $\sim 0.5$. This increases the frequency of $\lambda_2$'s that are completely negligible compared to $|W|$, and thus very nearly equal masses among the given pair. The axio-dilaton pairs' greater vertical extent throughout the triangle is due to the combination of the larger spread in $\lambda_2$ and $|W|$, and the fact that the intervals where they are supported partially overlap. Thus, more instances of close competition between $\lambda_2$ and $|W|$ occur than for $\lambda_1$ and $W$. 

These observations are helpful for building intuition, but a comparison of the \emph{degree} of degeneracy in the mass pairs of the two fields should involve dimensionless mass gaps, namely those scaled by the mean of the masses in each pair. Starting with the difference in the squared masses of the two members in the $i$th pair, $4\lambda_i |W|$, one finds the limiting form,
\begin{align}
&\frac{\Delta m_i/2}{m_{i,avg}}=\sqrt{\frac{\lambda_i |W|}{\lambda_i^2+|W|^2}}\\
\lambda_i<<|W|:\thickspace\thickspace &\frac{\Delta m_i/2}{m_{i,avg}}\thickspace\thickspace \rightarrow \frac{\lambda_i}{|W|}<<1\\
\lambda_i>>|W|: \thickspace\thickspace &\frac{\Delta m_i/2}{m_{i,avg}}\thickspace\thickspace \rightarrow \frac{|W|}{\lambda_i}<<1
\end{align}
That the result is the same for both limits simply reflects the fact that one can equally well view $\lambda_i$ as the degeneracy breaking term as one can $|W|$. A small $\lambda_i$ compared to $|W|$ yields a mass pair $m_{i\pm}\approx |W|\pm \epsilon$, and the reverse yields a mass pair $\approx \lambda_i\pm \epsilon$.

\begin{figure}
\centering
\includegraphics[width=.75\textwidth]{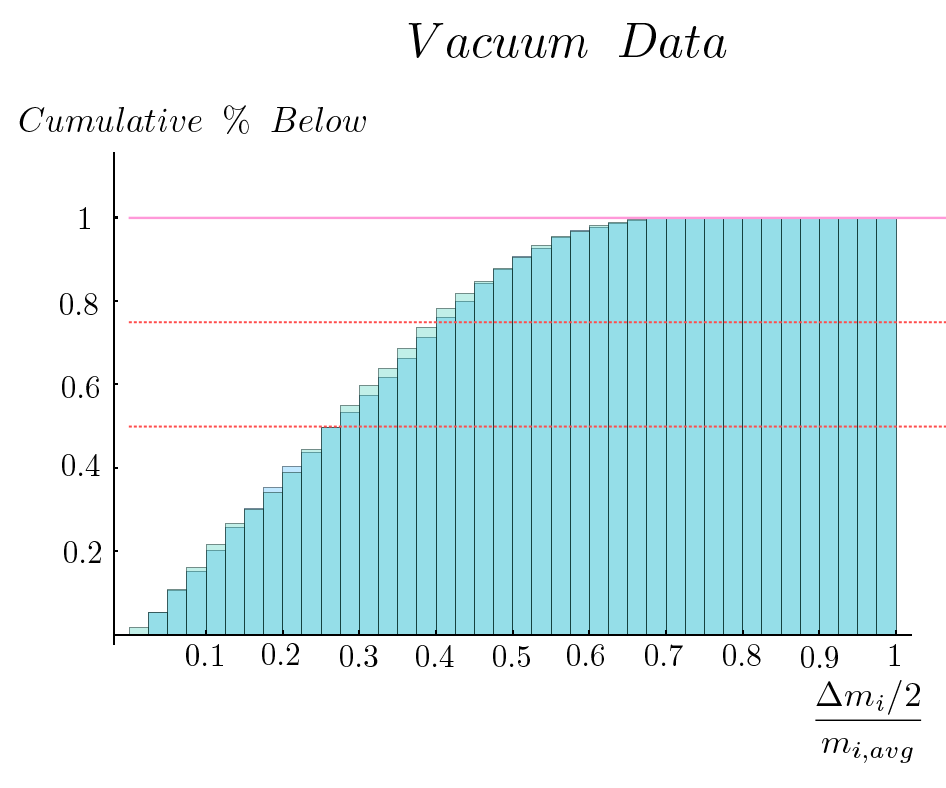}
\caption{The cumulative relative mass gaps for each pair of masses, $m_{i\pm}$. The pair associated with the axio-dilaton is shown in on top in the darker shade of blue, while that associated with the complex structure is shown behind in the lighter turquoise color. The two are nearly identical meaning that the relative degree of non-degeneracy between the masses in a given pair is distributed in the same manner across the vacua of our ensemble, regardless of  with which of the moduli the mass pair is associated.}  
\label{mass-gap-cdfs}
\end{figure}

Now, we may consider a probability density for each modulus as a function of the rescaled half mass gaps. For a given one of the moduli its value integrated over an interval $[a,b]$ would yield the probability of finding a vacuum for whom that modulus' associated masses each lie within $(b-a)m_{i,avg}$ of their mean, $m_{i, avg}$. We can then consider a cumulative density function obtained by integrating the probability density from $a=0$. In Figure \ref{mass-gap-cdfs} we display the histograms for our data corresponding to the cumulative density functions (their discrete analogs) for our sample of vacua.

Notice that the complex structure and the axio-dilaton's histograms are virtually indistinguishable. They both achieve $50\%$ within a threshold of $0.26$, and continue to rise together in step with $75\%$ of both fields having scaled gaps within the threshold of $0.40$. This assessment of relative degeneracy, or relative spread, is not evident from looking at the scatter plots alone.

The structure and patterns we've encountered in the masses clearly won't be replicated with an ordinary Random Matrix Model where the Hessian for each vacuum is taken to be  Wirshart -- a  Hermitian random matrix that is positive definite by construction. One essentially ``squares" a random (Wigner) matrix, $A$, which is not in general Hermitian, by multiplying it with its complex conjugate. The entries of the Wigner matrix are taken to be independent and identically distributed, that is drawn from an $\mathcal{O}(N^2)$ dimensional Gaussian. 

In light of the analytic form of the Hessian with which we begin, and the limiting behavior of one of its essential building blocks, $Z$, for near conifold vacua, we design a different Random Matrix Model. We've seen three (K{\"a}hler independent) parameters are ultimately in control. These are (1)  the proximity of a given vacuum to the conifold point, (2)  its value of $e^{\mathcal{K}/2}|Z_{01}|$ and (3) its value of $e^{\mathcal{K}/2}|W|$. We should be able to mimic the actual mass data with a random sample of these triples. The simplest case would be to treat each parameter independently. 

\begin{figure}
\centering
\includegraphics[width=\textwidth]{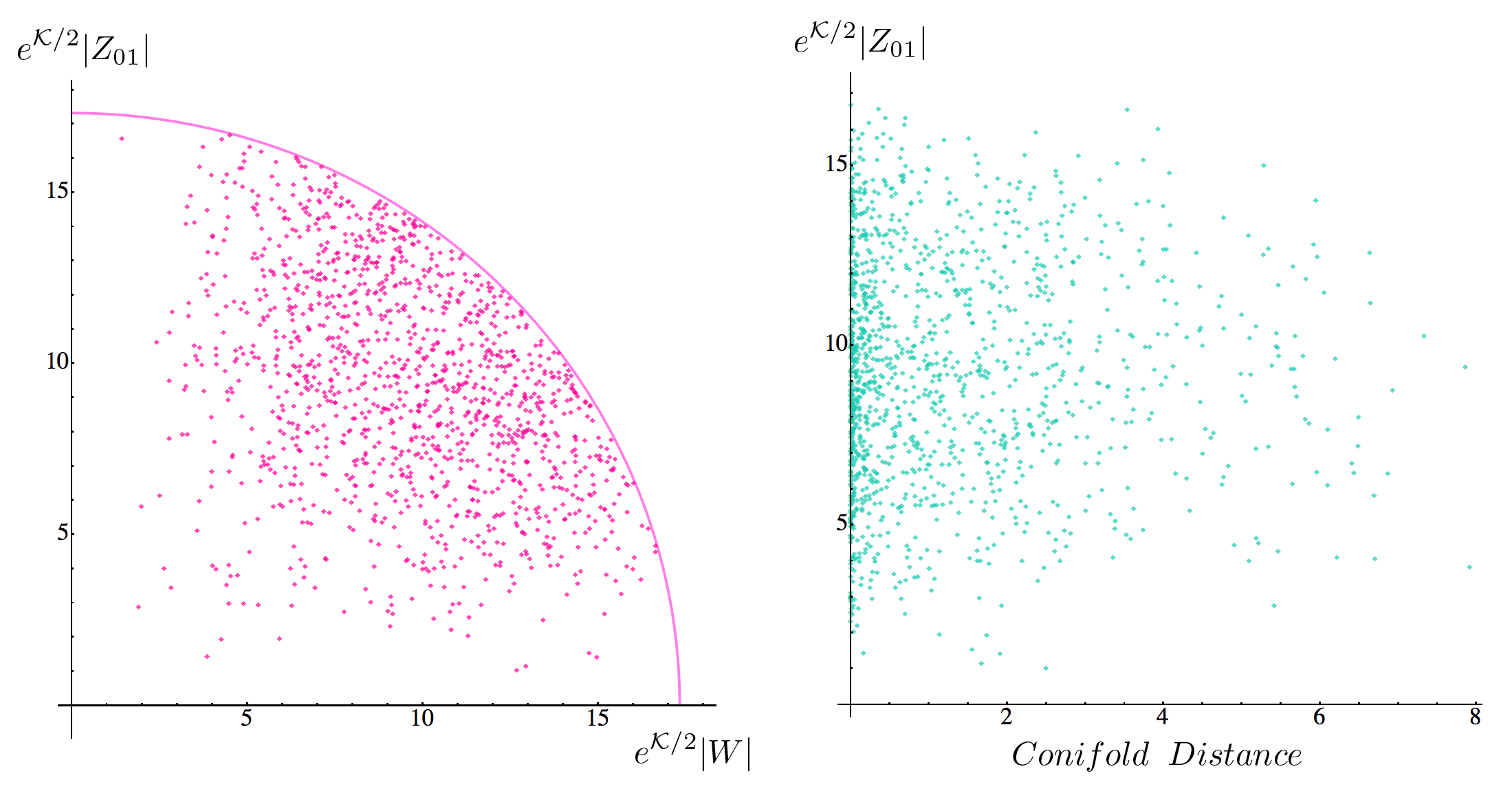}
\caption{The left panel shows a scatter plot of the (K{\"a}hler independent) magnitude of the $01$ entry of the $Z$ matrix in canonical coordinates against the vev of the superpotential. The sharp quarter-circle boundary is a manifestation of the tadpole condition, with the radius of the arc being $\sqrt{L_{max}}$ which for us is $\sqrt{300}$. Within the region allowed by the tadpole condition the data points exhibit no correlation. The panel on the right displays a plot of the $Z$ matrix entry against the remaining control parameter, the conifold distance. No correlation is evident between these either.}  
\label{WZ01-scatter}
\end{figure}

We saw earlier that control parameters (1) and (3) do not appear to depend on one another, as indicated by Figure \ref{w-dist}. A similar scatter plot for (1) and (2) is displayed in the right panel of Figure \ref{WZ01-scatter}, demonstrating their lack of correlation. The plot for (2) and (3) shows a sharp cutoff because the tadpole condition forbids these data points from leaving the quarter circle. It can be shown that the tadpole condition implies,
\begin{equation}
|Z_{01}|^2+|W|^2\leq L_{max}\label{tadpole-ineq}
\end{equation}
in Gaussian normal coordinates. 

The radius of the arc plotted in Figure \ref{WZ01-scatter} is indeed $\sqrt{L_{max}}=\sqrt{300}$.
Within this region however the data points vary independently. The empty bands along both axes are simply a reflection of the fact that the two distributions are peaked away from zero, with relatively little of their support coming from the interval $\approx [0,5]$. Just as with the other two scatter plots, the density of data points increases as either parameter is pushed towards the value where its distribution peaks while the other is held fixed. The lack of correlation within the region suggests we do the following. 

First obtain estimated probability densities for the K{\"a}hler invariant magnitudes in the canonically normalized fields, namely, $e^{\mathcal{K}/2}|W|$ and $e^{\mathcal{K}/2}|Z_{01}|$, as well as for the conifold distance. Draw a value from each distribution independently. If it has parameters (2) and (3) that violate inequality \ref{tadpole-ineq} dispose of it and redraw the triple until it is satisfied. Then compute the random eigenvalues, $\lambda_1^{2}$ and $\lambda_2^2$, by evaluating the Yukawa coupling at the randomly drawn conifold distance, and using it with the random $|Z_{01}|$ in eq. \ref{ZZbareigs}.

\begin{figure}
\centering
\includegraphics[width=\textwidth]{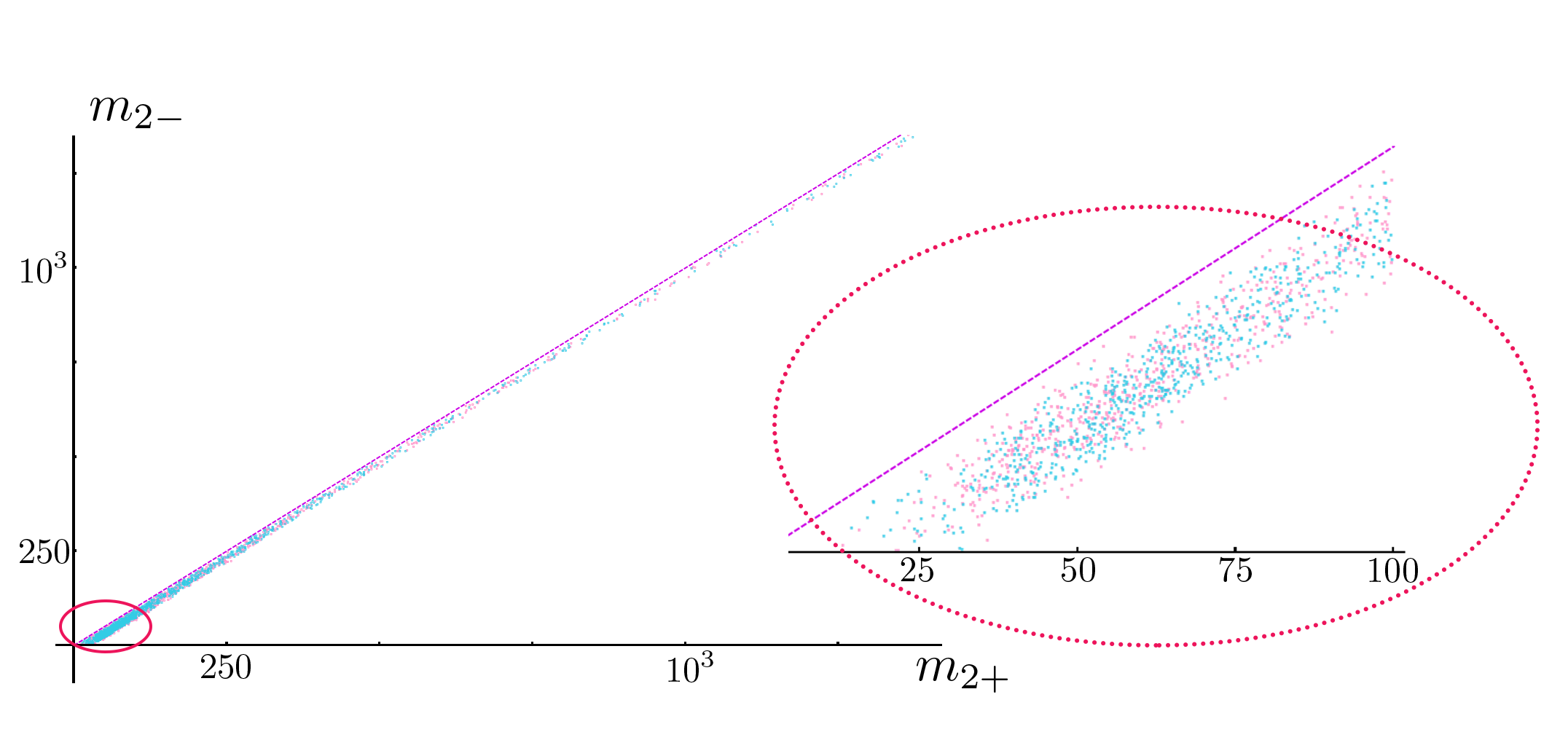}
\caption{The random matrix model data for the artificial complex structure mass pair are plotted in light blue over that of the actual vacuum data, shown in light pink. The panel to the right magnifies the portion of the plot where the mass distributions peak.}  
\label{zRMMmasses}
\end{figure}

\begin{figure}
\centering
\includegraphics[width=.6\textwidth]{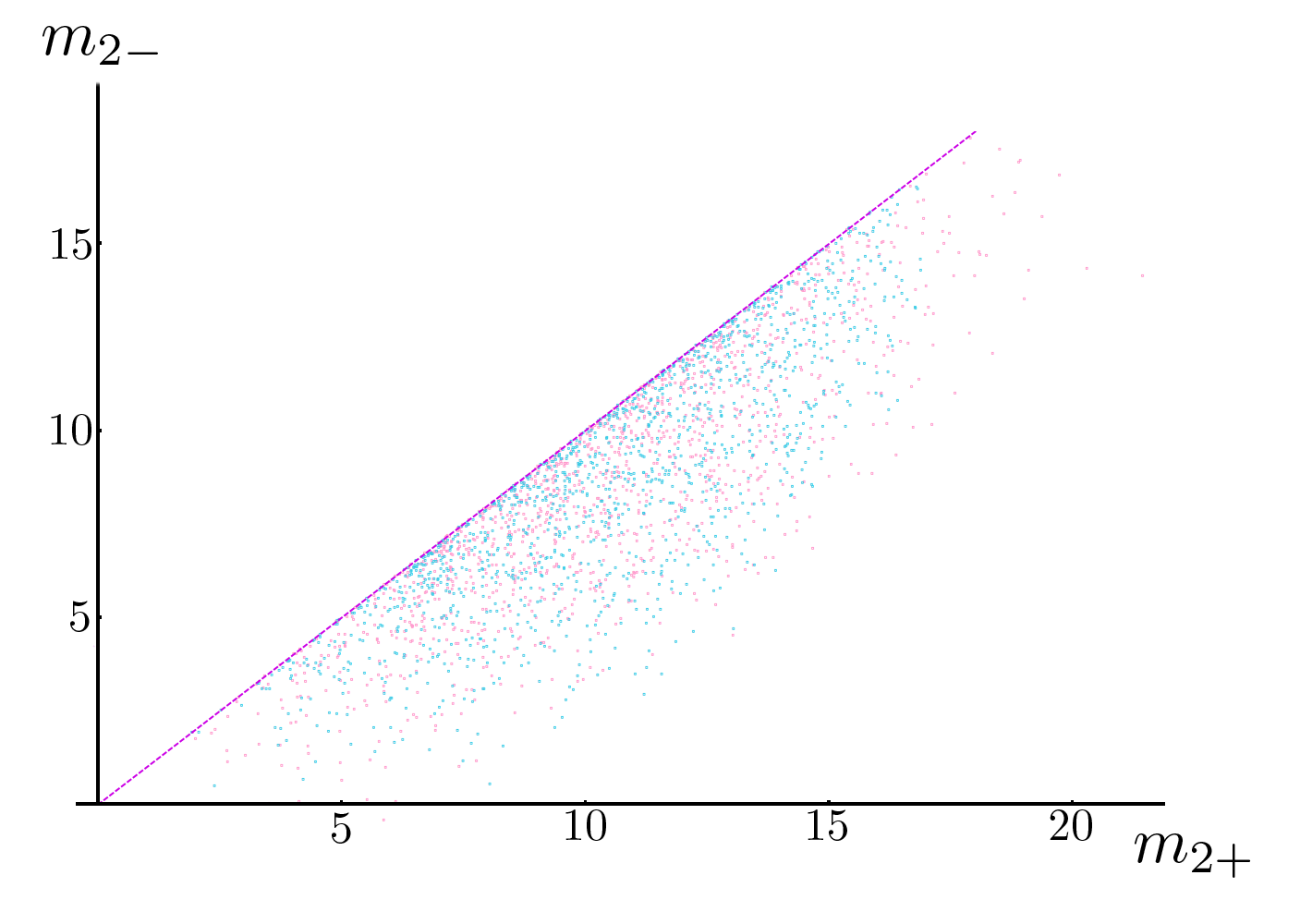}
\caption{The random matrix model data for the artificial axio-dilaton mass pair are plotted in light blue over that of the actual vacuum data, shown in light pink. }  
\label{axioRMMmasses}
\end{figure}

In Figure \ref{zRMMmasses} we display the resulting scatter plot for the $i=1$ random mass pair -- the artificial complex structure pair -- atop that from our sample of actual flux vacua for the full range of masses. The image in the ellipsoidal window zooms in on the range where complex structure mass distributions peak. The RMM does a good job in reproducing the data's features in both regimes: the peak and the tail of the mass distributions. The same is true for the RMM's performance with the axio-dilaton mass pair. The two are virually indistinguishable in their superposed scatter plots, shown in Figure \ref{axioRMMmasses}.

\begin{figure}
\centering
\includegraphics[width=.75\textwidth]{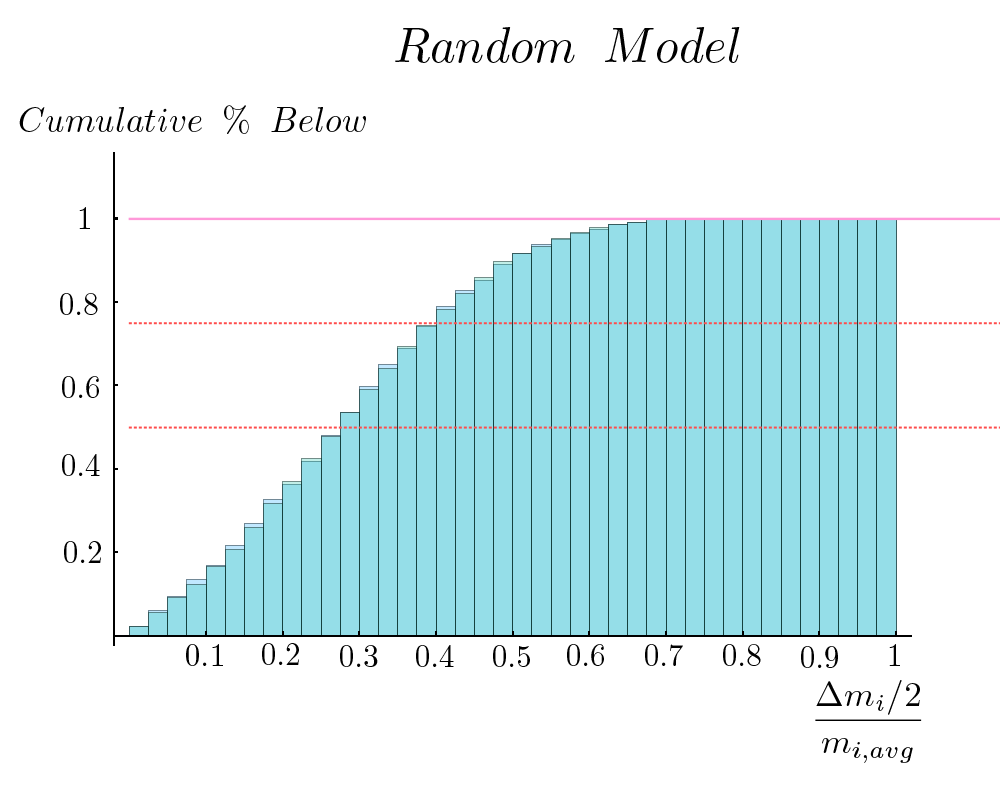}
\caption{The analogous histograms to those in Figure \ref{mass-gap-cdfs} for the random matrix model.}  
\label{random-cdfs}
\end{figure}

The scaled mass gap CDFs for the RMM also agree rather well with the actual data, and is shown in Figure \ref{random-cdfs}.
It is worth noting that without the step in the RMM procedure that eliminates draws that live outside the quarter circle allowed by the tadpole condition there is a noticeable overdensity of RMM axio-dilaton data points away from the diagonal in the analogous version of Figure \ref{axioRMMmasses}. The effect  on the complex structure's scatter plot of removing RMM tadpole condition is not perceptible. 

\subsection{Couplings}
The hierarchy present in the masses, due to the fact that our vacua accumulate where the Yukawa coupling is singular, persists through the third and fourth order couplings. Since the basis in which couplings ought to be reported factors into one half associated almost entirely with the complex structure and the other with the axio-dilaton, we naively expect each additional index associated with the former at the expense of the latter to involve the evaluation of increasingly more divergent terms near they singularity.

In particular, one expects four distinct scales to emerge among the distributions of third order coefficients, and five scales for the fourth order couplings. These correspond to the $3$-choose-$2$ and $4$-choose-$2$ ways one can differentiate the scalar potential. For instance, at third order we expect the $A_{i'j'k'}$ with all indices related to the complex structure (that is, equal to $1$ or $2$ in our convention for the basis of real canonically normalized fields) to be dominated by the term involving a derivative of the Yukawa coupling. The next highest scale expected would then be that with two complex structure and one axio-dilaton indices ($3$ or $4$ in our convention), followed by one complex structure and two axio-dilaton, and lastly that with all three axio-dilaton.  

First we establish that such a hierarchy of scales is realized, and then confirm the explanation in the preceding paragraph is valid by showing that the scale separation becomes ever more prominent as the vacuum-to-conifold distance diminishes. We then qualitatively investigate the correlations between the couplings at a given order, and indicate that they are \emph{not} the result of the coordinate transformation alone. We accomplish this with the use of another random construction. Specifically, we generate a set of random rank three symmetric tensors and transform each by the orthogonal matrices from the actual set of vacuum data. Though one might expect correlations to be built in by the special structure of the Hessian's eigenspace, the fact that none of the correlations present in the mirror quintic data is replicated by the random procedure indicates that they are not the consequence of diagonalization. 

\begin{figure}
\centering
\includegraphics[width=.9\textwidth]{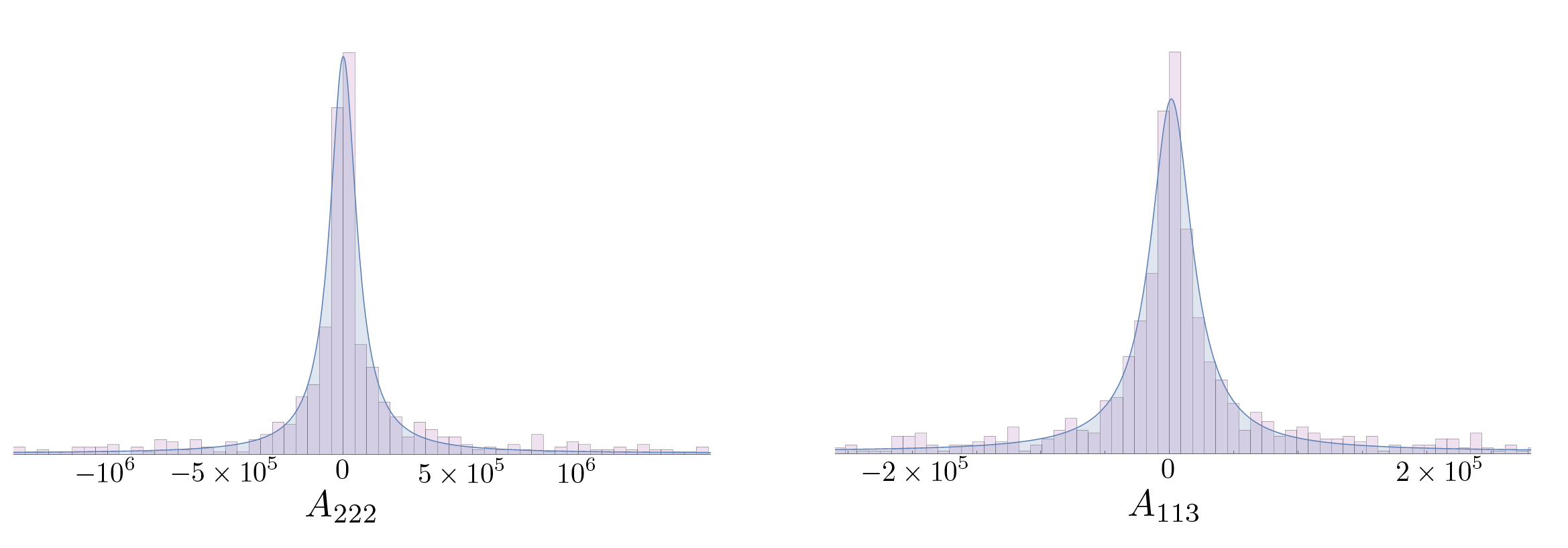}
\caption{A representative pair of distributions of the higher order couplings.}  
\label{cubic-distributions}
\end{figure}

Turning to the hierarchy among the magnitudes of the couplings, the scale separation can be shown visually by first imagining each of the entries in the third order couplings for a given vacuum, $A_{i'j'k'}$, as living in one of $64$ cells of a $4$-by-$4$-by-$4$ celled cube. We have one cube for each vacuum, and its entries take on positive or negative values (with equal likelihood, as indicated by the roughly Gaussian distributions centered at zero found for all coefficients. A representative sample of the histograms and estimated distributions can be found in Figure \ref{cubic-distributions}). Taking the labeling convention for the real and canonically normalized field coordinates defined in subsection \ref{taylor-coeffs-alg}, one $2$-by-$2$-by-$2$ subblock in, say, the front-bottom corner of this cube will involve all complex structure related indices. The subblock diagonally opposite it in the top-far corner will involve all axio-dilaton indices, and the two types of mixed index subblocks will live interspersed throughout the remaining $6$ off-diagonal subblocks.

\begin{figure}
\centering
   \begin{tabular}{cc}
   \includegraphics[width=.4\textwidth]{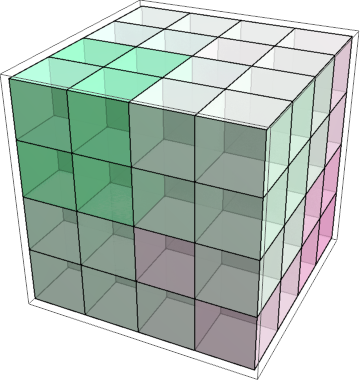}&\qquad\qquad\includegraphics[width=.4\textwidth]{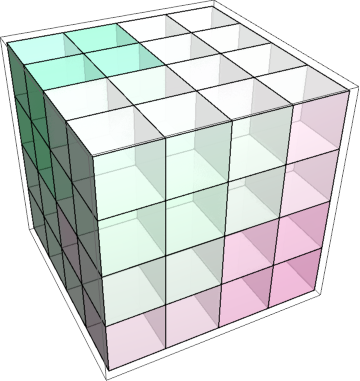}
    \end{tabular}
    \caption{The median across the ensemble of the magnitude of the transformed third order couplings, $A_{i'j'k'}$. Each cell represents one choice for the three indices. $i'=1$ or $2$ corresponds to the complex structure associated eigenvectors, $y_1$ and $y_2$, whereas $i'=3$ or $4$ corresponds to the axio-dilaton associated eigenvectors, $y_3$ and $y_4$. The scale is logarithmic with green representing the smallest median magnitude and pink representing the largest. The ``origin," so to speak, is located in the bottom right corner of the back face in the view of the cube in the left panel. The pink $2$-by-$2$-by-$2$ subcube in this corner contains the all-complex-structure subset of couplings since the indices are all either $1$ or $2$. Similarly the green corner diagonally opposite contains the all-axio-dilaton couplings. The four expected hierarchies based on the leading order behavior of the Yukawa coupling near the conifold can be seen by the partitioning of the cube into four types of subblocks each with cell colors in a different regime of the scale: pink, light pink/pale green, light green, and green. A view of the cube rotated about the vertical axis is shown on the right.}\label{cube}
   \end{figure}

Next, consider taking the magnitude of the value in each cell and then computing the median for each entry across the ensemble of cubes. The median is the more appropriate quantity because the distributions of the magnitudes are heavily skewed, just as the masses were. We may then represent the cube containing the ensemble's median values visually by coloring each cell according to a continuous scheme. The resulting hierarchy is, not surprisingly, best illustrated using a logarithmic scale. Two views of the resulting cube are shown in Figure \ref{cube}, with a color gradient of green to white to pink indicating smallest to largest. 
 
The cube arranges itself into the four 8-celled subcubes of different scale, which we've described. This is indicated by the green quadrant, which is flanked by much paler green (identical by symmetry) subcubes adjacent to it, the vibrant pink quadrant diagonally across from the green corner and lastly the (identical) subcubes with pale pink and green cells that share an edge with the pink corner. 

The smallest couplings (green) do indeed reside in the all-axio-dilaton subblock, which is located at the top left of the front face of the cube in the first view. The pink corner is in fact the all-complex structure subblock. Its neighboring subcubes -- those that share an edge with it (for instance those directly above and directly to the left of the pink corner in the front face of the second view) -- still have two complex structure indices because they are in the same 2-cell thick ``slice" of the cube, but have only one axio-dilaton index. The fact that the pale colors in these neighboring subcubes are pinker/less green than the pale subcubes that neighbor the green axio-dilaton corner means the $\xi$-$\xi$-$\sigma$ couplings are larger than the other mixed index cubic couplings, $\xi$-$\sigma$-$\sigma$. 
   
\begin{figure}
\centering
   \begin{tabular}{cc}
   \includegraphics[width=.25\textwidth]{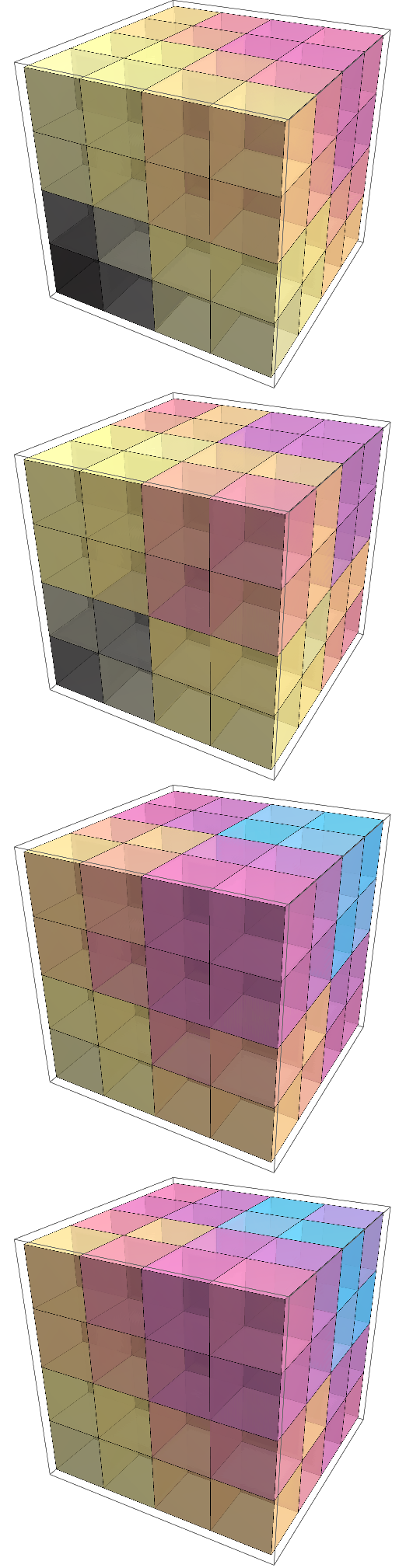}&\qquad\qquad\qquad\includegraphics[width=.25\textwidth]{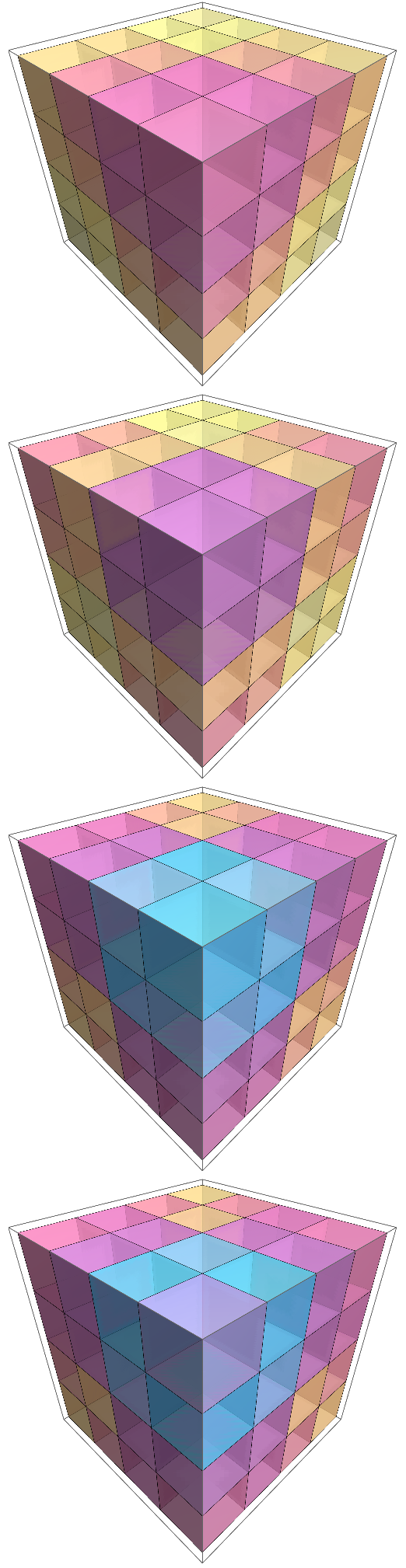}
    \end{tabular}
    \caption{The analogous data as that displayed in Figure \ref{cube} for the quartic couplings $A_{i'j'k'l'}$ with a logarithmic scale for the magnitudes represented with a CMYK color scheme (black being the largest, followed by yellow, magenta then cyan). The five scales expected due to the behavior of the Yukawa coupling near the conifold manifest themselves as the five different types of subcubes -- those with cells in the black, yellow, orange/pink, pink/purple, and blue. The origin of each of the four cubes in the stack is at the bottom left of the front face in the view on the left, making the all-complex-structure-couplings contained in the black corners of the top pair of cubes. The panel on the right shows a view of the hypercube rotated about the vertical axis, with the all-axio-dilaton couplings in the top front corner of the bottom pair of cubes in the stack, which are blue as expected.}
    \label{stack}
   \end{figure}

The hierarchy among the quartic couplings can be visualized in much the same way, only with a stack of four 64 celled cubes instead of a single one. We show two views of this hypercube in Figure \ref{stack}. The blocks are arranged top to bottom according to the first index, $i'$, in $A_{i'j'k'l'}$; the top having $i'=1$ and bottom having $i'=4$. The color scheme here is CMYK with cyan/blue representing the smallest magnitude, followed by purple, magenta, orange, yellow, gray and finally black indicating the largest. Notice that each cube in the stack partitions itself into quadrants of four distinct scales (just like the single cube of third order couplings). 

The top pair of blocks then has one additional index associated with the canonical complex structure coordinate, relative to the bottom pair with the canonical axio-dilaton. The largest magnitudes (the blackest cells) do in fact fill in the $\xi$-$\xi$-$\xi$ subcube of the top pair of blocks. These are the all-complex-structure quartic couplings. These corner subcubes each share an edge with (identical) yellow subcubes. Since neighboring subcubes differ by one index type these neighbors contain couplings with three complex structure indices and one axio-dilaton. The fact that it is yellow means $\xi$-$\xi$-$\xi$-$\sigma$ couplings rank second largest. Across from the black corner subcubes but within the same 2-cell thick slice we have the couplings that involve one more $\sigma$ in place of $\xi$, the $\xi$-$\xi$-$\sigma$-$\sigma$ couplings. The fact that they are orange indicates they are the third largest scale.

The remaining two scales in the hierarchy are displayed by the purple corners of the top pair of blocks in the stack and the blue corner cubes that are only present in the bottom two blocks in the stack. The purple corners of the top pair of blocks blocks do in fact lie diagonally opposite the black corners, making them cells containing $\xi$-$\sigma$-$\sigma$-$\sigma$ couplings. The subcube located in this same top back corner position in the bottom pair of blocks in the stack differs from the purple ones of the preceding top pair in the stack by the first index, making them the all-axio-dilaton couplings. The cells are indeed cyan/blue, making these couplings the smallest in scale. 

Now that we've confirmed the existence of the naively expected hierarchies we turn to their source -- the proximity to the conifold point. A priori it is possible that the divergent contributions due to the Yukawa coupling and its derivatives could have been tempered by some other mechanism as the conifold is approached. It is also important to assess the degree of variation in the expected scale separation. Just as the vev of the superpotential played the role of a random element complicating an otherwise clean analytic dependence on the conifold distance, here too we will have a layer of noise atop the signal. The significance of this noise, and importantly the degree to which it changes as the conifold is approached, is not obvious at the outset.

\begin{figure}
\centering
\includegraphics[width=\textwidth]{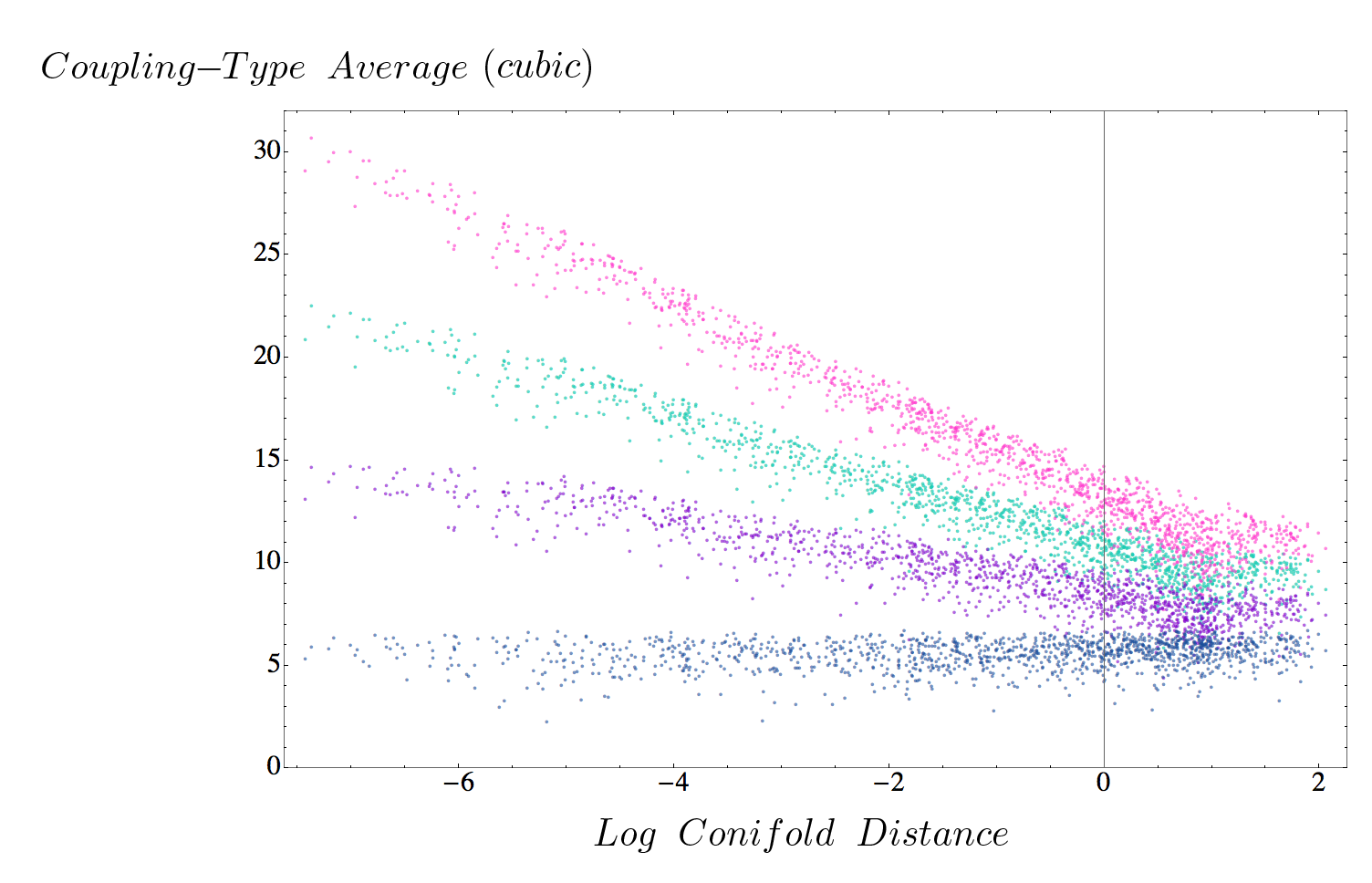}
\caption{A log-log scatter plot of the vacuum coupling subcube average magnitude versus conifold distance. Each subcube contains couplings of one of the four types: all-$\xi$ (shown in pink), $\xi$-$\xi$-$\sigma$ (green), $\xi$-$\xi$-$\sigma$ (purple), and all-$\sigma$ (navy). The fact that the data points organize themselves into approximately linear bands with increasingly negative slope for each $\xi$ at the expense of a $\sigma$ confirms the Yukawa coupling (through its successively more singular partial derivatives) is responsible for the hierarchy observed. The width of individual bands signals the presence/role of the random element, the vev of the superpotentail.}  
\label{cubic-log-log}
\end{figure}

We show the conifold distance dependence of the cubic couplings' scale separation visually as well. For each vacuum's four independent 8-cell subcubes we first compute each subcube's mean magnitude. The mean is the appropriate measure here since the entries in a single subcube for \emph{an individual} vacuum \emph{are} comparable. Each vacuum then has a list of four positive values-- the average magnitude of each of the four type of cubic couplings. We take the logarithm of each value in the list, as well that of the magnitude of the canonical vacuum coordinate, $|\xi_{vac}|$. The ensemble data for all four types of cubic couplings are displayed in the single log-log scatter plot in Figure \ref{cubic-log-log}, with different colors used for each of the four types.

Notice first that the colors separate into four approximately linear bands with negative slope. This indicates that each type of coupling has an inverse power law dependence on the canonical vacuum coordinate, $|\xi_{vac}|$. The data points with most negative slope, the pink band, are the ensemble of all-complex-structure cubic couplings. Each pink point is a different vacuum's mean $\xi$-$\xi$-$\xi$--type coupling magnitude. Below this band lies the second largest scale in the cubic couplings involving two $\xi$ and one $\sigma$, shown in teal, followed by purple and navy blue for the $\sigma$-$\sigma$-$\xi$ and the all-axio-dilaton couplings, respectively. The fact that the bands are approximately linear reflects the domination of the leading order term in the Yukawa coupling and derivatives thereof over other terms in the expressions for the cubic couplings. 

\begin{figure}
\centering
\includegraphics[width=\textwidth]{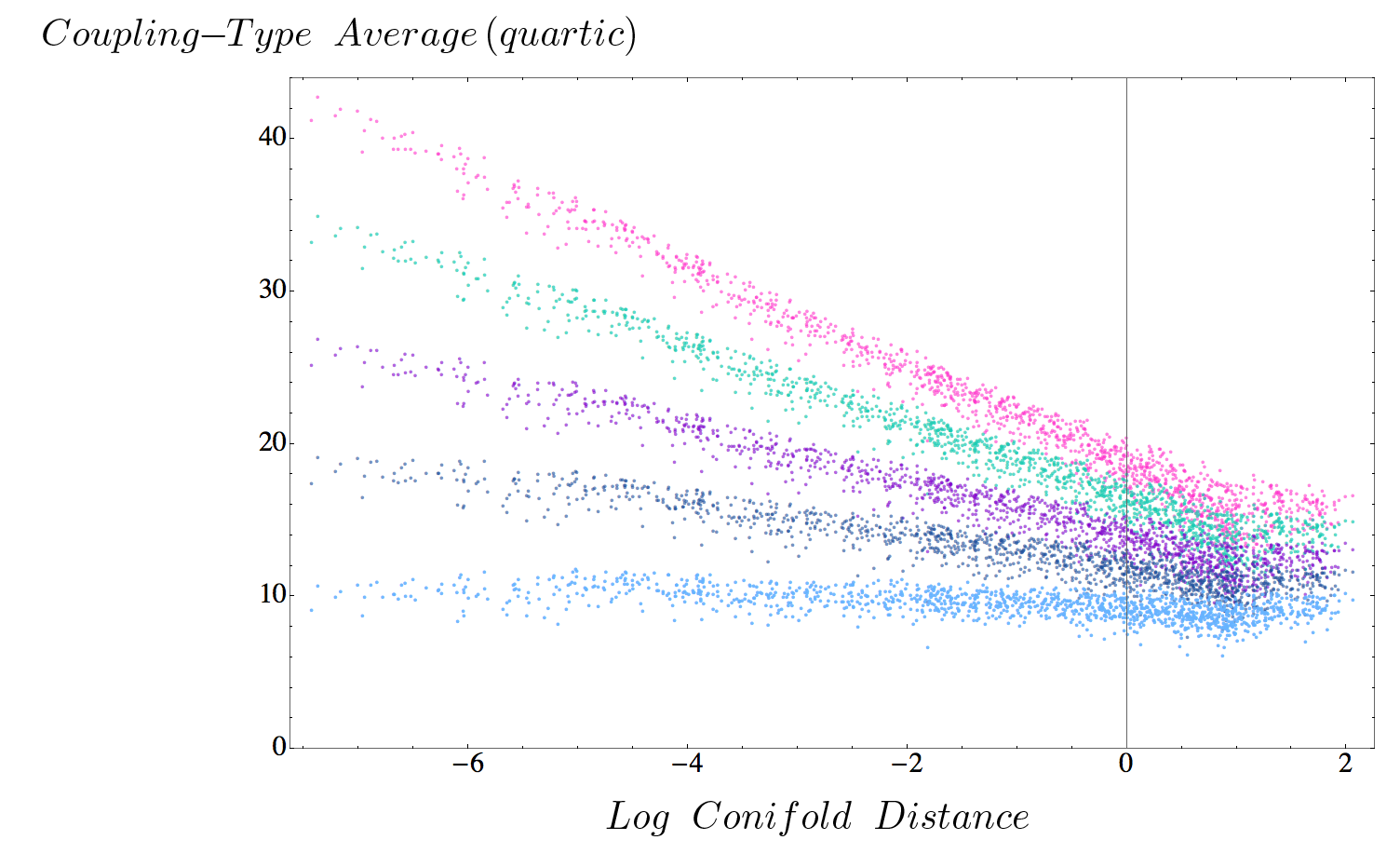}
\caption{The analogous plot as that in Figure \ref{cubic-log-log} for the quartic couplings. The colors are ordered in the same manner according to the number of $\xi$'s in the coupling-type, descending from all-$\xi$ (pink), with the addition of light-blue for the last of the five types, all-$\sigma$. The self organization of the vacuum data into the approximately linear bands of increasingly negative slope for every $\xi$ at the expense of a $\sigma$ confirms the validity of our explanation of the hierarchies based on the behavior of the Yukawa coupling near the conifold.}  
\label{quartic-log-log}
\end{figure}

The same analysis can be performed for the quartic couplings. We show the resulting log-log scatter plot for the five types of couplings in Figure \ref{quartic-log-log}, with the  same coloring scheme descending from the largest in pink ($\xi$-$\xi$-$\xi$-$\xi$--type), and the addition of a fifth color, light-blue, for the smallest (all-axio-dilaton type). The same reasoning indicates that the source of the hierarchy among the quartic couplings are the terms involving the most $\xi$ derivatives of Yukawa coupling evaluated near the conifold point. For both the cubic and the quartic scatter plots we may view the statistical variation within a given band as being supplied by the random element, the vev of the superpotential.
\begin{figure}
\centering
\includegraphics[width=\textwidth]{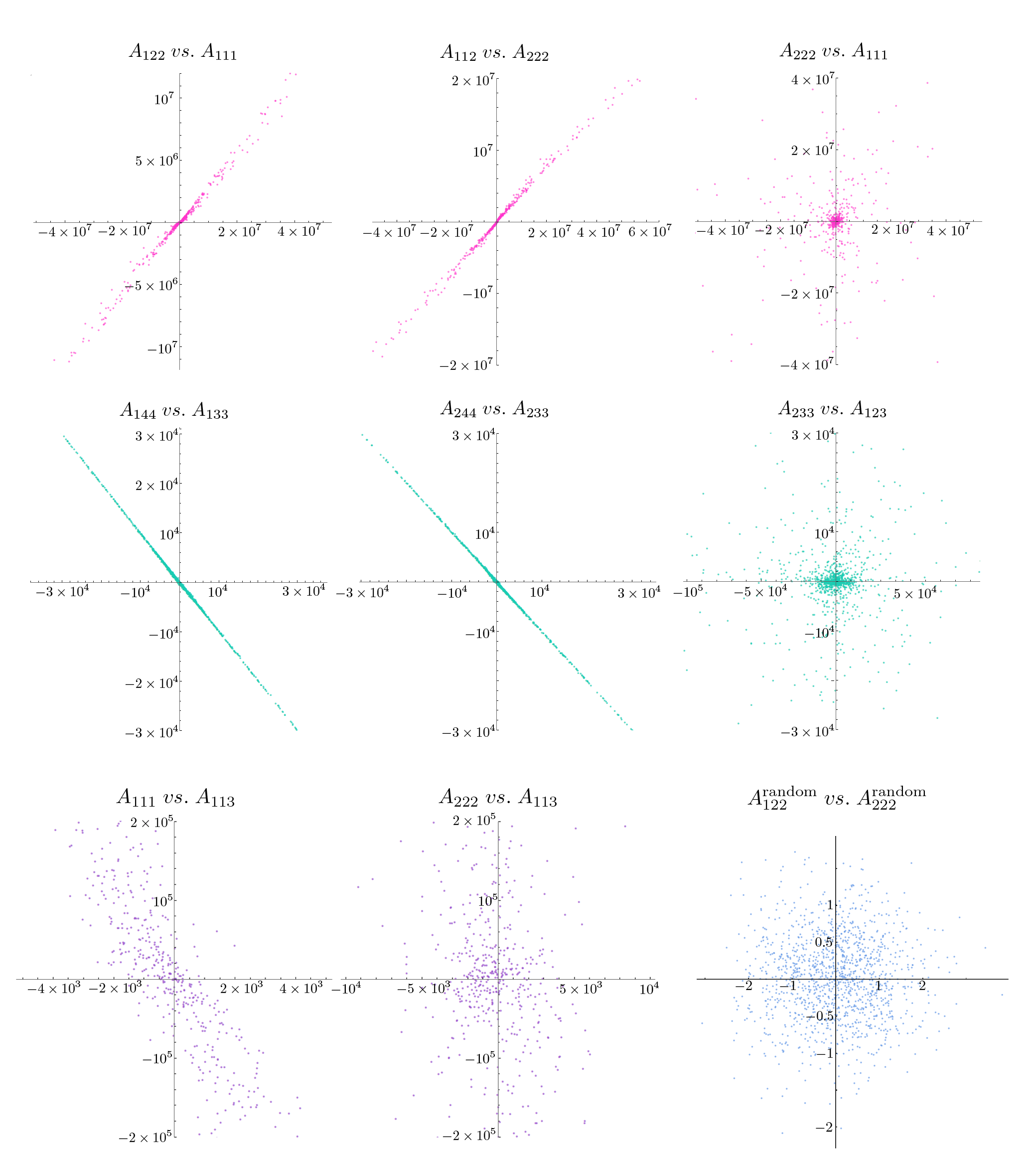}
\caption{A representative sample of the scatter plots of pairs of cubic and quartic couplings from the vacuum data  (pink, green, and purple), as well as from the random matrix model couplings (blue) designed as a diagnostic. Note that whereas the vacuum data exhibits sharply defined correlations between certain pairs of couplings, all the random matrix model pairs do not. This indicates that correlations are not merely built in by the diagnoalization of the Hessian in canonical coordinates.}  
\label{cubic-correlations}
\end{figure}
   
We conclude with a qualitative discussion of the remaining aspect of the structure among the couplings that is not captured by a Random Matrix Model, for example that of \cite{tumbling}. These are the pattern of nontrivial correlations we find between couplings. That is, the ensemble of $A_{i'j'k'}$ and $A_{i'j'k'l'}$ are \emph{not} accurately modeled by totally symmetric tensors whose entries are drawn separately from independent distributions. We've seen that the distribution of a particular cubic or quartic coupling is roughly Gaussian and is centered at zero. The hierarchies discussed mean that the spread of these distributions differ in scale, according to index type. For instance the $A_{112}$ distribution is comparable to $A_{222}$ in this regard, but not to, say, $A_{113}$, whose spread is smaller by comparison. 

The hierarchies and the non-flat distributions themselves need not have come with correlations between couplings. The fact that we find approximately linear scatter plots between particular pairs of couplings renders a random approach involving independent distributions -- uniform or otherwise -- a poor approximation to the actual coefficients. A representative sample of the nontrivial correlations for the cubic coupling data sets are shown eight of the nine panels in Figure \ref{cubic-correlations}, excluding that in the bottom right corner (in blue). 

The pink plots on the top row show that while the pairs $A_{111}$ with $A_{122}$, and $A_{112}$ with $A_{222}$ have an approximately constant ratio across the ensemble of vacua, there is no relationship between $A_{111}$ and $A_{222}$. We also find correlations in the couplings of medium scale, namely those that mix moduli type. For instance, $A_{144}$ and $A_{133}$ are approximately equal in magnitude, but opposite in sign, across models. This is shown in the teal plots in the middle row.

A reasonable hypothesis for the source of these correlations is the transformation performed to the field coordinates that simultaneously diagonalize the Hessian and canonically normalizes the kinetic terms. This seemingly mundane step in the processing of the raw coupling data might be suspected as being nontrivial at the level of correlations because of the special structure of the Hessian's eigenspace. We test this hypothesis by comparing the results of a modified Random Matrix Model designed entirely as a diagnostic for this purpose.

If it is the case that the transformation from the original noncanonical complex coordinates builds in the patterns of correlations we observe, then an ensemble of real and totally symmetric tensors with i.i.d. entries acted upon by the orthogonal transformation $O$ (defined in subsection \ref{calc-approach}) ought to exhibit correlations. Since we have 1358 $O$ matrices, we build the same number of random rank-3 tensors and perform the transformation, 
\begin{equation}
A^{rand}_{ijk}\rightarrow O^i_{i'} O^j_{j'} O^k_{k'}A^{rand}_{ijk}.
\end{equation}
The result is that the transformed random couplings are uncorrelated. We've included a single scatter plot of these RMM couplings as a representative example. This is the ninth panel in Figure \ref{cubic-correlations}.
\section{Discussion}\label{Discussion}

The initial expectation that string theory would result in a unique, or nearly unique, vacuum state whose low energy excitations would explain the familiar properties of particle physics has not been borne out by developments over the past few decades. Instead, a wealth of discoveries have revealed an ever greater abundance of mathematically consistent vacua, without any allied developments that single out one (or perhaps a few) such vacua as physically relevant. Because of this, significant attention has shifted to statistical properties of  these vacua and, more generally, to statistical properties of the easier to analyze surrogate, random field theories in high dimensional moduli spaces. 

In this paper, we have investigated the degree to which this latter surrogate faithfully models the space of low energy field theories arising from string compactifications. We reviewed arguments which suggest the relevance of random field theories--namely, the randomizing effects of arbitrary fluxes coupled with the broad spectrum of vacuum locations in moduli space associated with each such flux choice. We then tested this argument by focusing our attention on one particular compactification of the type IIB string, the famous mirror to the quintic hypersurface. We identified a class of 1358 low energy flux models built on this compactification, computed the scalar potential for the canonically normalized scalar fields in each such model, and considered the statistical distributions of the renormalizable coefficients in the Taylor expansions of the potentials. We confirmed previously known results for the second order coefficients -- mass terms --  and went on to study the third and fourth order terms. Our main result is that we found significant deviations from a random collection of coefficients, as illustrated in Figures \ref{cube}, \ref{stack} and \ref{cubic-correlations}, showing that some of the rich structure inherent in type IIB supergravity survives the randomizing influence of flux compactifications. 

The lesson, then, is that one must exercise care when invoking random field theories as a model for the space of low energy compactified string dynamics. More particularly, our results, and generalizations thereof to higher dimensional moduli spaces, provide a sharper ensemble for accurate statistical modeling of the features of low energy string theory.

Going forward, these results suggest a number of research directions. For ease of computation we have focused on a Calabi-Yau compactification with a single complex structure modulus. One would like to acquire an understanding of the distributions we have studied in more generic cases with higher dimensional moduli spaces. Explicit analysis of the sort we've undertaken here would be difficult. However, in the vicinity of a conifold locus -- where vacua generally accumulate -- we've reduced the statistical dependence to the three dominant control parameters introduced earlier. These each have natural higher dimensional generalizations and so it would be of interest to see if we can gain insight into more general Calabi-Yau compactificaitons guided by the results we found here, and thus avoiding direct calculation. We hope to return to this shortly. It would also be interesting to revisit the works \cite{tumbling} which have investigated the quantum stability of vacua in random high dimensional scalar field theories as a surrogate for the stability of the string landscape. Are those results modified by studying a collection of field theories whose distribution more closely aligns with that of low energy dynamics of string theory? We intend to return to this question as well.

\acknowledgments
It is a pleasure to thank Thomas Bachlechner, David Kagan, Ruben Monten, and especially Frederik Denef, for numerous insightful conversations and observations, which greatly assisted the completion of the research reported herein. We also gratefully acknowledge the support of the Department of Energy through grant DE-FG02-92-ER40699.

\appendix
\section{Near Conifold Period Expansion Coefficients}
\label{appendix}

Recall that our integral and symplectic basis for mirror quintic's period functions are denoted $\Pi_i$, with $i=0$ and $3$ an intersecting pair, and $i=1$ and $2$ the other. $\Pi_0$ is the only non-analytic period at the conifold point. It's partner, $\Pi_3$, is the period obtained by integrating the holomorphic $3$-form over the cycle that collapses. This period is nevertheless well-behaved (it simply vanishes at the conifold). The two periods associated with the other intersecting pair of cycles are also analytic. These are nonvanishing. The following expansions about the conifold point at $z=1$ hold,
\begin{align}
\Pi_1(z)&=\sum_{n=0}^{q}b_n (z-1)^n\\
\Pi_2(z)&=\sum_{n=0}^{q}c_n (z-1)^n\\
\Pi_3(z)&=\sum_{n=1}^{q}d_n (z-1)^n.
\end{align}
The values for the coefficients were computed in Mathematica by evaluating derivatives of the expressions for the $\Pi_i$ in terms of the Meijer-G functions (the $U_i$) at the conifold. We used $40$ digit accuracy in these computations. The values are listed in Table \ref{period-coeffs-123}.

\begin{table}[h!]
  \centering
  \begin{tabular}{|l|l|l|l|l|}
    \hline
 $n$ & $b_n$ & $c_n$ & $d_n$\\
      \hline
 $0$&  $ +1.293574 i$ &$ 6.19502 - 7.11466 i$&$ 0$\\
 $1$&  $ -0.150767i$ &$-1.016605 + 0.829217  i $&$ -0.355881i$\\
  $2$&  $ +0.0777445i$ &$0.570733 - 0.427595 i $&$ 0.249117i$\\
   $3$&  $ -0.0522815 i$ &$-0.401804 + 0.287548 i $&$ -0.194548i$\\
    $4$&  $+0.0393684  i $ &$0.312044 - 0.216526  i $&$  0.161285i$\\
     $5$&  $-0.0315669i $ &$-0.256050 + 0.173618 i $&$-0.138686i$\\
      $6$&  $+0.0263447 i $ &$0.217649 - 0.144896  i $&$0.122217 i$\\
       $7$&  $ -0.0226046 i$ &$-0.189607 + 0.124325 i $&$-0.109620i$\\
        $8$&  $0.0197941i $ &$0.168193 - 0.108868 i $&$ 0.0996353i$\\
    \hline
  \end{tabular}
      \caption{Expansion coefficients for period functions $\Pi_1$, $\Pi_2$ and $\Pi_3$.}
      \label{period-coeffs-123}
\end{table}

The remaining period, $\Pi_0$, is multiple-valued at the conifold point. The cycle it is associated with picks up one copy of its vanishing partner for each revolution around the conifold in moduli space. This fixes the form of $\Pi_0$ to \ref{Pi0expansion}. To match the branch cuts of the logarithm in the expansion with that of the relevant Mejer-G ($U_0$) in Mathematica we must negate the argument of the logarithm. Ultimately the expression we write for $\Pi_0$ is,
\begin{equation}
\Pi_0(z)=\Pi_3(z)\left(\frac{\log(-(z-1))}{2\pi i}-\frac{1}{2}\right)+f(z)
\end{equation}
with $f(z)$ analytic. 

Its expansion coefficients, $a_n$, were computed using a recursion relation based on the fact $\Pi_0$ satisfies the Picard-Fuchs equation. This is discussed at length in section \ref{periods-subsection}. Here we simply tabulate the resulting values. The first three $a_n$ are needed by the recursion to generate the rest. They are obtained numerically, as discussed in section \ref{periods-subsection}. All coefficients were calculated using 30--40 digit accuracy computations and are listed in Table \ref{an-coeffs}. 

\begin{table}[h!]
  \centering
  \begin{tabular}{|l|l|}
    \hline
   $n$ & $a_n$\\
       \hline
      $0$ & $1.07073$\\
      $1$ & $0.024708 - 0.177941i$\\
       $2$ & $-0.0115108 + 0.1245584  i$\\
        $3$ & $0.0065650 - 0.0972742i$\\
         $4$ & $-0.0042768 + 0.0806427 i$\\
          $5$ & $ 0.0030290 - 0.0693428  i$\\
           $6$ & $-0.0022701 + 0.0611087  i$\\
            $7$ & $0.0017719 - 0.0548102  i$\\ 
             $8$ & $ -0.0014261 + 0.0498177i$\\
    \hline
  \end{tabular}
    \caption{Expansion coefficients for analytic contribution to $\Pi_0$}
    \label{an-coeffs}
\end{table}
\end{document}